\documentclass[a4paper,twocolumn,10pt]{article} 

\usepackage[T2A]{fontenc}
\usepackage{amsmath}
\usepackage{mathtools}
\usepackage{amssymb}
\usepackage{bm}
\usepackage{graphicx}
\usepackage[dvipsnames]{xcolor}
\usepackage{array}
\usepackage{appendix}
\usepackage{authblk}
\usepackage{tabularx}
\usepackage[margin=1in]{geometry}
\usepackage{cite}
\usepackage{url}
\usepackage{microtype}
\usepackage{hyperref}
\hypersetup{
    colorlinks=true,
    pdfkeywords={spectroscopy, optical frequency standard, theory, vapor cell, Cesium, Cs, Rb, Rubidium},
    pdftitle={Theory of Sub-Doppler Spectroscopy},
    pdflang={en},
    pdfpagemode=FullScreen,
    linkcolor=Red,       % Internal links (sections, pages, etc.)
    citecolor=ForestGreen,   % Citations
    filecolor=magenta,  % Local files
    urlcolor=cyan       % External web links
}
\renewenvironment{abstract}{%
  \small
  \begin{center}%
    {\bfseries \abstractname\vspace{-0.5em}\vspace{0pt}}%
  \end{center}%
  \quotation}
  {\endquotation}
\newcolumntype{Y}{>{\centering\arraybackslash}X}
\newcommand{\SixJ}[6]{\begin{Bmatrix}\! #1 \!\! & \!\! #2 \!\! & \!\! #3 \!
                                    \\ \!#4 \!\! & \!\! #5 \!\! & \!\! #6\!\end{Bmatrix}}
\newcommand{\TreJ}[6]{\begin{pmatrix}\! #1 \!\! & \!\! #2 \!\! & \!\! #3 \!
                                    \\#4 \!\! & \!\! #5 \!\! & \!\! #6 \!\end{pmatrix}}                                    
\newcommand{\xiB}{\boldsymbol{\xi}}

\newcommand\bra[2][]{#1\langle {#2} #1\rvert}
\newcommand\ket[2][]{#1\lvert {#2} #1\rangle}

\renewcommand{\epsilon}{\varepsilon}
\begin{document}

\title{Theory of Sub-Doppler Spectroscopy of Alkali Metal Vapors in the Field of Counterpropagating Bichromatic Laser Beams: Probe Beam Regime} 

\author[1]{A. M. Mikhailov}
\author[1, *]{D. V. Brazhnikov}

\affil[1]{Institute of Laser Physics, Siberian Branch, Russian Academy of Sciences, Novosibirsk, 630090 Russia}
\affil[*]{e-mail: x-kvant@mail.ru}
\date{}

\twocolumn[
\maketitle
\begin{abstract}
Sub-Doppler resonances observed in miniature cells (V $\ll$1 cm$^3$) with Rb or Cs atoms in the field
of counterpropagating bichromatic laser beams are of interest for the creation of a miniature quantum frequency standard (QFS) of the optical range.
However, a theory of these resonances that takes into full account the real structure of energy levels in an atom has not yet been constructed, which would make it possible to obtain explicit analytical expressions for the absorption coefficient of the medium.
This paper addresses this gap using the density matrix formalism and the algebra of irreducible tensor operators.
Namely, we obtained an analytical expression for the absorption coefficient of the probe beam in the presence of a more intense counterpropagating beam (pump–probe regime).
Moreover, the calculations were carried out for arbitrary parameters of laser beam polarization, arbitrary angular momenta of optical dipole transitions in the atom and various relaxation processes.
The constructed theory allows us to correctly take into account the simultaneous action of various nonlinear optical effects, in contrast to the simplified three-level model of the atom, and also to investigate the partial contribution of each of them to the overall observed resonant circuit.
The analysis performed, in particular, made it possible to predict an increase in the amplitude of the sub-Doppler resonance when using counterpropagating light beams with opposite circular polarizations in relation to the configuration of counterpropagating beams with orthogonal linear polarizations, which is often used in practice.
\end{abstract}
]

\section{Introduction}\label{sec:1}

Laser spectroscopy (LS) of atoms and molecules is one of the main methods for studying the structure of matter and its interaction with physical fields \cite{Letokhov1977,Thompson1985,Rautian1991}.
LS is widely used in such highly demanded quantum devices as frequency and time standards \cite{Letokhov1977,Bagaev1986,Bagayev1997,Riehle2004}, as well as various quantum sensors: atomic magnetometers of ultra-weak \cite{Alexandrov2003,Budker2013,Petrenko2021,Makarov2025} and strong  \cite{Auzinsh2022,Sargsyan2022,Sargsyan2025} fields, sensors of inertial forces \cite{Geiger2020} and electric fields \cite{Bason2010,Ryabtsev2024}.
LS methods are used in various precision experiments in the field of fundamental physics: the search for dark matter \cite{Derevianko2014}, the search for the permanent electric dipole moment of the neutron \cite{Abel2020}, and also for measuring other subtle effects predicted by the theory of relativity \cite{Bagaev1972,Takamoto2020} and quantum field theory \cite{Hansch1975}.

Thermal motion of particles in atomic and molecular gases leads to significant Doppler broadening of the observed resonances, which degrades the accuracy of spectroscopic studies.
This broadening also does not allow achieving high stability of quantum frequency standards (QFS) in the optical range, which are in demand for solving many applied and fundamental problems.
With the development of laser technology, various nonlinear LS methods \cite{Letokhov1977,Rautian1991,Thompson1985} have begun to be actively developed, allowing for an increase in the spectral resolution of optical transitions in atoms and molecules by several orders of magnitude and a significant increase in the stability of optical QFS \cite{Bagaev1986,Bagayev1997,Riehle2004}.

In addition to large-scale laboratory QFSs, which possess extremely high stability and precision, miniature QFS variants are currently generating great interest. 
These can be used in mobile equipment, on satellites, and as part of other compact quantum sensors as highly stable laser sources.
However, in the optical region of the spectrum, it has long been impossible to achieve a degree of miniaturization of QFS comparable to that of miniature microwave-range QFS based on the phenomenon of coherent population trapping (CPT) \cite{Alzetta1976,Smirnov1989,Arimondo1996}.
Advanced examples of such microwave QFS have stability characterized by the Allan deviation ($\sigma_y$), near $10^{-11}$ for $1$~s of averaging and $\approx\,$$10^{-12}$ for $24$~h \cite{Zhang2019,Skvortsov2020}.

The main problem of miniaturization of QFS in the optical range was associated with the choice of the optimal spectroscopic method for excitation of nonlinear resonances and the search for suitable optical transitions in atoms or molecules that would allow achieving a high ratio of the resonance amplitude to its width in relation to miniature gas cells ($V$$\,\ll\,$$1$~cm$^3$).
For many years, attempts have been made to create miniature optical QFS (see, for example, \cite{Knappe2007,Gazazyan2007,Todorov2012}).
However, only in recent years has significant progress been made in this direction.

Currently, three approaches are being developed most successfully, each of which uses counterpropagating laser beams to excite nonlinear resonances in alkali metal vapors.
The first method is based on the “classical” and well-studied approach using saturated absorption resonance (SAR) in $^{133}$Cs vapor \cite{Klinger2024,Klinger2025} or $^{87}$Rb vapor \cite{Loh2016,Dyer2023,Zhang2024,Yun2025}. 
The second method uses two-photon spectroscopy of $^{87}$Rb vapor \cite{Hummon2018,Maurice2020,Callego2025}. 
The third method, unlike the first two, uses dual-frequency laser radiation and can also be used for both cesium atoms
\cite{Gusching2021,Gusching2023} and rubidium atoms \cite{Zhao2021,Velichansky2025,Chuchelov2026}.
According to this method, the difference in optical frequencies should coincide with the hyperfine splitting frequency in the ground state of the alkali atom: $\Delta_g$$\,\approx\,$$2\pi$$\times\,$$9.19$ GHz for $^{133}$Cs and $\Delta_g$$\,\approx\,$$2\pi$$\times\,$$6.83$ GHz for $^{87}$Rb.

All of the above methods have been tested using miniature atomic vapor cells and have demonstrated, with some exceptions, similar and very high levels of optical frequency stability in the range $(2$$-$$4)$$\,\times\,$$10^{-13}$ for $1$~s. 
These results mean that any of these methods could form the basis of future miniature optical atomic clocks whose stability would be one to two orders of magnitude higher than that of the best microwave miniature atomic clocks based on CPT. 
It is important to note that some of the listed methods have already been implemented in a miniature form factor with the volume of the physical part of the whole $35$~cm$^3$ \cite{Maurice2020}.
Moreover, using solid-state microresonators \cite{Newman2019}, the frequency stability can be transferred from the optical spectrum to the radio frequency range while maintaining the miniature dimensions of the QFS.

From a theoretical point of view, the single-frequency saturated absorption method \cite{Haroche1972,Shirley1973,Thomas1977,Baklanov1988} and the two-photon spectroscopy method \cite{Vasilenko1970,Cognac1973,Baklanov1978} have been studied quite well by now (see reviews \cite{Schawlow1982,Thompson1985}, monographs \cite{Letokhov1977,Rautian1991} and references therein). 
In particular, the saturated absorption method, which is most closely related to the present work, was comprehensively investigated not only on the basis of a simplified two-level spectroscopic model of the atom, but also for the case of real dipole transitions, taking into account the degeneracy of energy levels along the projection of the total angular momentum of the atom.
This made it possible to study a number of new nonlinear effects, manifested, for example, in the form of a change in the sign of the sub-Doppler resonance \cite{Pappas1980,Schmidt1994,Bloch1996,Brazhnikov2016,Saprykin2021,Zhao2024,Yun2025}  or the asymmetry of the line shape \cite{Brazhnikov2009}. 
Also, non-standard methods of pumping atoms have been proposed for observing SAR, for example, in the configuration of orthogonal light beams \cite{Izmailov2017}.

High-contrast nonlinear resonances observed by the method of dual-frequency sub-Doppler spectroscopy (DFSDS) were discovered experimentally relatively recently  \cite{Hafiz2016}. 
Therefore, the theory of these resonances is not as well developed as for the other two spectroscopy methods mentioned above.
Indeed, to date, only numerical calculations have been performed for a simplified spectroscopic model of the atom based on a three-level ($\Lambda$) scheme \cite{Hafiz2017} and for real optical transitions in the Cs atom \cite{Brazhnikov2019}. 
Some features of nonlinear resonances were studied for a real level set in a qualitative analysis in \cite{Velichansky2025}, without resorting to explicit analytical solutions.
Due to the complexity of the problem, analytical solutions were obtained only for the $\Lambda$-scheme \cite{Mikhailov2021}, which, however, cannot take into account the mutual influence of many nonlinear effects that simultaneously occur in the case of real optical transitions and “compete” with each other.
With this work, we address this gap.

In the presented work, using the formalism of the density matrix and the mathematical apparatus of irreducible tensor operators, a theory of DFSDS of alkali metal atoms is constructed. 
The theory assumes the presence of spectral resolution of the hyperfine structure of levels in the excited state, which best corresponds to D$_1$-lines in $^{87}$Rb and $^{133}$Cs.
From a practical point of view, two main regimes of interaction of atoms with the field are of interest: the “pump–probe” regime (or simply the probe beam regime), when the absorption of a weak probe beam is studied in the presence of a more intense counterpropagating pump beam, and the standing wave regime, in which both beams have close intensities. In this paper, for the sake of brevity, we have considered only one of the two regimes, namely the probe beam regime.
Thus, using perturbation theory (PT), we obtained analytical expressions for the resonant absorption coefficient of a probe light beam, taking into account the motion of atoms in the gas.
In this case, arbitrary elliptical polarizations of light beams and arbitrary angular momenta of optical dipole transitions are considered.

The constructed theory allows us to isolate and analyze contributions from individual nonlinear effects
in the general resonant circuit, which enables us to establish their role in the observation of sub-Doppler
resonances.
The obtained expressions are written in an invariant form, independent of the choice of coordinate system.
In addition, they can also be used to describe sub-Doppler resonances in the field of counterpropagating single-frequency light beams (with the corresponding parameters set to zero).
All this makes the obtained results quite universal.
 
\section{Statement of the Problem}\label{sec:2}

\subsection{Density Matrix Formalism}

\begin{figure}[!t]
    \centering
    \includegraphics[width=1\linewidth]{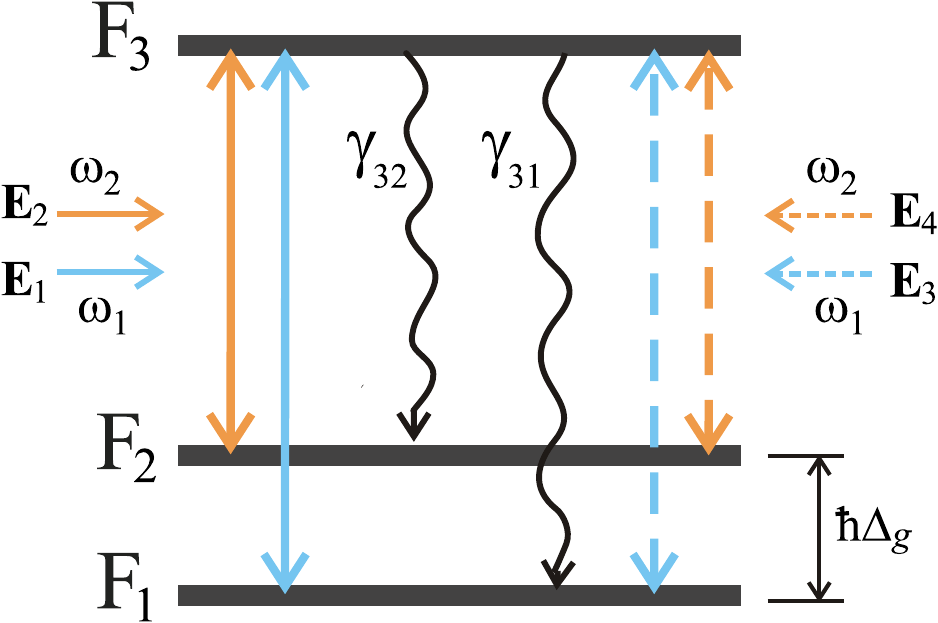}
    \caption{Optical dipole transitions in D$_1$-lines of alkali metal atoms.
    The bidirectional straight arrows indicate forced transitions under the action of laser radiation: solid ones – for waves running along the $z$ axis ,dashed—for waves running against the $z$ axis.
    The wavy arrows indicate radiative transitions as a result of spontaneous relaxation with the corresponding constants $\gamma_{31}$ and $\gamma_{32}$. $F_{1,2}$ is total angular momenta of the atom in the ground state (two hyperfine components), $F_3$ are the total angular momentum of an atom in an excited state (one of the hyperfine components).
    $\Delta_g$ is the hyperfine splitting frequency.
    For simplicity, the magnetic (Zeeman) energy sublevels are not shown in the figure.}
    \label{fig:1}
\end{figure}

Let us consider a typical laser field configuration often used in saturated absorption spectroscopy \cite{Letokhov1977,Rautian1991}.
It consists of two light beams propagating in opposite directions (Fig. \ref{fig:1}).
We consider one of the beams (probe) to be noticeably weaker in intensity than the other. 
Experimental methods for forming various configurations of the light field for conducting DFSDS were considered in \cite{Brazhnikov2020}.

The key difference between the problem considered here and the standard formulation of the problem in the SAR method is that each of the counterpropagating beams contains two mutually coherent monochromatic waves with frequencies $\omega_1$ and $\omega_2$. 
In experiments, such a field can be obtained in various ways. 
In most works employing the DFSDS method \cite{Hafiz2016,Gusching2021,Gusching2023,Brazhnikov2020,Zhao2021,Brazhnikov2021}, such a field was generated using an electro-optical modulator (EOM).
There are also works in which the required spectral composition of radiation was obtained by direct microwave modulation of the current of an external-cavity diode laser (ECDL) \cite{Velichansky2025,Zibrov2020}, double modulation—first of the pump current of a vertical-cavity surface-emitting laser (VCSEL) and then amplitude modulation of its radiation using an EOM \cite{Vishnyakov2024}, or modulation of the pump current of a distributed Bragg reflector (DBR) diode laser \cite{Mesenzova2024}.
The latest work is interesting because it used a polarimetric technique for recording sub-Doppler resonance to eliminate the unwanted influence of nonresonant optical side frequencies in the emission spectrum and to increase the signal-to-noise ratio.

So, let’s consider a laser field, which generally consists of four monochromatic waves:
\begin{multline} \label{eq:field}
    {\bf E}(z,t) = {\bf E}_1(z,t) + {\bf E}_2(z,t) + {\bf E}_3(z,t) + {\bf E}_4(z,t) =\\
    =\xiB_{(1)} E_1 e^{i(k_1z-\omega_1t)} +\xiB_{(1)} E_2e^{i(k_2z-\omega_2t)} + \\
        +\xiB_{(2)} E_3 e^{-i(k_1z+\omega_1t+\phi_1)}+\xiB_{(2)} E_4 e^{-i(k_2z+\omega_2t+\phi_2)} +\\+ c.c.
\end{multline}
Here $E_{1,2}$ and $E_{3,4}$ is real amplitudes corresponding to waves running along and against the $z$ axis, respectively.
In this case, the beginning of the $z$ axis compatible with gas cell inlet window.
counterpropagating waves may have constant phases that differ from those of waves traveling along the $z$ axis. 
This is reflected by the presence of phases $\phi_1$ and $\phi_2$ in (\ref{eq:field}). 
To simplify the problem, we consider the amplitudes of monochromatic waves to be independent of the transverse coordinates ($x$, $y$), i.e. the beams have uniform intensity in the cross section.

\begin{figure}[!t]
    \centering
    \includegraphics[width=0.5\linewidth]{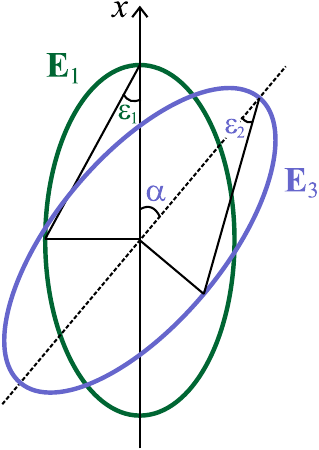}
    \caption{Parameterization of the polarization state of light using counterpropagating waves as an example ${\bf E}_1$ and ${\bf E}_3$.
    For another pair of counterpropagating waves (${\bf E}_2$ and ${\bf E}_4$) the polarization ellipses have the same location relative to the $x$ axis and the same ellipticity parameters $\epsilon_{1,2}$, but, in general, differ in size due to differences in wave amplitudes.}
    \label{fig:2}
\end{figure}

From (\ref{eq:field}) it follows that the waves ${\bf E}_{1,2}$ have the same polarization, which is described by the vector $\xiB_{(1)}$ while the polarization of waves ${\bf E}_{3,4}$, described by a vector $\xiB_{(2)}$. 
This is consistent with the experimental conditions for observing resonances using the DFSDS method.
Unit complex vectors $\xiB_{(1)}$ and $\xiB_{(2)}$ describe elliptical polarization ($\xiB_{(n)}$$\xiB_{(n)}^*$$\,=\,$$1$). 
Let us represent these vectors in a cyclic basis \cite{Varshalovich}:
\begin{equation}\label{eq:xi1}
\xiB_{(n)} = \sum_{q=0,\pm1}\xi_{(n)}^q {\bf e}_q\,,
\end{equation}
where $n$$\,=\,$$1,\,2$, while ${\bf e}_q$ -- covariant unit vectors of the cyclic basis associated with Cartesian relations: ${\bf e}_0$$\,=\,$${\bf e}_z$, ${\bf e}_{\pm1}$$\,=\,$$\mp$$\bigl($${\bf e}_x$$\pm$$i{\bf e}_y$$\bigr)$$/\surd2$.
In particular, when combining the main axis of the polarization ellipse of one of the beams (namely, for waves ${\bf E}_1$ and ${\bf E_2}$) with a Cartesian $x$ axis  (Fig. \ref{fig:2}), the contravariant components $\xi_{(n)}^q$ of the polarization vectors can be written as:
\begin{equation}\label{eq:xi2}
    \begin{aligned}
        &\xi_{(1)}^{-1} = -\sin{\left(\epsilon_1 - \frac{\pi}{4}\right)}\,,\\
        &\xi_{(1)}^{+1} = -\cos{\left(\epsilon_1 - \frac{\pi}{4}\right)}\,,\\
        &\xi_{(2)}^{-1} = -\sin{\left(\epsilon_2 - \frac{\pi}{4}\right)}\,e^{i\alpha}\,,\\
        &\xi_{(2)}^{+1} = -\cos{\left(\epsilon_2 - \frac{\pi}{4}\right)}\,e^{-i\alpha}\,.
    \end{aligned}
\end{equation}

The main axis of the wave polarization ellipses ${\bf E}_3$ and ${\bf E}_4$ in general, it is inclined at an angle $\alpha$ relative to the $x$ axis, which leads to the appearance of multipliers $e^{\pm i\alpha}$ in (\ref{eq:xi2}).
Note that since the absorption indices of the medium, which are scalar quantities, do not depend on the choice of coordinate system, then at an angle $\alpha$ one should also understand the angle between the main axes of the polarization ellipses (without reference to the direction of the $x$ axis ). Ellipticity parameters $\epsilon_{1,2}$ are given in the interval:  $-\pi/4\,$$\leq\,$$\epsilon$$\,\leq$$\,\pi/4$, while $\epsilon$$\,=\,$$0$ corresponds to the linear polarization of radiation, and $\epsilon$$\,=\,$$-\pi/4$ and $\pi/4$ correspond to left- and right-circular polarization, respectively. 
From (\ref{eq:xi1}) and (\ref{eq:xi2}) it is easy to show that the scalar product $\xiB_{(n)}$$\xiB_{(n)}$$\,=\,$$\cos(2\epsilon_{(n)})$ is an intrinsic characteristic of light polarization, independent of the coordinate system.

Our theory is based on the standard quantum mechanical formalism of the monatomic density matrix $\hat{\rho}$.
It obeys the Lindblad-type equation of motion \cite{Rautian1991,Blum2012}:
\begin{equation} \label{eq:DM1}
    \left(\frac{\partial }{\partial t}+\upsilon\frac{\partial}{\partial z}\right)\hat{\rho} = -\frac{i}{\hbar}\left[ \left( \widehat{H}_0+\widehat{V} \right), \hat{\rho} \right]+\widehat{\mathfrak{R}}\left\{\hat{\rho}\right\}.
\end{equation}
Here $\upsilon$ is projection of the velocity of the atom onto the quantization axis $z$, $\widehat{H}_0$ -- part of the general Hamiltonian describing the state of a free atom (in the absence of a laser field), $\widehat{V}$ -- the part of the Hamiltonian that describes the interaction of the atom with the laser field (interaction operator), $\widehat{\mathfrak{R}}$ -- an operator functional describing relaxation processes in an atom. 
Square brackets $[$$\dots,$$\,\dots$$]$ denote the commutation operation of two operators.

It is convenient to split the density matrix into several matrix blocks, which can be written using the Dirac bra- and ket-vectors in the following form:
\begin{equation} \label{eq:rhoblocks}
    \hat{\rho}=\sum_{i,j=1}^3\hat{\rho}_{ij}\ket{F_i}\bra{F_j}.
\end{equation}
Diagonal matrix blocks of the form $\hat{\rho}_{ii}$ describe the populations of Zeeman (magnetic) energy sublevels $i$-th hyperfine level with angular momentum $F_i$, as well as superpositions of these sublevels -- \textit{Zeeman coherences}.
Matrices $\hat{\rho}_{12}$ and $\hat{\rho}_{21}$ describe coherent superpositions of magnetic sublevels belonging to different hyperfine levels of the ground state-- \textit{icrowave (or low-frequency) coherences}.
These states are formed as a result of the quantum interference of the $F_1 \to F_3$ and $F_2 \to F_3$ dipole transitions induced by mutually coherent light waves.
Therefore, they oscillate in time at the microwave frequency $\omega_1 - \omega_2 \sim \Delta_g$ (see Fig. \ref{fig:1}).
Since the full density matrix is Hermitian -- i.e., $\hat{\rho}^\dag = \hat{\rho}$ -- the relation $\hat{\rho}_{12}^\dag = \hat{\rho}_{21}$ holds.
Finally, the matrices $\hat{\rho}_{13}$, $\hat{\rho}_{23}$, $\hat{\rho}_{31}$, and $\hat{\rho}_{32}$ describe superpositions of magnetic sublevels, one of which belongs to the ground-state hyperfine level $F_1$ or $F_2$, while the other belongs to the excited level $F_3$.
Such density matrix elements are called \textit{optical coherences} because they oscillate at optical frequencies ($\omega_1$ or $\omega_2$).

The Hamiltonian of a free atom has the form:
\begin{multline} \label{eq:H0}
    \widehat{H}_0=\sum_{j=1}^3\mathcal{E}_j\sum_{m=-F_j}^{F_j}\ket{F_j,m}\bra{F_j,m} = \\
=\begin{pmatrix}
\mathcal{E}_1\hat{1}_{g1} & \hat{0} & \hat{0}\\
 \hat{0} & \mathcal{E}_2\hat{1}_{g2} & \hat{0}\\
\hat{0} & \hat{0} & \mathcal{E}_3\hat{1}_{e}
    \end{pmatrix}.
\end{multline}
Here, $\mathcal{E}_j$ is the energy of the level with angular momentum $F_j$.
The matrix blocks $\hat{1}_{g1}$ and $\hat{1}_{g2}$ are identity matrices of dimensions $(2F_1+1)\times(2F_1+1)$ and $(2F_2+1)\times(2F_2+1)$, respectively, while $\hat{1}_{e}$ is the identity matrix for the excited state with dimension $(2F_3+1)\times(2F_3+1)$. 
The magnetic quantum numbers $m$ range from $-F_j$ to $F_j$; thus, in the absence of a magnetic field, each level $F_j$ comprises $2F_j+1$ magnetic sublevels with the same energy $\mathcal{E}_j$ -- that is, the energy level is degenerate with respect to the projection of the total angular momentum $\mathbf{F}$ onto the quantization axis $z$.
Further we will also use the following notation for the atomic optical transition frequencies: $\omega_{3j} = (\mathcal{E}_3 - \mathcal{E}_j)/\hbar$, where $j=1, 2$.

In the semiclassical approach, the light field is treated classically (non-quantized), in contrast to the internal degrees of freedom of the atom. In this case, within the dipole approximation and taking into account the expression for the field (\ref{eq:field}), the interaction operator $\widehat{V}$ takes the form:
\begin{multline} \label{eq:V}
    \widehat{V}(z,t) = -\hat{{\bf d}}{\bf E}(z,t) =
    -\hbar R_1\widehat{\tilde{V}}_{31}^{(1)}e^{-i(\omega_1t-k_1z)}-\\-\hbar R_2\widehat{\tilde{V}}_{32}^{(2)}e^{-i(\omega_2t-k_2z)}-\hbar R_3\widehat{\tilde{V}}_{31}^{(3)} e^{-i(\omega_1t+k_1z+\phi_1)}-\\
    -\hbar R_4\widehat{\tilde{V}}_{32}^{(4)} e^{-i(\omega_2t+k_2z+\phi_2)} + H.c.
\end{multline}
Only resonant terms were taken into account here (resonant approximation).
"H.c." denotes Hermitian conjugate terms, and $R_j$ are the Rabi frequencies characterizing the strength of the interaction between the $j$-th light wave and a specific optical transition in the atom:
\begin{equation} \label{eq:Rabi}
    R_{1,3} = \frac{d_{31}E_{1,3}}{\hbar},\quad R_{2,4} = \frac{d_{32}E_{2,4}}{\hbar}\,,
\end{equation}
where $d_{31}$ and $d_{32}$ -- the reduced matrix elements of the dipole moment operator for the corresponding optical transition, for which the property is satisfied: $d_{ij}$$\,=\,$$(-1)^{F_j-F_j}$$d_{ji}^*$.

Expression (\ref{eq:V}) was obtained using the Wigner-Eckart theorem \cite{Varshalovich} for the matrix elements of the atomic dipole moment operator:
\begin{equation}\label{Dipmomentoperat}
    \bra{F_i}\hat{{\bf d}}\ket{F_j} = \frac{d_{ij}}{\sqrt{3}}\widehat{{\bf T}}^{ij}_1,
\end{equation}
where ${\bf T}^{ij}_1$ are Wigner operators of rank $\varkappa=1$ (vector operators).
For an arbitrary rank, we have the following expression for the cyclic components of Wigner tensor operators \cite{Varshalovich}:
\begin{multline}\label{eq:T}
\widehat{T}_{\varkappa q}^{ij} = \sqrt{2\varkappa+1}\sum_{m_i,m_j}(-1)^{F_i-m_i}\times\\\times \TreJ{F_i}{\varkappa}{F_j}{-m_i}{q}{m_j} \ket{F_i,m_i}\bra{F_j,m_j}\,.
\end{multline}
Here, $q$ ranges from $-\varkappa$ to $\varkappa$, and the parentheses $(\dots)$ denote the Wigner $3j$-symbol.
The sum in (\ref{eq:T}) is taken over all magnetic quantum numbers for each angular momentum value; that is, $m_i$ ranges from $-F_i$ to $F_i$, while $m_j$ ranges from $-F_j$ to $F_j$.
The properties of the $3j$-symbol reflect the selection rules regarding the magnetic quantum number; specifically, the symbol is non-zero only if the equality $m_i - m_j = q$ holds.
The triangle rule $|F_i - F_j| \le \varkappa \le |F_i + F_j|$ also applies; consequently, for dipole transitions ($\varkappa = 1$) between levels with angular momenta $F_i$ and $F_j$, we obtain the selection rule for the quantum number $F$: $\Delta F = F_i - F_j = 0, \pm 1$.

In (\ref{eq:V}), dimensionless interaction operators were introduced, reflecting the specific features of the interaction between light waves and the atom that are associated with the light polarization parameters:
\begin{multline} \label{eq:V2}
    \widehat{\tilde{V}}_{31}^{(1)} = \frac{1}{\sqrt{3}} \xiB_{(1)}\widehat{{\bf T}}_1^{31}\,,\quad \widehat{\tilde{V}}_{32}^{(2)} = \frac{1}{\sqrt{3}} \xiB_{(1)}\widehat{{\bf T}}_1^{32}\,,\\
    \widehat{\tilde{V}}_{31}^{(3)} = \frac{1}{\sqrt{3}} \xiB_{(2)}\widehat{{\bf T}}_1^{31}\,,\quad \widehat{\tilde{V}}_{32}^{(4)} = \frac{1}{\sqrt{3}} \xiB_{(2)}\widehat{{\bf T}}_1^{32}\,.
\end{multline}

The operator $\widehat{\mathfrak{R}}$ appearing in (\ref{eq:DM1}) describes two types of relaxation associated with the internal degrees of freedom of the atom: spontaneous radiative relaxation at a rate $\gamma$, linked to the finite lifetime of the excited state with angular momentum $F_3$, and time of flight relaxation at a rate $\Gamma$, linked to the finite duration of the coherent interaction between the atom and the light field:
\begin{multline}\label{eq:Relax}
        \widehat{\mathfrak{R}}=\frac{\Gamma}{(2J_g+1)(2I_n+1)}\left(\hat{1}_{g1}+\hat{1}_{g2}\right)-\\-
        \gamma\hat{\rho}_{33}-\Gamma\left(\hat{\rho}_{11}+\hat{\rho}_{22}+\hat{\rho}_{33}\right)+\widehat{\Gamma}_{31} + \widehat{\Gamma}_{32}-\\-
        \left(\Gamma \hat{\rho}_{12}+\gamma_{eg} \hat{\rho}_{13}+\gamma_{eg} \hat{\rho}_{23} +H.c.\right)\,.
\end{multline}
The first term in this expression describes the isotropic distribution of atoms among the magnetic sublevels of the hyperfine energy levels $F_1$ and $F_2$ in the ground state—a distribution characteristic of an atom entering the light beam.
In formulating this term, account was taken of the fact that the total number of magnetic sublevels in the ground state is $2F_1 + 1 + 2F_2 + 1 = (2J_g + 1)(2I_n + 1)$, where $J_g$ is the total angular momentum of the atom's electron shell in the ground state and $I_n$ is the nuclear spin.
The second term ($\gamma\hat{\rho}_{33}$) describes the loss of atoms from the excited level due to spontaneous relaxation.
The third term relates to the relaxation of Zeeman coherences and populations of the magnetic sublevels of all three energy levels caused by the atom leaving the light beam (time of flight relaxation).
The terms involving the operators $\widehat{\Gamma}_{31}$ and $\widehat{\Gamma}_{32}$ describe the arrival of atoms at the two ground-state hyperfine levels $F_1$ and $F_2$ from the excited level $F_3$ via spontaneous relaxation.
Finally, the last term in parentheses describes the time of flight relaxation of microwave and optical coherences, as well as the spontaneous relaxation of optical coherences, with $\gamma_{eg} = \Gamma + \gamma/2$.

The operators describing the arrival of atoms at levels $F_1$ and $F_2$ due to spontaneous relaxation from level $F_3$ have the form \cite{Lazebnyi2015}:
\begin{equation}\label{eq:SponPrih}
    \widehat{\Gamma}_{3j} = \frac{\gamma\beta_{3j}(2F_3+1)}{3}\sum_{q=0,\pm1} \widehat{T}^{3j\dagger}_{1q}\hat{\rho}_{33}\widehat{T}^{3j}_{1q}\,,
\end{equation}
where $j = 1, 2$, and $\beta_{31}$ and $\beta_{32}$ are the branching coefficients:
\begin{equation}\label{eq:beta}
    \beta_{3j}=\left(2J_e+1\right)\left(2F_j+1\right)\SixJ{J_g}{F_j}{I_n}{F_3}{J_e}{1}^2\,,
\end{equation}
determining the relative probability of spontaneous radiative relaxation via each of the two channels: $F_3$$\,\to\,$$F_1$ and $F_3$$\,\to\,$$F_2$.
The curly brackets $\{\dots\}$ denote the Wigner $6j$-symbol \cite{Varshalovich}, and $J_e$ is the total angular momentum of the atom's electron shell in the excited state. Since atoms spontaneously relax from level $F_3$ only to levels $F_1$ and $F_2$, it is evident that $\beta_{31}$$+$$\beta_{32}$$\,=\,$$1$.

It has been shown theoretically and experimentally in \cite{Hafiz2017,Brazhnikov2019,Brazhnikov2020,Mikhailov2021} that, due to microwave coherence induced in an atomic medium by a dual-frequency light field, the absorption coefficient can exhibit slow spatial oscillations.
Specifically, for a cesium vapor cell, the period of these oscillations is $\approx\,$$16$~mm.
This effect also leads to slow spatial oscillations in the diagonal matrix blocks $\hat{\rho}_{nn}$; i.e., the following expansion holds \cite{Hafiz2017,Mikhailov2021}:
\begin{equation}\label{eq:Harmonics1}
    \hat{\rho}_{nn}(z) \approx \hat{\rho}_{nn}^{(0)}+\hat{\rho}_{nn}^{(+)}e^{2ik_{12}z}+\hat{\rho}_{nn}^{(-)}e^{-2ik_{12}z}\,.
\end{equation}
Here, $k_{12} = k_1 - k_2$.
However, these oscillations manifest either when the counterpropagating light beams have comparable intensities or in the regime where the response is nonlinear with respect to the probe beam.
Both cases lie beyond the scope of the present work; therefore, the non-zero harmonics in (\ref{eq:Harmonics1}) can be neglected.
This same result follows from a more rigorous analysis based on perturbation theory (see Chapter~\ref{sec:3}).

This work considers the stationary (steady-state) regime of the interaction between atoms and a light field.
Therefore, the matrix blocks $\hat{\rho}_{nn}$ in the expansion (\ref{eq:Harmonics1}) do not depend on time $t$.
However, such a dependence is explicitly present, for instance, in the microwave coherences $\hat{\rho}_{12}$ and $\hat{\rho}_{21}$, for which the following representation holds in the approximation of the lowest spatial harmonics \cite{Hafiz2017,Brazhnikov2019}:
\begin{multline}\label{eq:Harmonics2}
        \hat{\rho}_{12}(z,t)\approx e^{i\delta_{12}t}(\hat{\rho}_{12}^{(+)}e^{ik_{12}z}+\hat{\rho}_{12}^{(-)}e^{-ik_{12}z})\,,
\end{multline}
where $\delta_{12}$$\,=\,$$\omega_1$$-$$\omega_2$.

Finally, for optical coherences, in the approximation of the lowest spatial harmonics, we have the following expansion \cite{Hafiz2017,Brazhnikov2019}:
\begin{multline}\label{eq:Harmonics3}
        \hat{\rho}_{13}(z,t)\approx e^{i\omega_1t}(\hat{\rho}_{13}^{(-1)}e^{-ik_1z}+\hat{\rho}_{13}^{(+1)}e^{i(k_1z+\phi_1)}+\\+
        \hat{\rho}_{13}^{(-21)}e^{-i(2k_2-k_1)z}+\hat{\rho}_{13}^{(+21)}e^{i(2k_2-k_1)z+i\phi_1}),
\end{multline}  
\begin{multline}\label{eq:Harmonics4}
        \hat{\rho}_{23}(z,t)\approx e^{i\omega_2t}(\hat{\rho}_{23}^{(-2)}e^{-ik_2z}+\hat{\rho}_{23}^{(+2)}e^{i(k_2z+\phi_2)}+\\+
        \hat{\rho}_{23}^{(-12)}e^{-i(2k_1-k_2)z}+\hat{\rho}_{23}^{(+12)}e^{i(2k_1-k_2)z+i\phi_2})\,.
\end{multline}
The harmonics $\hat{\rho}_{13}^{(\pm1)}$ and $\hat{\rho}_{23}^{(\pm2)}$ are induced by the direct action of each of the waves $E_{1,2,3,4}$ on the corresponding dipole transitions (see Fig. \ref{fig:1}), whereas the combination harmonics $\hat{\rho}_{13}^{(\pm21)}$ and $\hat{\rho}_{23}^{(\pm12)}$ arise from the interaction of optical fields with the harmonics of the microwave coherences (\ref{eq:Harmonics2}).
Spatial harmonic expansions for the remaining microwave and optical coherences can be obtained by Hermitian conjugation of expressions (\ref{eq:Harmonics2})–(\ref{eq:Harmonics4}).
In this context, due to the Hermiticity of the total density matrix $\hat{\rho}$, the following relations hold: $\hat{\rho}_{21}^{(\pm)} = \hat{\rho}_{12}^{(\mp)\dagger}$, $\hat{\rho}_{31}^{(\pm1)} = \hat{\rho}_{13}^{(\mp1)\dagger}$, $\hat{\rho}_{32}^{(\pm1)} = \hat{\rho}_{23}^{(\mp1)\dagger}$, $\hat{\rho}_{31}^{(\pm21)} = \hat{\rho}_{13}^{(\mp21)\dagger}$, and $\hat{\rho}_{32}^{(\pm12)} = \hat{\rho}_{23}^{(\mp12)\dagger}$.

\subsection{Formalism of Polarization Moments}

To derive analytical expressions using the (PT) method for an atomic energy-level structure characterized by degeneracy with respect to the projection of the total angular momentum, it is convenient to represent the density matrix in the basis of irreducible tensor operators $\widehat{T}_{\varkappa q}^{ij}$—also known as the $\varkappa q$-representation of the density matrix \cite{Varshalovich,Dyakonov65,Auzinsh}:
\begin{equation} \label{eq:Irreduce1}
    \hat{\rho}_{ij} = \sum_{\varkappa,q}(-1)^{F_i-F_j-q}\rho_{\varkappa-q}^{ij}\widehat{T}_{\varkappa q}^{ij} = \sum_{\varkappa,q}\rho_{\varkappa q}^{ij}\widehat{T}_{\varkappa q}^{ji\dagger}\,.
\end{equation}
The coefficients of this expansion, $\rho_{\varkappa q}^{ij}$, are the cyclic components of tensors of rank $\varkappa$, known as polarization moments (PM).
The inverse of this transformation is given by:
\begin{equation}\label{eq:moments}
    \rho^{ij}_{\varkappa q}=\text{Tr}\{\widehat{\rho}_{ij}\widehat{T}^{ji}_{\varkappa q}\}\,.
\end{equation}
This formula allows one to obtain the form of any operator in the $\varkappa q$-representation used in the present work.
First, let us list a number of formulas from the algebra of irreducible tensor operators that we will employ.
These expressions can be derived from the corresponding formulas for sums of products of $3jm$-symbols \cite{Varshalovich,Lazebnyi2015}:
\begin{equation}
   \text{Tr}\{\widehat{T}^{ii}_{\varkappa q}\}=\sqrt{2F_i+1}\,\delta_{\varkappa 0}\delta_{q 0}\,,\label{eq:Tr1}
\end{equation}
\begin{equation}
    \text{Tr}\{\widehat{T}^{ij}_{\varkappa_1 q_1} \widehat{T}^{ji}_{\varkappa_2 q_2}\}=(-1)^{q_1+F_i+F_j}\delta_{\varkappa_1\varkappa_2}\delta_{q_1-q_2}\,,\label{eq:Tr2}
\end{equation}
\begin{multline}
    \text{Tr}\{\widehat{T}^{ij}_{\varkappa_1 q_1} \widehat{T}^{jk}_{\varkappa_2 q_2} \widehat{T}^{ki}_{\varkappa_3 q_3}\}=\\ =\Pi_{\varkappa_1\varkappa_2\varkappa_3}\TreJ{\varkappa_1}{\varkappa_2}{\varkappa_3}{-q_1}{-q_2}{-q_3}\SixJ{\varkappa_1}{\varkappa_2}{\varkappa_3}{F_k}{F_i}{F_j},\label{eq:Tr3}
\end{multline}
where $\delta_{ij}$ is the Kronecker delta.
For brevity, the symbol $\Pi_{\varkappa_1,\dots,\varkappa_n} = \sqrt{(2\varkappa_1+1)\dots(2\varkappa_n+1)}$ was introduced.

We will also need a formula for the $\varkappa q$-representation of the product of two operators \cite{Varshalovich,Lazebnyi2015}:
\begin{eqnarray}\label{eq:product}
    &&\left\{\widehat{A}^{ik}\widehat{B}^{kj}\right\}_{\varkappa q}=\nonumber\\
    &&=\sum_{\varkappa_1,\varkappa_2}(-1)^{F_i+F_j+\varkappa}\,\Pi_{\varkappa1\varkappa2}\SixJ{\varkappa_1}{\varkappa_2}{\varkappa}{F_i}{F_j}{F_k}\times\nonumber\\
    &&\qquad\qquad\qquad\times\left\{B^{kj}_{\varkappa_1}\otimes A^{ik}_{\varkappa_2}\right\}_{\varkappa q}\,.
\end{eqnarray}
Here, $\left\{B^{kj}_{\varkappa_1}\otimes A^{ik}_{\varkappa_2}\right\}_{\varkappa}$ denotes the irreducible tensor product of rank $\varkappa$.
In the derivation of (\ref{eq:product}) and subsequently, it is assumed that the quantum numbers $F_i$ take only integer values.
Use was also made of the symmetry property of $6j$-symbols regarding the permutation of any pair of columns and the property of the irreducible tensor product \cite{Varshalovich}:
\begin{multline}
    \left\{A^{ik}_{\varkappa_1}\otimes B^{kj}_{\varkappa_2}\right\}_{\varkappa q} =\\=
    (-1)^{\varkappa_1+\varkappa_2+\varkappa} \left\{B^{kj}_{\varkappa_2}\otimes A^{ik}_{\varkappa_1}\right\}_{\varkappa q}.
\end{multline}

Next, dimensionless interaction operators (\ref{eq:V2}) are introduced in the $\varkappa q$-representation.
Using (\ref{eq:moments}) and (\ref{eq:Tr2}), we find:
\begin{equation}
\begin{aligned}[b]\label{eq:Vkq}
    \left\{ \widehat{\tilde{V}}^{(1)}_{31} \right\}_{\varkappa} &\equiv \text{Tr}\left\{ \widehat{\tilde{V}}^{(1)}_{31} \widehat{T}^{13}_{\varkappa q} \right\} =\\&= \frac{1}{\sqrt{3}}\delta_{1\varkappa}(-1)^{F_1+F_3} \xiB_{(1)}\,,\\
     \left\{ \widehat{\tilde{V}}^{(2)}_{32} \right\}_{\varkappa} &= \frac{1}{\sqrt{3}}\delta_{1\varkappa}(-1)^{F_2+F_3} \xiB_{(1)}\,,\\
     \left\{ \widehat{\tilde{V}}^{(3)}_{31} \right\}_{\varkappa} &= \frac{1}{\sqrt{3}}\delta_{1\varkappa}(-1)^{F_1+F_3} \xiB_{(2)}\,,\\
     \left\{ \widehat{\tilde{V}}^{(4)}_{32} \right\}_{\varkappa} &= \frac{1}{\sqrt{3}}\delta_{1\varkappa}(-1)^{F_2+F_3} \xiB_{(2)}\,.
\end{aligned}
\end{equation}
Thus, in the $\varkappa q$-representation, the dimensionless interaction operators coincide, up to a constant, with the polarization vectors of light waves in the cyclic basis.
For the Hermitian conjugate operators, we have analogous expressions:
\begin{eqnarray}\label{eq:Vkqconj}
    &&\left\{ \widehat{\tilde{V}}^{(1)}_{13} \right\}_{\varkappa} = \frac{1}{\sqrt{3}}\delta_{1\varkappa} \xiB_{(1)}^*\,,\nonumber\\
     &&\left\{ \widehat{\tilde{V}}^{(2)}_{23} \right\}_{\varkappa} = \frac{1}{\sqrt{3}}\delta_{1\varkappa}\xiB_{(1)}^*\,,\nonumber\\
     &&\left\{ \widehat{\tilde{V}}^{(3)}_{13} \right\}_{\varkappa} = \frac{1}{\sqrt{3}}\delta_{1\varkappa}\xiB_{(2)}^*\,,\nonumber\\
     &&\left\{ \widehat{\tilde{V}}^{(4)}_{23} \right\}_{\varkappa} = \frac{1}{\sqrt{3}}\delta_{1\varkappa}\xiB_{(2)}^*\,.
\end{eqnarray}

For the second-order calculation in perturbation theory, we will need the $\varkappa q$-representations of the following products of interaction operators, which can be obtained using formulas (\ref{eq:product}), (\ref{eq:Vkq}), and (\ref{eq:Vkqconj}):
    \begin{multline*}
        \left\{ \widehat{\tilde{V}}^{(1)}_{31} \widehat{\tilde{V}}^{(1)}_{13} \right\}_{\mathclap{\varkappa}} = (-1)^{F_1+F_3+\varkappa} \SixJ{1}{1}{\varkappa}{F_3}{F_3}{F_1} \times \\ \times \left\{ \xiB_{(1)}^*  \otimes  \xiB_{(1)} \right\}_{\varkappa},
    \end{multline*} 
    \begin{multline*}
        \left\{ \widehat{\tilde{V}}^{(2)}_{32} \widehat{\tilde{V}}^{(2)}_{23} \right\}_{\mathclap{\varkappa}} =(-1)^{F_2+F_3+\varkappa} \SixJ{1}{1}{\varkappa}{F_3}{F_3}{F_2}\times \\ \times\left\{ \xiB_{(1)}^*  \otimes  \xiB_{(1)} \right\}_{\varkappa},
    \end{multline*} 
    \begin{multline*}
        \left\{ \widehat{\tilde{V}}^{(3)}_{31} \widehat{\tilde{V}}^{(3)}_{13} \right\}_{\mathclap{\varkappa}} =(-1)^{F_1+F_3+\varkappa} \SixJ{1}{1}{\varkappa}{F_3}{F_3}{F_1}\times \\ \times\left\{ \xiB_{(2)}^*  \otimes  \xiB_{(2)} \right\}_{\varkappa},
    \end{multline*} 
    \begin{multline*}
        \left\{ \widehat{\tilde{V}}^{(4)}_{32} \widehat{\tilde{V}}^{(4)}_{23} \right\}_{\mathclap{\varkappa}} =(-1)^{F_2+F_3+\varkappa} \SixJ{1}{1}{\varkappa}{F_3}{F_3}{F_2}\times \\ \times\left\{ \xiB_{(2)}^*  \otimes  \xiB_{(2)} \right\}_{\varkappa},
    \end{multline*} 
    \begin{multline*}
        \left\{ \widehat{\tilde{V}}^{(1)}_{13} \widehat{\tilde{V}}^{(1)}_{31}\right\}_{\mathclap{\varkappa}} =(-1)^{F_1+F_3+\varkappa} \SixJ{1}{1}{\varkappa}{F_1}{F_1}{F_3}\times \\ \times\left\{ \xiB_{(1)} \otimes \xiB_{(1)}^* \right\}_{\varkappa},
    \end{multline*} 
    \begin{multline*}
        \left\{ \widehat{\tilde{V}}^{(2)}_{23} \widehat{\tilde{V}}^{(2)}_{32} \right\}_{\mathclap{\varkappa}} =(-1)^{F_2+F_3+\varkappa} \SixJ{1}{1}{\varkappa}{F_2}{F_2}{F_3}\times \\ \times\left\{ \xiB_{(1)} \otimes  \xiB_{(1)}^* \right\}_{\varkappa},
    \end{multline*} 
    \begin{multline*}
        \left\{ \widehat{\tilde{V}}^{(3)}_{13} \widehat{\tilde{V}}^{(3)}_{31} \right\}_{\mathclap{\varkappa}} =(-1)^{F_1+F_3+\varkappa} \SixJ{1}{1}{\varkappa}{F_1}{F_1}{F_3}\times \\ \times\left\{ \xiB_{(2)} \otimes  \xiB_{(2)}^* \right\}_{\varkappa},
    \end{multline*} 
    \begin{multline} \label{eq:VVkq1}
        \left\{ \widehat{\tilde{V}}^{(4)}_{23} \widehat{\tilde{V}}^{(4)}_{32} \right\}_{\mathclap{\varkappa}} =(-1)^{F_2+F_3+\varkappa} \SixJ{1}{1}{\varkappa}{F_2}{F_2}{F_3}\times \\ \times\left\{ \xiB_{(2)} \otimes \xiB_{(2)}^* \right\}_{\varkappa}.
    \end{multline}
These expressions contain products of operators $\widehat{\tilde{V}}^{(j)}_{nm}$ only with their Hermitian conjugates $\widehat{\tilde{V}}^{(j)}_{mn} = \widehat{\tilde{V}}^{(j)\dagger}_{nm}$.
However, for the third-order perturbation theory calculation, we will also need cross-products of interaction operators corresponding to different waves:
\begin{multline*}
    \left\{\! \widehat{\tilde{V}}^{(1)}_{13} \widehat{\tilde{V}}^{(2)}_{32}\! \right\}_{\mathclap{\varkappa}}\! =\! (-1)^{F_1+F_3+\varkappa}\! \SixJ{1}{1}{\varkappa}{F_1}{F_2}{F_3}\times \\ \times\!\left\{\! \xiB_{(1)}\!\! \otimes \! \xiB_{(1)}^* \! \right\}_{\varkappa},
\end{multline*}
\begin{multline*}
    \left\{\! \widehat{\tilde{V}}^{(2)}_{23} \widehat{\tilde{V}}^{(1)}_{31}\! \right\}_{\mathclap{\varkappa}}\! =\!(-1)^{F_2+F_3+\varkappa}\! \SixJ{1}{1}{\varkappa}{F_1}{F_2}{F_3}\times \\ \times\!\left\{\! \xiB_{(1)}\!\! \otimes \! \xiB_{(1)}^* \! \right\}_{\varkappa},
\end{multline*}
\begin{multline*}
    \left\{\! \widehat{\tilde{V}}^{(3)}_{13} \widehat{\tilde{V}}^{(4)}_{32}\! \right\}_{\mathclap{\varkappa}}\! =\!(-1)^{F_1+F_3+\varkappa}\! \SixJ{1}{1}{\varkappa}{F_1}{F_2}{F_3}\times \\ \times \!\left\{\! \xiB_{(2)}\!\! \otimes \! \xiB_{(2)}^* \! \right\}_{\varkappa},
\end{multline*}
\begin{multline} \label{eq:VVkq2}
    \left\{\! \widehat{\tilde{V}}^{(4)}_{23} \widehat{\tilde{V}}^{(3)}_{31}\! \right\}_{\mathclap{\varkappa}}\! =\!(-1)^{F_2+F_3+\varkappa}\! \SixJ{1}{1}{\varkappa}{F_1}{F_2}{F_3}\times \\ \times\!\left\{\! \xiB_{(2)}\!\! \otimes \! \xiB_{(2)}^* \! \right\}_{\varkappa}.
\end{multline}
% \begin{equation}\label{eq:VVkq2}
% \begin{aligned}[b]
%     \left\{\! \widehat{\tilde{V}}^{(1)}_{13} \widehat{\tilde{V}}^{(2)}_{32}\! \right\}_{\mathclap{\varkappa}}\! =\! (-1)^{F_1+F_3+\varkappa}\! \SixJ{1}{1}{\varkappa}{F_1}{F_2}{F_3}\!\left\{\! \xiB_{(1)}\!\! \otimes \! \xiB_{(1)}^* \! \right\}_{\varkappa},\\
%     \left\{\! \widehat{\tilde{V}}^{(2)}_{23} \widehat{\tilde{V}}^{(1)}_{31}\! \right\}_{\mathclap{\varkappa}}\! =\!(-1)^{F_2+F_3+\varkappa}\! \SixJ{1}{1}{\varkappa}{F_1}{F_2}{F_3}\!\left\{\! \xiB_{(1)}\!\! \otimes \! \xiB_{(1)}^* \! \right\}_{\varkappa},\\
%     \left\{\! \widehat{\tilde{V}}^{(3)}_{13} \widehat{\tilde{V}}^{(4)}_{32}\! \right\}_{\mathclap{\varkappa}}\! =\!(-1)^{F_1+F_3+\varkappa}\! \SixJ{1}{1}{\varkappa}{F_1}{F_2}{F_3}\!\left\{\! \xiB_{(2)}\!\! \otimes \! \xiB_{(2)}^* \! \right\}_{\varkappa},\\
%     \left\{\! \widehat{\tilde{V}}^{(4)}_{23} \widehat{\tilde{V}}^{(3)}_{31}\! \right\}_{\mathclap{\varkappa}}\! =\!(-1)^{F_2+F_3+\varkappa}\! \SixJ{1}{1}{\varkappa}{F_1}{F_2}{F_3}\!\left\{\! \xiB_{(2)}\!\! \otimes \! \xiB_{(2)}^* \! \right\}_{\varkappa}.
% \end{aligned}
% \end{equation}
It follows from formulas (\ref{eq:VVkq1}) and (\ref{eq:VVkq2}) -- specifically from the properties of the $6j$-symbols appearing in them -- that, in contrast to the $\varkappa q$-representation for the interaction operators (\ref{eq:Vkq}) and (\ref{eq:Vkqconj}), in this case the non-zero terms include not only multipole moments of rank $\varkappa=1$ (dipoles) but also those of rank $\varkappa=0$ (monopoles) and $\varkappa=2$ (quadrupoles).

Finally, let us write the spontaneous relaxation operators $\widehat{\Gamma}_{jj}(\hat{\rho})$ in the $\varkappa q$-representation, based on expressions (\ref{eq:SponPrih}), (\ref{eq:Irreduce1}), and (\ref{eq:moments}) and the formulas from \cite{Varshalovich}:
\begin{eqnarray}\label{eq:sponprihkq}
    &&\left\{\widehat{\Gamma}_{3j}\right\}_{\varkappa q}=-\gamma\beta_{3j} (2F_3+1)(-1)^{F_j+F_3+\varkappa}\times\nonumber\\
    &&\qquad\qquad\qquad\times\SixJ{F_3}{F_3}{\varkappa}{F_j}{F_j}{1}\rho^{33}_{\varkappa q}\,.
\end{eqnarray}

\section{Analytical Solution}\label{sec:3}

To simplify the presentation, the system of equations for the density matrix elements (optical Bloch equations) is provided in Appendix A.
Here, we consider the solution to this system using perturbation theory up to the third order in the light field. 
It is known that in the approximation of spatially unbounded light beams ($\Gamma \to 0$), the small parameter used in perturbation theory for the equations governing the polarization moments (multipoles) of the ground and excited states is the saturation parameter; at exact resonance, this equals $R_j^2/\gamma_{eg}^2$.
However, in the presence of time of flight relaxation ($\Gamma \ne 0$), the small parameter becomes the "flight" saturation parameter \cite{Lazebnyi2015}, which equals $R_j^2/\Gamma\gamma_{eg}$ at exact resonance. 
This requires the condition $\gamma \ge \Gamma$ to be satisfied.
Physically, this condition implies, that during the time an excited atom spends in the laser beam ($\tau \sim \Gamma^{-1}$), it  will emit one or more photons via spontaneous relaxation.
Since the spontaneous relaxation rate for alkali metal atoms is on the order of a few MHz, this condition is typically satisfied with a wide margin in practice.

\subsection{Zeroth Order}

Moreover, in a gas of thermal atoms, the populations of these sublevels are equal to each other with high accuracy due to the fact that the condition is satisfied $k_BT \gg \hbar\Delta_g$ (where $k_B$ is the Boltzmann constant and $T$ is the temperature of the atoms in the gas).

To determine the form of the non-zero matrix blocks from the system of equations (\ref{eq:sysro110})–(\ref{eq:sysro12m}) given in Appendix A, one must set $R_3 = R_4 = 0$ in the zeroth order with respect to the pump beam (recall that $R_1 = R_2 = 0$ must also be set in these expressions, given that we are considering the linear regime with respect to the probe beam). 
Then, the only non-zero elements of the density matrix are those describing the populations of the ground-state Zeeman sublevels that are independent of $z$, i.e., the zeroth harmonics, namely:
\begin{align}
    &\hat{\rho}^{(0)}_{11}(0) = \hat{\Upsilon}_{11}\,,\label{r110zero0}\\[2ex]
    &\hat{\rho}^{(0)}_{22}(0) = \hat{\Upsilon}_{22}\,.\label{r220zero0}
\end{align}
Here and in the following, the notation $\hat{\rho}(n)$ is used for the $n$-th order solution in perturbation theory.
The matrices $\hat{\Upsilon}_{11}$ and $\hat{\Upsilon}_{22}$ are proportional to identity matrices and are given by formula (\ref{eq:Upsilon}) in Appendix A.

In the language of the $\varkappa q$-representation, only scalar ($\varkappa=0$) multipoles are non-zero in the atomic ground state.
Expressions for them can be obtained from (\ref{r110zero0}) and (\ref{r220zero0}) using formulas (\ref{eq:moments}), (\ref{eq:Tr1}), and (\ref{eq:Upsilon}):
\begin{align}
    &\left\{\hat{\rho}^{(0)}_{11}(0)\right\}_{\varkappa q} = \frac{\sqrt{2F_1+1}}{(2J_g+1)(2I_n+1)}\delta_{\varkappa 0}\delta_{q 0}\,,\label{r110zero}\\[2ex]
    &\left\{\hat{\rho}^{(0)}_{22}(0)\right\}_{\varkappa q} = \frac{\sqrt{2F_2+1}}{(2J_g+1)(2I_n+1)}\delta_{\varkappa 0}\delta_{q 0}\,.\label{r220zero}
\end{align}

From equations (\ref{eq:sysro110})–(\ref{eq:sysro12m}) and (\ref{eq:sysro13m1})–(\ref{eq:sysro23p12}) presented in Appendix A, it can be shown that the solutions for the diagonal matrix blocks $\hat{\rho}_{ii}$ ($i=1, 2, 3$), as well as for the low-frequency coherences $\hat{\rho}_{12}$ and $\hat{\rho}_{21}$, will contain contributions only from even orders of perturbation theory, including the zeroth order.
In contrast, the solutions for the off-diagonal matrix blocks, the optical coherences, will contain contributions only from odd orders of PT.

From (\ref{r110zero}) and (\ref{r220zero}), as well as from formulas (\ref{eq:Irreduce1}) and (\ref{eq:Tr1}), it follows that the normalization condition holds for the density matrix in the zeroth approximation:
\begin{multline}
    \text{Tr}\left\{\hat{\rho}^{(0)}_{11}(0)\right\} + \text{Tr}\left\{\hat{\rho}^{(0)}_{22}(0)\right\} = \\=
    \frac{2F_1+1}{(2J_g+1)(2I_n+1)} +\\+ \frac{2F_2+1}{(2J_g+1)(2I_n+1)} = 1,
\end{multline}
in other words, the total population is conserved, which is evidently a consequence of the closed nature of the system of optical transitions under consideration in the atom.

\subsection{First Order}

At the first order of perturbation theory, non-zero solutions exist only for optical coherences, and specifically only for certain spatial harmonics in the representations (\ref{eq:Harmonics3}) and (\ref{eq:Harmonics4}) and their complex conjugates.
In particular, from equation (\ref{eq:sysro13m1}) given in Appendix A, as well as from the form of the solution for the zero harmonic (\ref{r110zero0}), we find:
\begin{multline}\label{eq:solro13m11}
    \hat{\rho}_{13}^{(-1)}(1) = iR_1L_1^{(-)}\Bigl(\widehat{\tilde{V}}_{13}^{(1)}\hat{\rho}_{33}^{(0)}(0)-\hat{\rho}_{11}^{(0)}(0)\widehat{\tilde{V}}_{13}^{(1)}\Bigr)-\\
    -iR_2L_1^{(-)}\hat{\rho}_{12}^{(-)}(0)\widehat{\tilde{V}}_{23}^{(2)} = -i R_1 L_1^{(-)} \hat{\rho}_{11}^{(0)}(0)\widehat{\tilde{V}}_{13}^{(1)} = \\ =
     \frac{-iR_1 L_1^{(-)}}{(2J_g+1)(2I_n+1)}\,\widehat{\tilde{V}}_{13}^{(1)},
\end{multline}
or, taking (\ref{eq:Vkqconj}) into account, in the $\varkappa q$-representation:
\begin{multline}
     \left\{\hat{\rho}^{(-1)}_{13}(1)\right\}_{\varkappa} =  \frac{-i R_1 L_1^{(-)}}{(2J_g+1)(2I_n+1)}\left\{\widehat{\tilde{V}}_{13}^{(1)}\right\}_\varkappa = \\ =
     \frac{-i R_1 L_1^{(-)}}{\sqrt{3}(2J_g+1)(2I_n+1)} \xiB^*_{(1)}\delta_{\varkappa 1}.
\end{multline}\label{r13m1one}
Similarly for the remaining non-zero spatial harmonics of the optical coherences:
\begin{align}
     \left\{\hat{\rho}^{(+1)}_{13}(1)\right\}_{\varkappa} &= -\frac{i R_3 L_1^{(+)}\xiB^*_{(2)}\delta_{\varkappa 1}}{\sqrt{3}(2J_g+1)(2I_n+1)} ,\label{r13p1one}\\
     \left\{\hat{\rho}^{(-2)}_{23}(1)\right\}_{\varkappa} &= -\frac{i R_2 L_2^{(-)}\xiB^*_{(1)}\delta_{\varkappa 1}}{\sqrt{3}(2J_g+1)(2I_n+1)} ,\label{r23m2one}\\
     \left\{\hat{\rho}^{(+2)}_{23}(1)\right\}_{\varkappa} &= -\frac{i R_4 L_2^{(+)}\xiB^*_{(2)}\delta_{\varkappa 1}}{\sqrt{3}(2J_g+1)(2I_n+1)} ,\label{r23p2one}\\
     \left\{\hat{\rho}^{(+1)}_{31}(1)\right\}_{\varkappa} &= \frac{i (-1)^{F_1+F_3} R_1 L_1^{(-)*}\xiB_{(1)}\delta_{\varkappa 1}}{\sqrt{3}(2J_g+1)(2I_n+1)} ,\label{r31p1one}\\
     \left\{\hat{\rho}^{(-1)}_{31}(1)\right\}_{\varkappa} &= \frac{i (-1)^{F_1+F_3} R_3 L_1^{(+)*}\xiB_{(2)}\delta_{\varkappa 1}}{\sqrt{3}(2J_g+1)(2I_n+1)} ,\label{r31m1one}\\
     \left\{\hat{\rho}^{(+2)}_{32}(1)\right\}_{\varkappa} &= \frac{i (-1)^{F_2+F_3} R_2 L_2^{(-)*}\xiB_{(1)}\delta_{\varkappa 1}}{\sqrt{3}(2J_g+1)(2I_n+1)} ,\label{r32p2one}\\
     \left\{\hat{\rho}^{(-2)}_{32}(1)\right\}_{\varkappa} &= \frac{i (-1)^{F_2+F_3} R_4 L_2^{(+)*}\xiB_{(2)}\delta_{\varkappa 1}}{\sqrt{3}(2J_g+1)(2I_n+1)}. \label{r32m2one}
\end{align}

Since these optical coherences describe the atomic polarization induced on each of the two dipole transitions independently by each of the four optical waves, only vector multipoles ($\varkappa=1$) appear in these expressions.

In formulas (\ref{r13m1one})–(\ref{r32m2one}), complex Lorentzians were introduced that account for Doppler frequency shifts for moving atoms:
\begin{equation}
    \begin{aligned}\label{eq:lorenziansL}
        &L_1^{(\pm)} = \frac{1}{\gamma_{eg} + i \left( \omega_1 - \omega_{31} \pm k_1\upsilon \right)},\\
        &L_2^{(\pm)} = \frac{1}{\gamma_{eg} + i \left( \omega_2 - \omega_{32} \pm k_2\upsilon \right)}.
    \end{aligned}
\end{equation}

In experiments, the optical frequencies $\omega_1$ and $\omega_2$ are typically first-order sidebands resulting from amplitude or frequency modulation of the laser radiation at the carrier frequency $\omega_0$.
To observe the resonance curve, the frequency $\omega_0$ is scanned, while $\omega_1$ and $\omega_2$ remain offset from $\omega_0$ by a fixed amount determined by the microwave generator frequency.
Under these conditions, instead of the frequency detunings $\omega_1 - \omega_{31}$ and $\omega_2 - \omega_{32}$ appearing in (\ref{eq:lorenziansL}), it is convenient to introduce alternative detunings defined by the following expressions (see also Fig. \ref{fig:1}):
\begin{equation}
    \delta = \omega_0 - \frac{\omega_{31}+\omega_{32}}{2}\,,\quad \delta_R = \omega_1 - \omega_2 - \Delta_g\,.
\end{equation}
Here, $\delta$ is the detuning of the radiation carrier frequency from the mean frequency of the two dipole transitions, and $\delta_R$ is the Raman (two-photon) detuning. 
We will subsequently examine the resonant absorption profiles specifically as functions of $\delta$. In this case, the optical frequency detunings in the Lorentzians (\ref{eq:lorenziansL}) can be replaced by the following expressions: $\omega_1 - \omega_{31} = \delta + \frac{1}{2}\delta_R$ and $\omega_2 - \omega_{32} = \delta - \frac{1}{2}\delta_R$.
We will continue to use these same substitutions.

\subsection{Second Order}

From equation (\ref{eq:sysro110}) given in Appendix A, as well as from the solutions for the first-order corrections to the optical coherences, which can be obtained by analogy with solution (\ref{eq:solro13m11}), we find the following expression for the second-order correction to the density matrix of level $F_1$:
\begin{multline}\label{eq:r1102}
        \hat{\rho}_{11}^{(0)}(2) = \tau\,\widehat{\Gamma}_{31}\left\{\hat{\rho}_{33}^{(0)}(2)\right\} + \\+ i\tau\biggl( R_1 \widehat{\tilde{V}}_{13}^{(1)}\hat{\rho}_{31}^{(+1)}(1) +
        R_3 \widehat{\tilde{V}}_{13}^{(3)}\hat{\rho}_{31}^{(-1)}(1) -\\- R_1 \hat{\rho}_{13}^{(-1)}(1) \widehat{\tilde{V}}_{31}^{(1)} -
        R_3 \hat{\rho}_{13}^{(+1)}(1) \widehat{\tilde{V}}_{31}^{(3)} \biggr) = \\= 
        \tau\,\widehat{\Gamma}_{31}\left\{\hat{\rho}_{33}^{(0)}(2)\right\} - \frac{2\gamma_{eg}\tau R_1^2 S_1^{(-)}}{(2J_g+1)(2I_n+1)} \widehat{\tilde{V}}_{13}^{(1)} \widehat{\tilde{V}}_{31}^{(1)} -\\-
        \frac{2\gamma_{eg}\tau R_3^2 S_1^{(+)}}{(2J_g+1)(2I_n+1)} \widehat{\tilde{V}}_{13}^{(3)} \widehat{\tilde{V}}_{31}^{(3)}.
\end{multline}
Taking into account (\ref{eq:VVkq1}) and (\ref{eq:sponprihkq}), this expression in the $\varkappa q$-representation takes the form:
\begin{multline}\label{eq:r11022}
    \left\{ \hat{\rho}_{11}^{(0)}(2) \right\}_\varkappa =(-1)^{F_1+F_3+\varkappa + 1}\SixJ{F_3}{F_3}{\varkappa}{F_1}{F_1}{1} \times \\ \times \gamma\tau\beta_{31} (2F_3+1)\left\{\hat{\rho}_{33}^{(0)}(2)\right\}_\varkappa +\\+ 
    \frac{(-1)^{F_1+F_3+\varkappa + 1}}{(2J_g+1)(2I_n+1)}\SixJ{1}{1}{\varkappa}{F_1}{F_1}{F_3}\times \\ \times \biggl( 2\gamma_{eg}\tau R_1^2 S_1^{(-)}\left\{ \xiB_{(1)}\otimes\xiB_{(1)}^* \right\}_\varkappa +\\+ 2\gamma_{eg}\tau R_3^2 S_1^{(+)}\left\{ \xiB_{(2)}\otimes\xiB_{(2)}^* \right\}_\varkappa \biggr).
\end{multline}
In these expressions and in the following, we use the following notation for real Lorentzian forms:
\begin{eqnarray}\label{eq:lorenziansS}
    &&S_1^{(\pm)} = \frac{1}{\gamma_{eg}^2 + \left( \delta + \frac{1}{2}\delta_R \pm k_1\upsilon \right)^2},\nonumber\\
    &&S_2^{(\pm)} = \frac{1}{\gamma_{eg}^2 + \left( \delta - \frac{1}{2}\delta_R \pm k_2\upsilon \right)^2}.
\end{eqnarray}

Although the solution (\ref{eq:r11022}), written in the $\varkappa q$-representation, appears more cumbersome, it possesses a number of undeniable advantages.
Specifically, a comparison of (\ref{eq:r11022}) and (\ref{eq:r1102}) reveals that the $\varkappa q$-representation explicitly highlights the polarization properties of the solution—contained in the terms $\left\{ \xiB_{(j)}\otimes\xiB_{(j)}^* \right\}_\varkappa$.
Furthermore, the characteristics of various types of dipole transitions—explicitly described in this expression via $6j$-symbols—become clear. 
Thus, it is evident that the $\varkappa q$-representation significantly simplifies the theoretical analysis of the problem, enabling one to identify the physical origins of specific features observed in the resonant absorption profile.
We shall return to this point when analyzing the corresponding spectroscopic signals.
We also note that, as previously mentioned, the expressions (\ref{eq:r1102}) and (\ref{eq:r11022}) contain a "time-of-flight" saturation parameter of the form $\gamma_{eg}\tau R_j^2 S_1^{(\pm)}$ and under exact resonance conditions ($\delta=0$, $\delta_R=0$, $k_1\upsilon=0$, $k_2\upsilon=0$), this parameter takes on the familiar form $R_j^2/\Gamma\gamma_{eg}$.

The first term in (\ref{eq:r11022}) describes the transfer of the total population from the excited level $F_3$ to the level $F_1$ due to spontaneous relaxation (for $\varkappa=0$), as well as the spontaneous transfer of anisotropy (for $\varkappa \neq 0$), i.e., polarization moments of the first and second ranks.
As is known \cite{Rautian1994,Akulshin1998,Taichenachev1999,Goren2003,Zigdon2008}, this process can significantly affect the absorption level in a medium of resonant atoms and will be discussed later in the analysis of the sub-Doppler resonance line shape (Section~\ref{sec:5.2}).
The second and third terms in (\ref{eq:r11022}) are associated with the process of stimulated light absorption by the atom.

By analogy with the derivation of formula (\ref{eq:r1102}), using equations (\ref{eq:sysro220})–(\ref{eq:sysro12m}) given in Appendix A, we write down the remaining non-zero second-order PT corrections in the $\varkappa q$-representation, taking into account formulas (\ref{eq:VVkq1}), (\ref{eq:VVkq2}), and (\ref{eq:sponprihkq}):
\begin{multline} \label{eq:r12m2}
        \left\{ \hat{\rho}_{12}^{(-)}(2) \right\}_\varkappa = \frac{(-1)^{F_1+F_3+\varkappa+1}}{(2J_g+1)(2I_n+1)}\SixJ{1}{1}{\varkappa}{F_1}{F_2}{F_3} \times \\ \! \times \! 
        \tau R_1 R_2 N^{(-)} \biggl(\! L_1^{(-)}\! +\! L_2^{(-)*}\! \biggr)\! \left\{ \xiB_{(1)}\otimes\xiB_{(1)}^* \right\}_\varkappa,
\end{multline}
\begin{multline} \label{eq:r12p2}
        \left\{\!\hat{\rho}_{12}^{(+)}(2)\! \right\}_\varkappa \! = \! \frac{(-1)^{F_1+F_3+\varkappa+1}e^{i\phi_{12}}}{(2J_g+1)(2I_n+1)}\! \SixJ{1}{1}{\varkappa}{F_1}{F_2}{F_3}\times \\ \! \times \!
        \tau R_3 R_4 \!  N^{(+)} \biggl(\! L_1^{(+)}\! +\! L_2^{(+)*} \!\biggr)\! \left\{ \xiB_{(2)}\otimes\xiB_{(2)}^* \right\}_\varkappa,
\end{multline}
\begin{multline} \label{eq:r2202}
        \left\{ \hat{\rho}_{22}^{(0)}(2) \right\}_\varkappa = (-1)^{F_2+F_3+\varkappa + 1}\SixJ{F_3}{F_3}{\varkappa}{F_2}{F_2}{1} \times \\ \times \gamma\tau\beta_{32} (2F_3+1)\left\{\hat{\rho}_{33}^{(0)}(2)\right\}_\varkappa +\\+
        \frac{(-1)^{F_2+F_3+\varkappa + 1}}{(2J_g+1)(2I_n+1)}\SixJ{1}{1}{\varkappa}{F_2}{F_2}{F_3} \times \\ \times 
        \biggl( 2\gamma_{eg}\tau R_2^2 S_2^{(-)}\left\{ \xiB_{(1)}\otimes\xiB_{(1)}^* \right\}_\varkappa + \\ +
        2\gamma_{eg}\tau R_4^2 S_2^{(+)}\left\{ \xiB_{(2)}\otimes\xiB_{(2)}^* \right\}_\varkappa \biggr),
\end{multline}
\begin{multline} \label{eq:r3302}
        \left\{ \hat{\rho}_{33}^{(0)}(2) \right\}_\varkappa = \\ = \frac{(-1)^{F_1+F_3+\varkappa}}{(2J_g+1)(2I_n+1)(1+\gamma\tau)} \SixJ{1}{1}{\varkappa}{F_3}{F_3}{F_1} \times \\ \times \Biggl( 2\gamma_{eg}\tau R_1^2 S_1^{(-)}\left\{ \xiB_{(1)}^*\otimes\xiB_{(1)} \right\}_\varkappa +\\+
        2\gamma_{eg}\tau R_3^2 S_1^{(+)}\left\{ \xiB_{(2)}^*\otimes\xiB_{(2)} \right\}_\varkappa \Biggr) +\\+ 
        \frac{(-1)^{F_2+F_3+\varkappa}}{(2J_g+1)(2I_n+1)(1+\gamma\tau)}\SixJ{1}{1}{\varkappa}{F_3}{F_3}{F_2} \times \\ \times \Biggl( 2\gamma_{eg}\tau R_2^2 S_2^{(-)}\left\{ \xiB_{(1)}^*\otimes\xiB_{(1)} \right\}_\varkappa +\\+
        2\gamma_{eg}\tau R_4^2 S_2^{(+)}\left\{ \xiB_{(2)}^*\otimes\xiB_{(2)} \right\}_\varkappa \Biggr),
\end{multline}
Expressions for $\hat{\rho}_{21}^{(+)}(2)$ and $\hat{\rho}_{21}^{(-)}(2)$ can be obtained by Hermitian conjugation of (\ref{eq:r12m2}) and (\ref{eq:r12p2}), respectively.
In writing the solutions, the phase difference $\phi_{12} = \phi_1 - \phi_2$ was introduced, as well as complex Lorentzians associated with the Raman frequency detuning:
\begin{equation}\label{eq:NLorentz}
    N^{(\pm)} = \frac{1}{1 + i \tau \left( \delta_R \pm k_{12}\upsilon \right)}\,.
\end{equation}

Note that the expressions (\ref{eq:r2202})–(\ref{eq:r12p2}) do not include solutions for the harmonics $\hat{\rho}_{nn}^{(\pm)}$, the equations for which are given in Appendix A.
This is because these harmonics are non-zero only starting from the fourth order of smallness in perturbation theory, which lies beyond the scope of the present work.

\subsection{Third Order}
Next, to calculate the spectroscopic signal—specifically, the medium's polarization vector at the corresponding frequency—to third order in the transition dipole moment, we require only the optical coherences.
Therefore, we write down the solutions only for the harmonics of the optical coherences in the $\varkappa q$-representation:
\begin{multline} \label{eq:r13m13}
        \left\{ \hat{\rho}_{13}^{(-1)}(3) \right\}_1 = i(-1)^{F_1+F_3}L_1^{(-)}\sum_{\varkappa=0}^2\Pi_\varkappa \times \\ \times
        \Bigg(R_1\SixJ{1}{1}{\varkappa}{F_1}{F_1}{F_3} \left\{\xiB_{(1)}^*\otimes\left\{\hat{\rho}^{(0)}_{11}(2)\right\}_\varkappa\right\}_1 - \\ - 
        R_1\SixJ{1}{1}{\varkappa}{F_3}{F_3}{F_1} \left\{\left\{\hat{\rho}^{(0)}_{33}(2)\right\}_\varkappa\otimes\xiB_{(1)}^*\right\}_1 + \\ +
        R_2\SixJ{1}{1}{\varkappa}{F_1}{F_2}{F_3} \left\{\xiB_{(1)}^*\otimes\left\{\hat{\rho}^{(-)}_{12}(2)\right\}_\varkappa\right\}_1\Bigg),
\end{multline}
\begin{multline} \label{eq:r13m213}
        \left\{ \hat{\rho}_{13}^{(-21)}(3) \right\}_1 = i(-1)^{F_1+F_3} M_1^{(-)} \sum_{\varkappa=0}^2\Pi_\varkappa \times \\ \times 
        R_2\SixJ{1}{1}{\varkappa}{F_1}{F_2}{F_3} \left\{\xiB_{(1)}^*\otimes\left\{\hat{\rho}^{(+)}_{12}(2)\right\}_\varkappa\right\}_1,
\end{multline}
\begin{multline} \label{eq:r23m23}
    \left\{ \hat{\rho}_{23}^{(-2)}(3) \right\}_1 = i(-1)^{F_2+F_3}L_2^{(-)}\sum_{\varkappa=0}^2\Pi_\varkappa \times \\ \times \Bigg(R_2\SixJ{1}{1}{\varkappa}{F_2}{F_2}{F_3} \left\{\xiB_{(1)}^*\otimes\left\{\hat{\rho}^{(0)}_{22}(2)\right\}_\varkappa\right\}_1 -\\-
    R_2\SixJ{1}{1}{\varkappa}{F_3}{F_3}{F_2} \left\{\left\{\hat{\rho}^{(0)}_{33}(2)\right\}_\varkappa\otimes\xiB_{(1)}^*\right\}_1 +\\+
    R_1\SixJ{1}{1}{\varkappa}{F_1}{F_2}{F_3} \left\{\xiB_{(1)}^*\otimes\left\{\hat{\rho}^{(+)}_{21}(2)\right\}_\varkappa\right\}_1\Bigg),
\end{multline}
\begin{multline} \label{eq:r23m123}
        \left\{ \hat{\rho}_{23}^{(-12)}(3) \right\}_1 = i(-1)^{F_2+F_3} M_2^{(-)} \sum_{\varkappa=0}^2\Pi_\varkappa \times \\ \times
        R_1\SixJ{1}{1}{\varkappa}{F_1}{F_2}{F_3} \left\{\xiB_{(1)}^*\otimes\left\{\hat{\rho}^{(-)}_{21}(2)\right\}_\varkappa\right\}_1.
\end{multline}
Here, $M_{1,2}^{(-)}$ are complex Lorentzians defined by the following formulas:
\begin{equation}
    \begin{aligned}[b] \label{eq:MLorentz}
    &M_1^{(\pm)} = \frac{1}{\gamma_{eg} + i \left[ \delta + \frac{1}{2}\delta_R \pm (2k_2-k_1)\upsilon \right)},\\
    &M_2^{(\pm)} = \frac{1}{\gamma_{eg} + i \left[ \delta - \frac{1}{2}\delta_R \pm (2k_1-k_2)\upsilon \right)}.
    \end{aligned}
\end{equation}

\section{Propagation of Light in a Rarefied Gas of Resonant Atoms}\label{sec:4}

\subsection{Reduced Wave Equation}

As a result of absorption in the medium, the amplitude $E_j$ of each monochromatic light component in (\ref{eq:field}) depends slowly (compared to the wave phase) on the coordinate $z$. This process is governed by the reduced wave equation for the vector amplitude of the $j$-th wave:
\begin{equation}\label{eq:Propagation1}
    \frac{\partial {\bf E}_j}{\partial z} = 2\pi i k \tilde{{\bf P}}_j(z)\,,
\end{equation}
where ${\bf E}_j = E_j \xiB_{(j)}$ and $\tilde{{\bf P}}_j(z)$ is the spatially slowly varying vector amplitude of the medium's polarization corresponding to the polarization at the frequency of the $j$-th wave.
In equation (\ref{eq:Propagation1}), we use a single notation for the wave number $k_1 = k_2 = k$ preceding the vector $\tilde{{\bf P}}_j(z)$, since the difference between $k_1$ and $k_2$ is negligible in this case.

Let us consider the absorption of a probe beam consisting of waves ${\bf E}_1$ and ${\bf E}_2$.
For the moment, we place no restrictions on its intensity relative to the counterpropagating pump beam.
Since experiments measure the light intensity $I$, the propagation law (\ref{eq:Propagation1}) can be written for the intensity of wave ${\bf E}_1$ using the relation $I_j = (c/2\pi)|{\bf E}_j|^2$:
\begin{equation}\label{eq:Propagation2}
    \frac{\partial I_1}{\partial z} = 2kcE_1\,\text{Im}\left\{\xiB_{(1)}\tilde{{\bf P}}_1^*(z)\right\}=-\chi_{p1}\,I_1\,,
\end{equation}
where the absorption coefficient $\chi_{p1}$ for the wave ${\bf E}_1$ was introduced (measured in cm$^{-1}$).
In accordance with the definition of polarization as the average dipole moment per unit volume of the medium, we have:
\begin{equation}\label{eq:Polarization}
    {\bf P}_1(z,t) = n_a\langle{\bf d}_1(z,t)\rangle_\upsilon = \tilde{{\mathbf{P}}}_1(z) e^{-i\omega_1t+ik_1z},
\end{equation}
where $n_a$ is the gas atom concentration, assumed to be low enough that particle collisions do not play a significant role.
Otherwise, accounting for collisions would substantially complicate the theoretical analysis (see, e.g., \cite{Rautian1991}).
The angle brackets $\langle\dots\rangle_\upsilon$ in (\ref{eq:Polarization}) denote averaging over the (Maxwellian) velocity distribution of the gas particles, and ${\bf d}_{1}$ is the quantum-mechanical average of the atomic dipole operator associated with the action of the wave ${\bf E}_1$, calculated in the standard way: ${\bf d}(z,t)$$\,=\,$$\text{Tr}$$\left\{\hat{{\bf d}}\hat{\rho}(z,t)\right\}$.
Then, taking into account (\ref{Dipmomentoperat}), (\ref{eq:Harmonics3}), and (\ref{eq:moments}), we arrive at the expression:
\begin{multline} \label{eq:Polarization2}
    \tilde{\bf P}_1^*(z) = \frac{n_a d_{31}}{\sqrt{3}}\Biggl(\Bigl\langle \left\{\hat{\rho}_{13}^{(-1)}\right\}_1 \Bigr \rangle_\upsilon  +  \\
    +\Bigl\langle \left\{\hat{\rho}_{13}^{(-21)}\right\}_1 \Bigr \rangle_\upsilon \,e^{2ik_{12}z}\biggr)\,.
\end{multline}

For the second component of the probe beam, i.e., the wave ${\bf E}_2$, similar expressions can be obtained:
\begin{equation}\label{eq:Propagation3}
    \frac{\partial I_2}{\partial z} = 2kcE_2\,\text{Im}\left\{\xiB_{(1)}\tilde{{\bf P}}_2^*(z)\right\}=-\chi_{p2}\,I_2\,,
\end{equation}
where $\chi_{p2}$ is the absorption coefficient of the wave ${\bf E}_2$, and the polarization induced by this wave is:
\begin{multline}\label{eq:Polarization3}
    \tilde{\bf P}_2^*(z) = \frac{n_a d_{32}}{\sqrt{3}}\Biggl(\Bigl\langle \left\{\hat{\rho}_{23}^{(-2)}\right\}_1 \Bigr \rangle_\upsilon  +  \\
    +\Bigl\langle \left\{\hat{\rho}_{23}^{(-12)}\right\}_1 \Bigr \rangle_\upsilon \,e^{-2ik_{12}z}\biggr)\,.
\end{multline}

The total intensity of the probe beam, i.e., $I_p = (c/2\pi)|{\bf E}_1 + {\bf E}_2|^2$, contains an interference term proportional to $\text{Re}\{{\bf E}_1{\bf E}_2^*\}$.
However, this term oscillates at an ultra-high frequency ($\approx \Delta_g$) and averages to zero at the photodetector. 
Therefore, it can be disregarded. 
In this case, the total intensity of the probe beam is simply the sum of the intensities of its two constituent waves, $I_p = I_1 + I_2$.
Taking this into account, and considering equations (\ref{eq:Propagation2}) and (\ref{eq:Propagation3}), the propagation law for $I_p$ takes a simple form:
\begin{equation}\label{eq:PropagationProbe}
    \frac{\partial I_p}{\partial z} = -\chi_{p1}I_1 -\chi_{p2}I_2\,.
\end{equation}

Substituting (\ref{eq:Polarization2}) and (\ref{eq:Polarization3}) into (\ref{eq:Propagation2}) and (\ref{eq:Propagation3}), respectively, we find expressions for the absorption coefficients:
\begin{multline}\label{eq:Chi11}
    \chi_{p1} = -\frac{\sqrt{3}n_a\lambda^2\beta_{31}\gamma \, (2F_3+1)}{4\pi R_1}\times\\ \times
    \text{Im}\Biggl[ \xiB_{(1)}\Biggl( \Big\langle \left\{ \hat{\rho}_{13}^{(-1)}(1) \right\}_1 \Big\rangle_\upsilon + \Big\langle \left\{ \hat{\rho}_{13}^{(-1)}(3) \right\}_1 \Big\rangle_\upsilon +\\+
    \Big\langle \left\{ \hat{\rho}_{13}^{(-21)}(3) \right\}_1 \Big\rangle_\upsilon e^{2ik_{12}z}\Biggr) \Biggr],
\end{multline} 
\begin{multline}\label{eq:Chi21}
    \chi_{p2} = -\frac{\sqrt{3}n_a\lambda^2\beta_{32}\gamma \, (2F_3+1)}{4\pi R_2}\times\\\times
    \text{Im}\Biggl[ \xiB_{(1)}\Biggl( \Big\langle \left\{ \hat{\rho}_{23}^{(-2)}(1) \right\}_1 \Big\rangle_\upsilon +\Big\langle \left\{ \hat{\rho}_{23}^{(-2)}(3) \right\}_1 \Big\rangle_\upsilon +\\+
    \Big\langle \left\{ \hat{\rho}_{23}^{(-12)}(3) \right\}_1 \Big\rangle_\upsilon e^{-2ik_{12}z}\Biggr) \Biggr].
\end{multline}
The following relations were taken into account in deriving these expressions:
\begin{equation}\label{eq:IntensRabi}
    I_1 = \frac{c\hbar^2}{2\pi d_{31}^2}R_1^2\,,\quad I_2 = \frac{c\hbar^2}{2\pi d_{32}^2}R_2^2,
\end{equation}
as well as the relationship between the square of the transition dipole moment matrix element (relative line strength) and the spontaneous relaxation rate \cite{Sobelman1972}:
\begin{equation}\label{eq:Dipelement}
    d_{3j}^2 = \frac{3\hbar\lambda^3\beta_{3j}\gamma\,(2F_3+1)}{32\pi^3}\,.
\end{equation}

Next, we adopt an approximation in which the probe beam intensity is sufficiently low that the medium's absorption coefficient is independent of that beam—a linear approximation with respect to the probe field. 
In this case, we assume $I_1 \approx I_2 \ll I_3, I_4$, and that all nonlinear effects arise solely from the counterpropagating pump beam (waves ${\bf E}_3$ and ${\bf E}_4$), which modifies the optical properties of the medium.
We also assume the resonant medium is sufficiently short that the terms $e^{\pm 2ik_{12}z}$ in (\ref{eq:Chi11}) and (\ref{eq:Chi21}) do not change significantly over the length of the cell.
This approximation is entirely adequate for the development of a compact phase-conjugate mirror utilizing a miniature atomic cell; such cells have been used, for example, in experiments observing sub-Doppler resonances via the DFSDS technique \cite{Brazhnikov2019,Gusching2021,Gusching2023}.

We assume the optical density of the medium for the pump beam to be low, such that the change in its intensity along the medium can be neglected. 
This approximation is particularly easy to satisfy in practice in the "pump-probe" regime.
This is due, firstly, to the short length of the atomic cell and, secondly, to the fact that the pump beam transfers a significant fraction of the atoms into a CPT state, thereby rendering the medium transparent to the beam itself. 
Since the probe beam is sufficiently weak, it does not destroy the atomic state induced by the pump beam.
Note that, in the general case, accounting for the dependence of the pump wave intensity on the coordinate $z$ in a gas of resonant atoms requires the simultaneous solution of the Maxwell-Bloch equations \cite{Castin1995,Voloshin2022,Yudin2022}.
However, doing so would significantly complicate the problem at hand and preclude obtaining analytical solutions.

All the approximations made allow us to treat the coordinate $z$ not as a variable, but as an additional parameter (alongside other parameters, such as Rabi frequencies or wave polarization parameters). 
Equations (\ref{eq:Propagation2}) and (\ref{eq:Propagation3}) can then be easily integrated, yielding solutions in the form of the Bouguer–Lambert–Beer law:
\begin{equation}\label{eq:SolutionProb}
    I_1(L) = I_{10}\,e^{-\chi_{p1}L}\,,\quad I_2(L) = I_{20}\,e^{-\chi_{p2}L}\,,
\end{equation}
where $I_{10}$ and $I_{20}$ are the intensities of waves ${\bf E}_1$ and ${\bf E}_2$ at the entrance to a cell of length $L$. Note that this solution holds even when the medium is optically dense for the probe beam—i.e., when $\chi_{p1}L, \chi_{p2}L \gtrsim 1$.

Next, we focus on analyzing the expressions for the absorption coefficients $\chi_{p1}$ and $\chi_{p2}$, rather than on analyzing the intensities of the waves ${\bf E}_1$ and ${\bf E}_2$ at the cell output. Expressions for the latter can always be obtained using formula (\ref{eq:SolutionProb}). Based on the form of the expressions for the absorption coefficients (\ref{eq:Chi11}) and (\ref{eq:Chi21}), they can be represented in the following general form ($j=1, 2$):
\begin{equation}\label{eq:Chi}
    \chi_{pj} = \chi_{pj}^{\rm lin} + \chi_{pj}^{\rm nlin1} + \chi_{pj}^{\rm nlin2}(z)\,,
\end{equation}
where $I_{10}$ and $I_{20}$ are the intensities of waves ${\bf E}_1$ and ${\bf E}_2$ at the entrance to a cell of length $L$.
Note that this solution holds even when the medium is optically dense for the probe beam, i.e., when $\chi_{p1}L, \chi_{p2}L \gtrsim 1$.

\subsection{Linear Absorption of the Probe Beam}

In expressions (\ref{eq:Chi11}) and (\ref{eq:Chi21}), the terms associated with first-order field corrections, namely, $\left\{ \hat{\rho}_{13}^{(-1)}(1) \right\}_1$ and $\left\{ \hat{\rho}_{23}^{(-2)}(1) \right\}_1$, account for linear light absorption in the medium. Using the explicit expressions for these corrections, (\ref{r13m1one}) and (\ref{r23m2one}), we arrive at the following result:
\begin{align}
    \chi_{p1}^{\rm lin} = \frac{n_a\lambda^2\gamma_{eg}\beta_{31}\gamma (2F_3+1)}{4\pi (2J_g+1)(2I_n+1)} \langle S_1^{(-)} \rangle_\upsilon,\label{eq:Chi1Lin1}\\
    \chi_{p2}^{\rm lin} = \frac{n_a\lambda^2\gamma_{eg}\beta_{32}\gamma (2F_3+1)}{4\pi (2J_g+1)(2I_n+1)} \langle S_2^{(-)} \rangle_\upsilon,\label{eq:Chi2Lin1}
\end{align}
where the Lorentzians $S_{1,2}^{(-)}$ are given in (\ref{eq:lorenziansS}).
Upon scanning the optical frequency detuning $\delta$, these expressions describe a broad Doppler profile.
To demonstrate this, we perform an averaging over the Maxwellian velocity distribution:
\begin{multline}\label{eq:S1maveraging}
    \langle S_1^{(-)} \rangle_\upsilon = \frac{1}{\sqrt{\pi}\upsilon_0}\int\limits_{\mathclap{-\infty}}^{\mathclap{\infty}}\frac{e^{-\upsilon^2/\upsilon_0^2}}{\gamma_{eg}^2+\left[ \delta + \frac{1}{2}\delta_R-k\upsilon \right]^2}d\upsilon = \\= 
    \frac{\sqrt{\pi}}{k\upsilon_0\gamma_{eg}}\,W_r\left(\frac{\gamma_{eg}-i[\delta+\frac{1}{2}\delta_R]}{k\upsilon_0}\right).
\end{multline}
A similar expression can be written for $\langle S_2^{(-)} \rangle_\upsilon$.
Here, $\upsilon_0 = \sqrt{2k_BT/m_a}$ is the most probable thermal speed of the atoms in the gas, where $m_a$ is the atomic mass.
The function $W_r$ is the real part of the complex error function, i.e.:
\begin{equation}\label{eq:Wr}
    W_r(p) = \text{Re}\left\{ e^{-p^2} \Bigl[1-\text{Erf}(-ip)\Bigr]\right\}
\end{equation}
In our case, $p = (\delta + \delta_R/2 + i\gamma_{eg})/k\upsilon_0$. 
The function $\text{Erf}(-ip)$ is the error function:
\begin{equation}\label{eq:Erf}
    \text{Erf}(x) = \frac{2}{\sqrt{\pi}}\int\limits_0^x e^{-t^2} dt\,.
\end{equation}

When scanning the detuning $\delta$, the function $W_r$ describes a broad absorption profile known as the Voigt profile.
However, in spectroscopic studies, this profile is often approximated by a Gaussian function, also known as the Doppler profile.
This approximation is readily justified by noting that, in the vast majority of experiments employing the frequency-modulated saturated-absorption spectroscopy (DFSDS) technique, the condition $\delta_R \ll \gamma_{eg} \ll k\upsilon_0$ holds.
Consequently, in the range of $\delta$ values of greatest interest—specifically $\delta < k\upsilon_0$, where the sub-Doppler resonance is concentrated—one can make the following approximation:
\begin{equation}\label{eq:Wr2}
    W_r(p) \approx e^{-\left(\delta/k\upsilon_0\right)^2}\,,
\end{equation}
and the expressions for the linear absorption coefficients (\ref{eq:Chi1Lin1}) and (\ref{eq:Chi2Lin1}) will take the form:
\begin{multline}\label{eq:Chi1Lin2}
    \chi_{p1}^{\rm lin} \approx \frac{n_a\lambda^2\beta_{31}\gamma (2F_3+1)}{4\sqrt{\pi} \Delta_D (2J_g+1)(2I_n+1)} e^{-\delta^2/\Delta_D^2} =\\= 
    a_0 s_{13}e^{-\delta^2/\Delta_D^2},
\end{multline}
\begin{multline} \label{eq:Chi2Lin2}
    \chi_{p2}^{\rm lin} \approx \frac{n_a\lambda^2\beta_{32}\gamma (2F_3+1)}{4\sqrt{\pi} \Delta_D (2J_g+1)(2I_n+1)} e^{-\delta^2/\Delta_D^2} =\\= 
    a_0 s_{23}e^{-\delta^2/\Delta_D^2},
\end{multline}
These formulas describe broad Doppler profiles with a half-width at the $e^{-1}$ level equal to $\Delta_D = k\upsilon_0$.
The amplitudes of these profiles are related by the ratio $\chi_{p1}^{\rm lin}/\chi_{p2}^{\rm lin} = s_{31}/s_{32}$, where $s_{ij}$ is the relative strength of the dipole transition $F_i \to F_j$ with respect to all other transitions of the multiplet component:
\begin{equation}\label{eg:strength}
    s_{ij} = \frac{(2F_i+1)(2F_j+1)}{2I_n+1}\SixJ{J_g}{F_i}{I_n}{F_j}{J_e}{1}^2
\end{equation}
These strengths are normalized such that $\sum_{ij}s_{ij} = 1$.
Specific values of these coefficients for the D$_1$ and D$_2$ lines are listed in Tables \ref{tab:1} and \ref{tab:2} in Appendix B.

In formulas (\ref{eq:Chi1Lin2}) and (\ref{eq:Chi2Lin2}), the constant $a_0$ was introduced for convenience, in accordance with the definition:
\begin{equation}\label{eq:acoef}
    a_0 = \frac{n_a\lambda^2\gamma}{4\sqrt{\pi} \Delta_D}\times\frac{2J_e+1}{2J_g+1}\,.
\end{equation}
The second fraction in this expression shows the factor by which the D$_2$ line is stronger than the D$_1$ line in alkali metal atoms.
Specifically, since $J_e = 1/2$ for the D$_1$ line and $J_e = 3/2$ for the D$_2$ line, the D$_2$ line is twice as strong as the D$_1$ line.

Thus, in the general case, the waves ${\bf E}_1$ and ${\bf E}_2$ undergo different absorption in the medium.
The expressions (\ref{eq:Chi1Lin2}) and (\ref{eq:Chi2Lin2}) also imply the obvious fact that, in the linear regime, the absorption coefficients for the pump beam do not depend on its polarization parameters.

\subsection{Nonlinear Effects in Probe Beam Absorption}

Since we are considering the linear regime with respect to the probe beam, all nonlinear optical effects in the probe beam absorption arise exclusively from the action of the pump beam.
Clearly, this effect manifests itself only when both beams (probe and pump) interact with the same atoms in the gas. Due to the Doppler frequency shift, this condition is met only within the range $\delta \lesssim \gamma_{eg}$.
Consequently, the broad Doppler profile of the probe beam absorption remains unaffected by nonlinear effects -- a fact we will demonstrate below when deriving the relevant expressions.

The nonlinear terms $\chi_{pj}^{\rm nlin1}$ from (\ref{eq:Chi}) are related to the third-order corrections to the optical coherences $\left\{ \hat{\rho}_{13}^{(-1)}(3) \right\}_1$ and $\left\{ \hat{\rho}_{23}^{(-2)}(3) \right\}_1$ in formulas (\ref{eq:Chi11}) and (\ref{eq:Chi21}). 
Substituting the solutions for these corrections according to formulas (\ref{eq:r13m13}) and (\ref{eq:r23m23}), we obtain:
\begin{multline} \label{eq:chi1nlin1}
    \chi_{p1}^{\rm nlin1} = (-1)^{F_1+F_3+1}\frac{\sqrt{3}n_a\lambda^2\beta_{31}\gamma(2F_3+1)}{4\pi} \times \\ \times 
    \sum_{\varkappa=0}^2 \Pi_\varkappa
    \text{Re}\Bigg[  \Big\langle\xiB_{(1)}\Big\{ \xiB_{(1)}^* \otimes \big\{ \hat{\rho}_{11}^{(0)}(2) \big\}_\varkappa \Big\}_1 L_1^{(-)}\Big\rangle_\upsilon \times \\ \times \SixJ{1}{1}{\varkappa}{F_1}{F_1}{F_3} - \Big\langle\xiB_{(1)}\Big\{ \big\{ \hat{\rho}_{33}^{(0)}(2) \big\}_\varkappa \otimes \xiB_{(1)}^*\Big\}_1 L_1^{(-)}\Big\rangle_\upsilon \times \\ \times \SixJ{1}{1}{\varkappa}{F_3}{F_3}{F_1} +\Big\langle\xiB_{(1)}\Big\{ \xiB_{(1)}^* \otimes \big\{ \hat{\rho}_{12}^{(-)}(2) \big\}_\varkappa \Big\}_1 L_1^{(-)}\Big\rangle_\upsilon \times \\ \times \frac{R_2}{R_1}\SixJ{1}{1}{\varkappa}{F_1}{F_2}{F_3} \Bigg],
\end{multline}
\begin{multline} \label{eq:chi2nlin1}
    \chi_{p2}^{\rm nlin1} = (-1)^{F_2+F_3+1}\frac{\sqrt{3}n_a\lambda^2\beta_{32}\gamma(2F_3+1)}{4\pi} \times \\ \times 
     \sum_{\varkappa=0}^2\Pi_\varkappa\text{Re}\Bigg[ \Big\langle\xiB_{(1)}\Big\{ \xiB_{(1)}^* \otimes \big\{ \hat{\rho}_{22}^{(0)}(2) \big\}_\varkappa \Big\}_1 L_2^{(-)}\Big\rangle_\upsilon \times \\ \times \SixJ{1}{1}{\varkappa}{F_2}{F_2}{F_3} - \Big\langle\xiB_{(1)}\Big\{ \big\{ \hat{\rho}_{33}^{(0)}(2) \big\}_\varkappa \otimes \xiB_{(1)}^*\Big\}_1 L_2^{(-)}\Big\rangle_\upsilon \times \\ \times \SixJ{1}{1}{\varkappa}{F_3}{F_3}{F_2} + \Big\langle\xiB_{(1)}\Big\{ \xiB_{(1)}^* \otimes \big\{ \hat{\rho}_{21}^{(+)}(2) \big\}_\varkappa \Big\}_1 L_2^{(-)}\Big\rangle_\upsilon \times \\ \times \frac{R_1}{R_2}\SixJ{1}{1}{\varkappa}{F_1}{F_2}{F_3} \Bigg].
\end{multline} 
These formulas reflect the nonlinear action of the probe beam, as they include the Rabi frequencies $R_1$ and $R_2$.
In the linear regime with respect to the probe beam, which is the case of interest here, all such terms must be discarded after substituting the expressions for the second-order corrections (\ref{eq:r11022})–(\ref{eq:r12p2}).
In particular, the contributions from $\left\{\hat{\rho}_{12}^{(-)}(2)\right\}_\varkappa$ and $\left\{\hat{\rho}_{21}^{(+)}(2)\right\}_\varkappa$ will vanish in (\ref{eq:chi1nlin1}) and (\ref{eq:chi2nlin1}); these terms are associated with the generation of microwave (low-frequency) coherence between the hyperfine levels of the ground state driven by the probe beam, which is a nonlinear effect.

Thus, taking into account everything said above, we arrive at the following expressions for the absorption coefficients $\chi_{p1}^{\rm nlin1}$ and $\chi_{p2}^{\rm nlin1}$ in the linear regime with respect to the probe beam ($j = 1, 2$):
\begin{equation}\label{eq:chinlin1}
    \chi_{pj}^{\rm nlin1} = \chi_{pj}^{\rm nlin1}({\rm SAR}) + \chi_{pj}^{\rm nlin1}({\rm OP})\,,
\end{equation}
где:
\begin{multline} \label{eq:chi1nlin1SAR}
    \chi_{p1}^{\rm nlin1}({\rm SAR}) \approx \frac{\sqrt{3}\,a_0 s_{13} \gamma_{eg}\tau}{\delta^2+\gamma_{eg}^2}\sum_{\varkappa=0}^2 \biggl[A_{13}^2(\varkappa)R_3^2+\\+
    \frac{A_{31}^2(\varkappa)R_3^2}{1+\gamma\tau}-\frac{A_{31}(\varkappa)A_{32}(\varkappa)R_4^2}{1+\gamma\tau}\biggl]\Psi_{12}(\varkappa),
\end{multline}
\begin{multline} \label{eq:chi2nlin1SAR}
    \chi_{p2}^{\rm nlin1}({\rm SAR}) \approx \frac{\sqrt{3}\,a_0 s_{23} \gamma_{eg}\tau}{\delta^2+\gamma_{eg}^2}\sum_{\varkappa=0}^2 \biggl[A_{23}^2(\varkappa)R_4^2+\\+
    \frac{A_{32}^2(\varkappa)R_4^2}{1+\gamma\tau}-\frac{A_{32}(\varkappa)A_{31}(\varkappa)R_3^2}{1+\gamma\tau}\biggl]\Psi_{12}(\varkappa),
\end{multline}
\begin{multline} \label{eq:chi1nlin1OP}
    \chi_{p1}^{\rm nlin1}({\rm OP}) \approx \frac{\sqrt{3}\,a_0 s_{13}\beta_{31}(2F_3+1)\gamma\gamma_{eg}\tau^2}{(1+\gamma\tau)(\delta^2+\gamma_{eg}^2)}\times\\\times
    \sum_{\varkappa=0}^2 \Big[R_3^2 B_{11}(\varkappa)+R_4^2 B_{12}(\varkappa)\Big]\Psi_{12}(\varkappa),
\end{multline}
\begin{multline} \label{eq:chi2nlin1OP}
    \chi_{p2}^{\rm nlin1}({\rm OP}) \approx \frac{\sqrt{3}\,a_0 s_{23}\beta_{32}(2F_3+1)\gamma\gamma_{eg}\tau^2}{(1+\gamma\tau)(\delta^2+\gamma_{eg}^2)}\times\\ \times
    \sum_{\varkappa=0}^2 \Big[R_3^2 B_{21}(\varkappa)+R_4^2 B_{22}(\varkappa)\Big]\Psi_{12}(\varkappa).
\end{multline}
The following coefficients were introduced here ($i=1, 2$; $j=1, 2$):
\begin{equation} \label{eq:Acoef}
    A_{ij}(\varkappa) = \sqrt{\Pi_\varkappa}\,\SixJ{1}{1}{\varkappa}{F_i}{F_i}{F_j},
\end{equation}
\begin{multline} \label{eq:Bcoef}
    B_{ij}(\varkappa) = (-1)^{F_j+F_3}\Pi_\varkappa\SixJ{1}{1}{\varkappa}{F_i}{F_i}{F_3}\times\\
    \SixJ{F_3}{F_3}{\varkappa}{F_i}{F_i}{1}\SixJ{1}{1}{\varkappa}{F_3}{F_3}{F_j},
\end{multline}
\begin{equation} \label{eq:Psicoef}
    \Psi_{12}(\varkappa) = \xiB_{(1)}\Big\{ \big\{ \xiB_{(2)} \otimes \xiB_{(2)}^* \big\}_\varkappa \otimes \xiB_{(1)}^* \Big\}_1.
\end{equation}
The values of these coefficients for transitions in the D$_1$ lines of $^{87}$Rb and $^{133}$Cs atoms are listed in Tables \ref{tab:3}–\ref{tab:6} in Appendix B.

In deriving (\ref{eq:chi1nlin1SAR})–(\ref{eq:chi2nlin1OP}), we used the following approximate formulas by analogy with (\ref{eq:S1maveraging})–(\ref{eq:Wr2}):
\begin{multline}
    \langle S_1^{(-)} S_1^{(+)} \rangle_\upsilon \approx \langle S_1^{(-)} S_2^{(+)} \rangle_\upsilon \approx \\ \approx \frac{\sqrt{\pi}}{2\Delta_D\gamma_{eg}\left(\delta^2+\gamma_{eg}^2\right)}.
\end{multline}

The nonlinear terms (\ref{eq:chi1nlin1SAR})–(\ref{eq:chi2nlin1OP}) describe homogeneously broadened sub-Doppler resonances in the form of Lorentzian absorption profiles with a half-width of $\gamma_{eg}$.
The sign of these resonances and their relative weight within the total nonlinear signal are determined by the coefficients $A_j$, $B_{ij}$, and $\Psi_{12}$ (see Appendix B).
A more detailed analysis of these resonances is presented in Section \ref{sec:5.2}.
We note here only that the use of the $\varkappa q$-representation makes it possible to separate the features of the nonlinear resonances associated with the angular momenta of the dipole transitions (coefficients $A_j$, $B_{ij}$) from those associated exclusively with the polarization parameters of the light waves ($\Psi_{12}$).

In expression (\ref{eq:chinlin1}), the terms designated as "SAR" are associated with the saturated absorption resonance resulting from the action of counterpropagating pump waves.
Specifically, the first term in the square brackets of (\ref{eq:chi1nlin1SAR}), proportional to $R_3^2$, relates to the change in absorption of the probe wave ${\bf E}_1$ caused by the action of wave ${\bf E}_3$ on the ground state of atoms with angular momentum $F_1$ (see Fig.~\ref{fig:1}).
The second term in the square brackets is associated with the action of wave ${\bf E}_3$ on the excited state of the atoms (level $F_3$).
Finally, the third term in the square brackets of (\ref{eq:chi1nlin1SAR}), proportional to $R_4^2$, relates to the action of wave ${\bf E}_4$ on the excited state of the atoms as they undergo the stimulated transition $F_2$$\,\to\,$$F_3$.
The terms in expression (\ref{eq:chi2nlin1SAR}) have an analogous physical meaning. 
Thus, the "SAR" terms reflect nonlinear effects associated with stimulated atomic transitions that involve the redistribution of populations among Zeeman sublevels, as well as the formation of Zeeman coherences in the ground (levels $F_1$ and $F_2$) and excited ($F_3$) atomic states.

In the square brackets of expressions (\ref{eq:chi1nlin1SAR}) and (\ref{eq:chi2nlin1SAR}), the fractional terms can be neglected provided that $\gamma\tau \gg 1$.
This requirement is fully consistent with typical experimental conditions; physically, it implies that the lifetime of the atom in the excited energy level is significantly shorter than the time it spends within the light beam.
Neglecting these terms in formulas (\ref{eq:chi1nlin1SAR}) and (\ref{eq:chi2nlin1SAR}) is equivalent to neglecting the terms involving $\hat{\rho}_{33}^{(0)}(2)$ in the absorption coefficients (\ref{eq:chi1nlin1}) and (\ref{eq:chi2nlin1}).
However, the effect of populating the excited state $F_3$ does not vanish entirely; instead, it persists through the transfer of atoms to the ground-state levels $F_1$ and $F_2$ via spontaneous relaxation. 
According to formulas (\ref{eq:r11022}) and (\ref{eq:r11022}), this process contributes to the terms $\hat{\rho}_{11}^{(0)}(2)$ and $\hat{\rho}_{22}^{(0)}(2)$, which appear in the expressions for the absorption coefficients (\ref{eq:chi1nlin1}) and (\ref{eq:chi2nlin1}), respectively.

We will examine the process by which atoms populate the ground-state levels via spontaneous relaxation, which is distinct from the aforementioned process of stimulated photon absorption ("SAR"), and designate it as "OP" (optical pumping).
As seen from (\ref{eq:chi1nlin1OP}) and (\ref{eq:chi2nlin1OP}), both counterpropagating waves, ${\bf E}_3$ and ${\bf E}_4$, affect each of the probe-beam waves, ${\bf E}_1$ and ${\bf E}_2$, because spontaneous relaxation from level $F_3$ can occur to either level $F_1$ or level $F_2$ (in proportion to the respective branching ratios).
In formulas (\ref{eq:chi1nlin1OP}) and (\ref{eq:chi2nlin1OP}), the terms with $\varkappa=0$ describe the change in total population at levels $F_1$ and $F_2$, respectively, whereas the terms with $\varkappa \ne 0$ account for the process of spontaneous transfer of non-scalar polarization moments (orientation and alignment) from the excited state to the ground state—a process we also refer to as anisotropy transfer \cite{Rautian1994,Taichenachev1999,Goren2003} (for its role in the formation of sub-Doppler resonance, see Section \ref{sec:5.2}).

In the approximation $\gamma\tau \! \gg \! 1$, expressions (\ref{eq:chi1nlin1SAR})–(\ref{eq:chi2nlin1OP}) take a simpler form:
\begin{multline}
    \chi_{p1}^{\rm nlin1}({\rm SAR}) \approx \frac{\sqrt{3}\,a_0 s_{13}\gamma_{eg}\tau R_3^2}{\delta^2+\gamma_{eg}^2}\times \\ \times
    \sum_{\varkappa=0}^2 A_{13}^2(\varkappa)\Psi_{12}(\varkappa)\,,\label{eq:chi1nlin1SAR2}
\end{multline}
\begin{multline}
    \chi_{p2}^{\rm nlin1}({\rm SAR}) \approx \frac{\sqrt{3}\,a_0 s_{23}\gamma_{eg}\tau R_4^2}{\delta^2+\gamma_{eg}^2}\times \\ \times
    \sum_{\varkappa=0}^2 A_{23}^2(\varkappa)\Psi_{12}(\varkappa)\,,\label{eq:chi2nlin1SAR2}
\end{multline}
\begin{multline}
    \chi_{p1}^{\rm nlin1}({\rm OP}) \approx \frac{\sqrt{3}\,a_0 s_{13}\beta_{31}(2F_3+1)\gamma_{eg}\tau}{\delta^2+\gamma_{eg}^2}\times \\ \times
    \sum_{\varkappa=0}^2 \Big[R_3^2 B_{11}(\varkappa)+R_4^2 B_{12}(\varkappa)\Big]\Psi_{12}(\varkappa)\,,\label{eq:chi1nlin1OP2}
\end{multline}
\begin{multline}
    \chi_{p2}^{\rm nlin1}({\rm OP}) \approx \frac{\sqrt{3}\,a_0 s_{23}\beta_{32}(2F_3+1)\gamma_{eg}\tau}{\delta^2+\gamma_{eg}^2}\times \\ \times
    \sum_{\varkappa=0}^2 \Big[R_3^2 B_{21}(\varkappa)+R_4^2 B_{22}(\varkappa)\Big]\Psi_{12}(\varkappa)\,.\label{eq:chi2nlin1OP2}
\end{multline}

The remaining nonlinear terms from (\ref{eq:Chi}) that depend on $z$ are associated with the terms $\left\{ \hat{\rho}_{13}^{(-21)}(3) \right\}_1$ and $\left\{ \hat{\rho}_{23}^{(-12)}(3) \right\}_1$ in formulas (\ref{eq:Chi11}) and (\ref{eq:Chi21}).
From a practical standpoint, the total intensity of the probe beam after the atomic cell, which is governed by equation (\ref{eq:PropagationProbe}), is of interest.
Therefore, it is convenient to analyze the aforementioned terms in the combined form: $\chi_{p1}^{\rm nlin2}(z)\,I_1 + \chi_{p2}^{\rm nlin2}(z)\,I_2$. 
Taking into account the solutions (\ref{eq:r13m213}) and (\ref{eq:r23m123}), we find:
\begin{equation}\label{eq:nlin2z}
    \chi_{p1}^{\rm nlin2}(z)\,I_1 + \chi_{p2}^{\rm nlin2}(z)\,I_2 = 2\,\chi_{p12}^{\rm inter}(z)\,\sqrt{I_1 I_2}\,, 
\end{equation}
where a separate absorption coefficient is introduced for this nonlinear effect:
\begin{multline}\label{eq:chiinter}
\chi_{p12}^{\rm inter}(z) \approx \frac{a_0 \gamma_{eg} \sqrt{3 s_{13} s_{23}}}{(\gamma_{eg}^2+\delta^2)(\Gamma^2+\delta_R^2)} R_3 R_4 \times\\\times \Big\{ \Gamma\cos\big[\theta(z)\big] + \delta_R\sin\big[\theta(z)\big] \Big\} \times \\ \times \sum_{\varkappa=1}^2 C(\varkappa)\Psi_{12}(\varkappa).
\end{multline}
Here $\theta(z) = \phi_{12} + 2k_{12}z$, and the coefficient $C(\varkappa)$ is determined by the expression:
\begin{equation}
    C(\varkappa) = \Pi_\varkappa \SixJ{1}{1}{\varkappa}{F_1}{F_2}{F_3}^2\,, 
\end{equation}
The values of this coefficient for transitions in the $D_1$ lines of $^{87}$Rb and $^{133}$Cs atoms are listed in Table \ref{tab:7} in Appendix B.
To derive expression (\ref{eq:chiinter}), we employed the approximations $\delta_R \ll \gamma_{eg}$ and $k_1 \approx k_2$ within all the Lorentzian functions $L$, $M$, and $N$ appearing in (\ref{eq:lorenziansL}), (\ref{eq:NLorentz}), and (\ref{eq:MLorentz}); this significantly simplified the averaging over the atomic velocity distribution.
Under this approximation, the effect of the Raman detuning manifests itself solely in the coefficient $\chi_{p12}^{\rm inter}$. 
Otherwise, this effect would also be reflected in the coefficients $\chi_{pj}^{\rm nlin1}(\rm SAR)$ and $\chi_{pj}^{\rm nlin1}(\rm OP)$, but we do not consider this more general case here.

It should be noted that, unlike the spectroscopic atomic model based on the $\Lambda$-scheme \cite{Mikhailov2021}, the angle $\alpha$ does not appear in the phase $\theta(z)$ in our case.
This is because, with real energy levels, a multitude of such $\Lambda$-schemes are formed in the atom under the influence of a dual-frequency light field; the $\sigma^+$ and $\sigma^-$ transitions within these schemes may either share the same phase or differ by a phase of $\pi$, depending on the properties of the Clebsch-Gordan coefficients.
In a $\Lambda$-scheme, the angle $\alpha$ acts as the effective phase between the $\sigma^+$ and $\sigma^-$ transitions.
Therefore, if multiple distinct $\Lambda$-schemes are formed in the atom, the phase $\alpha$ is not common to all of them and cannot be incorporated into a single, unified phase $\theta(z)$.

Expression (\ref{eq:chiinter}) is distinguished by a number of interesting and important features from the previously obtained nonlinear absorption coefficients (\ref{eq:chi1nlin1SAR}) and (\ref{eq:chi2nlin1OP}). 
In particular, $\chi_{p12}^{\rm inter}$ contains the product of Rabi frequencies $R_3R_4$ associated with counterpropagating pump waves, indicating that this absorption coefficient is responsible for a nonlinear interference effect.
This is further evidenced by the presence of a narrow Lorentzian profile of the form $\propto\,$$(\Gamma^2$$+$$\delta_R^2)^{-1}$ with a half-width $\Gamma$$\,\ll\,$$\gamma_{eg}$, which is related to the lifetime of the ground-state atomic anisotropy.
Furthermore, only first- and second-rank multipoles contribute to (\ref{eq:chiinter}) because, for the case $F_2$$\,=\,$$F_1$$\,+\,$$1$ under consideration, the equality $C(0)$$\,=\,$$0$ holds (see Appendix B).

The term $\chi_{p12}^{\rm inter}$ is associated with the microwave coherences $\hat{\rho}_{12}$ and $\hat{\rho}_{21}$ generated in the atom through simultaneous interaction with two mutually coherent waves, ${\bf E}_3$ and ${\bf E}_4$.
It is precisely this term, $\chi_{p12}^{\rm inter}$, that describes the formation of a CPT state in the atom under the influence of a pump beam involving magnetic sublevels of different hyperfine components of the ground state ($F_1$ and $F_2$). 
The Raman frequency detuning plays a more significant role for this term than for the other nonlinear effects described above; specifically, when $\delta_R \gg \Gamma$, the term $\chi_{p12}^{\rm inter}$ can be entirely neglected.
However, as we shall see in the next section, with an appropriate choice of the phase $\theta$ and the beam polarization parameters entering into $\psi_{12}$, the term $\chi_{p12}^{\rm inter}$ makes a constructive contribution to the formation of a sub-Doppler resonance appearing as a peak in the probe beam absorption.

It also follows from the expression in square brackets in (\ref{eq:chiinter}) that, when scanning the Raman detuning ($\delta_R$), a resonance with a significantly asymmetric line shape may be observed in the probe beam absorption; this shape is a superposition of a Lorentzian absorption profile $\propto\,$$(\Gamma^2$$+$$\delta_R^2)^{-1}$ and a Lorentzian dispersion profile $\propto\,$$\delta_R$$(\Gamma^2$$+$$\delta_R^2)^{-1}$.
This asymmetry depends on the phase $\theta$; specifically, the resonance is symmetric when $\sin(\theta)$$\,=\,$$0$.
The sign of this resonance, electromagnetically induced transparency (EIT) or electromagnetically induced absorption (EIA), depends on both the sign of $\cos(\theta)$ and the polarization parameters of the counterpropagating beams, which influence the sign of the sum appearing in (\ref{eq:chiinter}).
These features pertain to the line shape of the CPT resonance in the field of counterpropagating bichromatic waves, rather than to the sub-Doppler resonances investigated in the present work.
Therefore, we will not discuss this issue in detail (this problem was addressed within the $\Lambda$-scheme framework in \cite{Brazhnikov2024}).

From a practical standpoint, the case of equal probe-wave intensities, $I_1 = I_2 = I_p/2$, is of primary interest, as this relates to the very method used to generate the mutually coherent waves ${\bf E}_1$ and ${\bf E}_2$ in experiments.
For the sake of definiteness, we shall focus our analysis specifically on this case.
In that instance, the law (\ref{eq:PropagationProbe}) takes a simpler form:
\begin{equation}\label{eq:PropagationProbe2}
    \frac{\partial I_p}{\partial z} = -\chi_{p}I_p\,.
\end{equation}
Here, $\chi_p$ is the absorption coefficient for the probe beam—that is, for the combination of the two waves ${\bf E}_1$ and ${\bf E}_2$.
It is convenient to decompose this coefficient into several terms and analyze them individually:
\begin{equation}\label{eq:ChiTotal}
    \chi_p = \chi_{p}^{\rm lin} + \chi_{p}^{\rm nlin}\,.
\end{equation}
Here, the linear absorption coefficient of the probe beam is:
\begin{equation}\label{eq:ChiLinTotal}
    \chi_{p}^{\rm lin} = \frac{1}{2}\Bigl[\chi_{p1}^{\rm lin} + \chi_{p2}^{\rm lin}\Bigr]\,,
\end{equation}
where $\chi_{p1}^{\rm lin}$ and $\chi_{p2}^{\rm lin}$ are defined by expressions (\ref{eq:Chi1Lin2}) and (\ref{eq:Chi2Lin2}). 
The nonlinear absorption coefficient encompasses several nonlinear optical effects:
\begin{equation}\label{eq:ChiNonLinTotal}
    \chi_p^{\rm nlin} =\chi_{p}^{\rm nlin1}({\rm SAR})+\chi_{p}^{\rm nlin1}({\rm OP}) + \chi_{p12}^{\rm inter}(z).
\end{equation}
Here:
\begin{align}\label{eq:ChiNonLinSAROP}
    &\!\!\chi_p^{\rm nlin}({\rm SAR})\! =\! \frac{1}{2}\Bigl[\chi_{p1}^{\rm nlin1}({\rm SAR})\! +\! \chi_{p2}^{\rm nlin1}({\rm SAR})\Bigr],\\
    &\chi_p^{\rm nlin}({\rm OP})\! =\! \frac{1}{2}\Bigl[\chi_{p1}^{\rm nlin1}({\rm OP})\! +\! \chi_{p2}^{\rm nlin1}({\rm OP})\Bigr].
\end{align}
The absorption coefficients appearing in these expressions are determined by formulas (\ref{eq:chi1nlin1SAR2})–(\ref{eq:chi2nlin1OP2}).

\section{Line Shape of the Resonance Curve}\label{sec:5}

Before analyzing the various linear and nonlinear optical effects described by the derived formulas, we will plot the dependence $\chi_p(\delta)$ for specific polarization configurations of counterpropagating light beams commonly used in atomic laser spectroscopy: linear parallel ($lin\,$$||$$\,lin$) and linear orthogonal ($lin\,\perp\,lin$) polarizations, identical circular ($\sigma\sigma$) polarizations, and circular polarizations with opposite handedness ($\sigma^+\sigma^-$).
Additionally, we will consider, as examples, two arbitrary cases of elliptical polarizations ($\epsilon_1$$\alpha$$\epsilon_2$).

In this work, for the sake of definiteness, we consider the Cs D$_1$ line: $I_n = 7/2$, $J_g = J_e = 1/2$, $F_1 = 3$, $F_2 = 4$, $\lambda = 894.6$ nm, $\gamma = 2\pi \times 4.56$ MHz \cite{Steck}. The following parameters were chosen for the experimental conditions: $T = 40^\circ$C, $n_a = 2 \times 10^{11}$ cm$^{-3}$, $\Delta_D = 2\pi \times 220$ MHz, and $\Gamma = 10^{-2}\gamma$, which corresponds to a laser beam diameter of $\approx 1$ mm. For simplicity, we assume the intensities of the counterpropagating pump waves (${\bf E}_3$ and ${\bf E}_4$) to be equal, as is the case in the vast majority of experiments using the frequency-modulated saturated absorption spectroscopy (DFSDS) method. In this case, the Rabi frequencies are related as follows:
\begin{equation}\label{eq:RabiRatio}
    R_4 = \sqrt{\frac{s_{23}}{s_{13}}} \, R_3\,.
\end{equation}

Experimental data \cite{Hafiz2016} indicate that the sub-Doppler resonance associated with the excitation of cesium atoms to the $F_3 = 4$ level exhibits the highest amplitude-to-width ratio—a parameter of importance for metrological applications \cite{Gusching2021}. 
Therefore, the calculations presented here focus on this specific value of the excited-state total angular momentum (though analogous calculations could be performed for $F_3 = 3$ if required).

\begin{figure}[!t]
    \centering
    \includegraphics[width=1\linewidth]{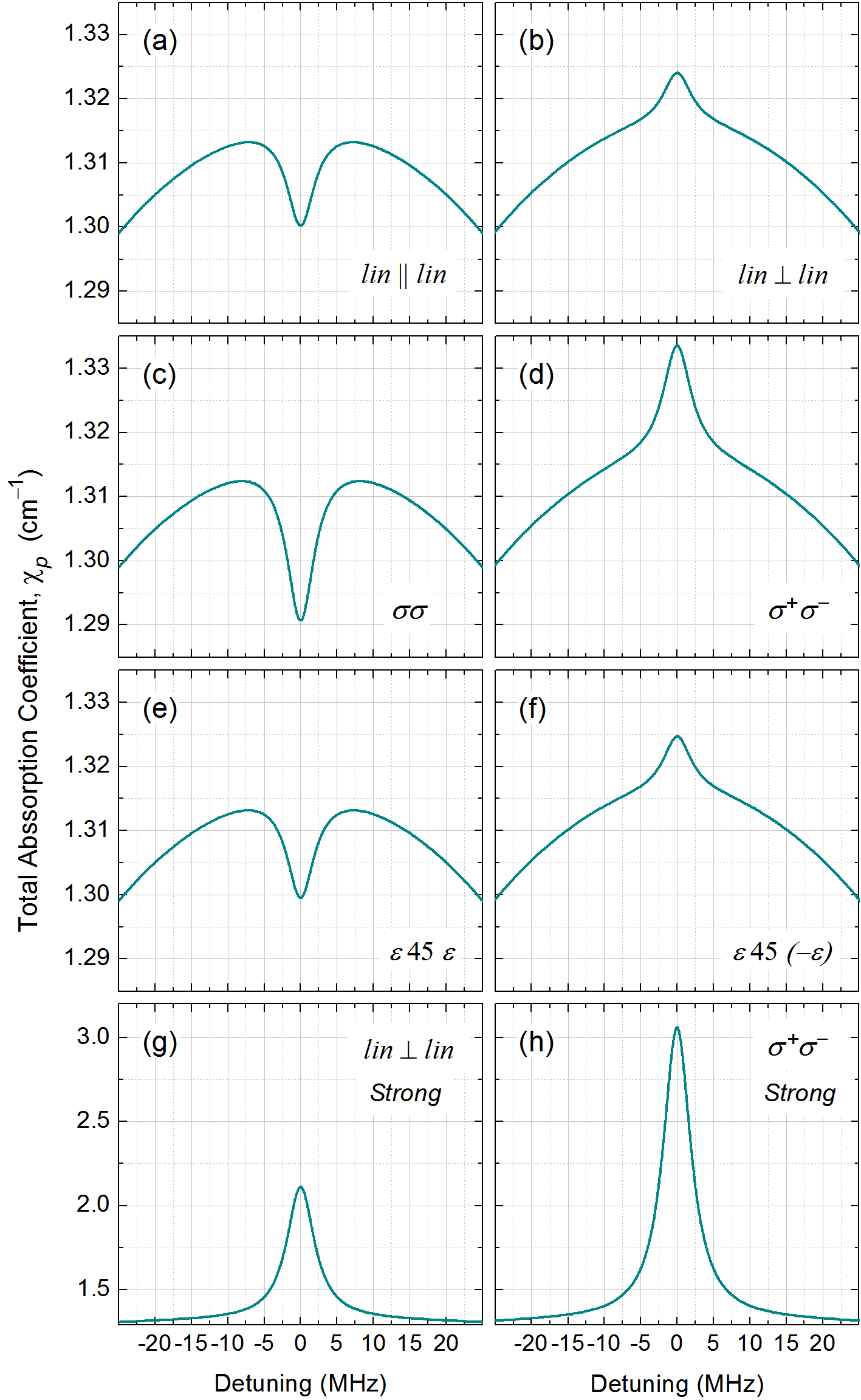}
    \caption{Resonance profiles of the probe-beam absorption coefficient ($\chi_p$), calculated using formula (\ref{eq:ChiTotal}) for various polarizations of the counterpropagating beams: (a) parallel linear, (b) orthogonal linear, (c) identical circular, (d) opposite circular, (e) identical elliptical, $\epsilon_1$$\,=\,$$\epsilon_2$$\,=\,$$25^\circ$, $\alpha$$\,=\,$$45^\circ$, (f) opposite elliptical, $\epsilon_1$$\,=\,$$-\epsilon_2$$\,=\,$$25^\circ$, $\alpha$$\,=\,$$45^\circ$. Panels (g) and (h) show the same configurations as (b) and (d), but at higher pump-wave intensity, where perturbation theory is, strictly speaking, inapplicable. For all plots, the relation $R_4$$\,\approx\,$$0.85\,$$R_3$ holds according to formula (\ref{eq:RabiRatio}), with $R_3$$\,=\,$$0.05$$\,\gamma$ for (a)–(f) and $R_3$$\,=\,$$0.5$$\,\gamma$ for (g) and (h).}
    \label{fig:3}
\end{figure}

It can be seen from Fig.~\ref{fig:3} that a sub-Doppler resonance in the form of an absorption peak is observed for orthogonal linear polarizations (Fig.~\ref{fig:3}b), whereas a dip is observed for parallel linear polarizations (Fig.~\ref{fig:3}b), resembling the saturated absorption resonance seen in a standard single-frequency observation scheme.
This behavior of the resonance curve is consistent with experimental data \cite{Hafiz2017,Velichansky2025}. 
A similar sign reversal of the nonlinear resonance occurs when switching from the $\sigma\sigma$ configuration (Fig.~\ref{fig:3}c) to the $\sigma^+\sigma^-$ configuration (Fig.~\ref{fig:3}d).

It is important to note that the DFSDS method has scarcely been investigated previously for the circular polarizations of counterpropagating beams. 
However, as a comparison of Figs. \ref{fig:3}b and \ref{fig:3}d reveals, with opposite circular polarizations, the resonance can exhibit an even greater amplitude than when orthogonal linear polarizations are used.
Figures \ref{fig:3}e and \ref{fig:3}f demonstrate that a sign reversal of the resonance can also occur with various elliptical beam polarizations.
Nevertheless, for the sake of brevity, we will henceforth focus on analyzing the field configuration formed by counterpropagating light beams with linear or circular polarizations, as these are easier to implement and control with high precision in practice.

Figures~\ref{fig:3}g and \ref{fig:3}h illustrate the behavior of the resonance curve at higher pump beam intensities, a regime where the PT approximation is, strictly speaking, no longer applicable. 
Nevertheless, a trend toward a sharp increase in the sub-Doppler resonance amplitude with increasing pump wave intensity is evident, which is in good agreement with experimental results \cite{Hafiz2016,Brazhnikov2019,Gusching2021}.

\subsection{Broad Doppler Contour}\label{sec:5.1}

\begin{figure}[!t]
    \centering
    \includegraphics[width=1\linewidth]{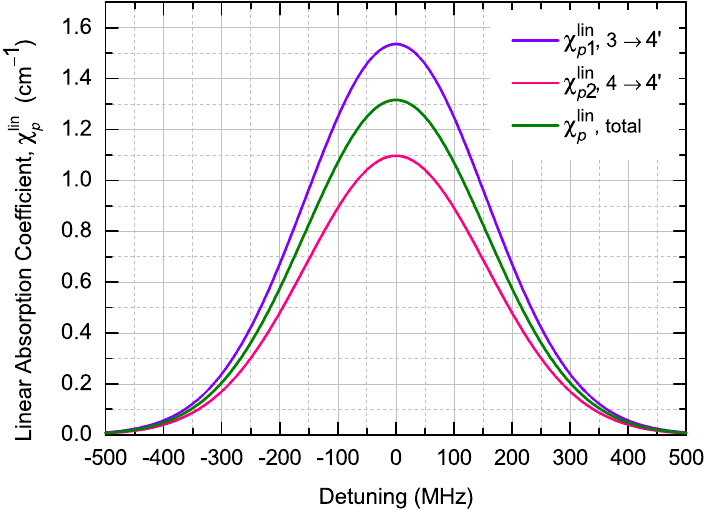}
    \caption{Resonance profiles describing the linear absorption of the probe beam according to formula (\ref{eq:ChiLinTotal}). The upper (purple) and lower (pink) curves correspond to the absorption coefficients for the probe waves ${\bf E}_1$ and ${\bf E}_2$, respectively, while the middle (green) curve describes the total absorption coefficient of the entire probe beam.}
    \label{fig:4}
\end{figure}

The first term in (\ref{eq:ChiTotal}), which accounts for the linear absorption of the probe beam, is independent of its polarization parameters, as follows from expressions (\ref{eq:Chi1Lin2}) and (\ref{eq:Chi2Lin2}).
This term describes a broad Doppler profile with an $e^{-1}$ half-width of $\Delta_D$, shown in Fig.~\ref{fig:4} (the middle green curve).
The lower (pink) and upper (purple) curves separately show the linear absorption coefficients for waves ${\bf E}_1$ and ${\bf E}_2$—namely, $\chi_{p1}^{\rm lin}$ and $\chi_{p2}^{\rm lin}$, respectively.
As noted above, the relative contribution of each of these two components to the total linear absorption of the probe beam is determined by the relative transition strength $s_{13}$ or $s_{23}$ from (\ref{eg:strength}). 
Note that the relative amplitudes of the depicted Doppler profiles are consistent with experimental data \cite{Sagle1996,Kozlova2011}.
Since the linear absorption of the probe beam is independent of its polarization parameters, the total profile shown in Fig.~\ref{fig:4} (the middle curve) constitutes the common background signal for all the resonance curves presented in Fig.~\ref{fig:3}.

\subsection{Sub-Doppler Resonance}\label{sec:5.2}

To analyze nonlinear effects in probe-beam absorption, it is convenient to decompose the absorption coefficients $\chi_{pj}^{\rm nlin1}({\rm SAR})$ and $\chi_{pj}^{\rm nlin1}({\rm OP})$ from Eqs. (\ref{eq:chi1nlin1SAR2})–(\ref{eq:chi1nlin1OP2}) into a scalar part $\chi_{pj,\,\rm sc}^{\rm nlin1}$, corresponding to contributions solely from terms with $\varkappa=0$, and a tensor part $\chi_{pj,\,\rm ten}^{\rm nlin1}$, which accounts for contributions from terms with $\varkappa \neq 0$:
\begin{equation}
    \chi_{p1,\,\rm sc}^{\rm nlin1}({\rm SAR}) \approx -\frac{\,a_0 s_{13}\gamma_{eg}\tau R_3^2}{\delta^2+\gamma_{eg}^2} A_{13}^2(0)\,,\label{eq:chi1nlin1SAR3Scal}
\end{equation}
\begin{multline}
    \chi_{p1,\rm ten}^{\rm nlin1}({\rm SAR}) \approx \frac{\sqrt{3}a_0 s_{13}\gamma_{eg}\tau R_3^2}{\delta^2+\gamma_{eg}^2}\times\\\times
    \sum_{\varkappa=1}^2 A_{13}^2(\varkappa)\Psi_{12}(\varkappa)\,,\label{eq:chi1nlin1SAR3Ten}
\end{multline}
\begin{equation}
    \chi_{p2,\rm sc}^{\rm nlin1}({\rm SAR}) \approx -\frac{a_0 s_{23}\gamma_{eg}\tau R_4^2}{\delta^2+\gamma_{eg}^2} A_{23}^2(0),\label{eq:chi2nlin1SAR3Scal}
\end{equation}
\begin{multline}
    \chi_{p2,\rm ten}^{\rm nlin1}({\rm SAR}) \approx \frac{\sqrt{3}a_0 s_{23}\gamma_{eg}\tau R_4^2}{\delta^2+\gamma_{eg}^2}\times\\\times
    \sum_{\varkappa=1}^2 A_{23}^2(\varkappa)\Psi_{12}(\varkappa),\label{eq:chi2nlin1SAR3Ten}
\end{multline}
\begin{multline}
\chi_{p1,\rm sc}^{\rm nlin1}({\rm OP}) \approx \frac{a_0 s_{13}\beta_{31}(2F_3+1)\gamma_{eg}\tau}{\delta^2+\gamma_{eg}^2}\times\\\times
\Big[R_3^2 |B_{11}(0)| + R_4^2|B_{12}(0)|\Big],\label{eq:chi1nlin1OP3Scal}
\end{multline}
\begin{multline}
    \chi_{p1,\rm ten}^{\rm nlin1}({\rm OP}) \approx \frac{\sqrt{3}a_0 s_{13}\beta_{31}(2F_3+1)\gamma_{eg}\tau}{\delta^2+\gamma_{eg}^2}\times\\\times
    \sum_{\varkappa=1}^2 \Big[R_3^2 B_{11}(\varkappa)+R_4^2 B_{12}(\varkappa)\Big]\Psi_{12}(\varkappa),\label{eq:chi1nlin1OP3Ten}
\end{multline}
\begin{multline}
    \chi_{p2,\rm sc}^{\rm nlin1}({\rm OP}) \approx \frac{a_0 s_{23}\beta_{32}(2F_3+1)\gamma_{eg}\tau}{\delta^2+\gamma_{eg}^2}\times\\\times
    \Big[R_3^2|B_{21}(0)| + R_4^2|B_{22}(0)|\Big],\label{eq:chi2nlin1OP3Scal}
\end{multline}
\begin{multline}
    \chi_{p2,\rm ten}^{\rm nlin1}({\rm OP}) \approx \frac{\sqrt{3}a_0 s_{23}\beta_{32}(2F_3+1)\gamma_{eg}\tau}{\delta^2+\gamma_{eg}^2}\times\\\times
    \sum_{\varkappa=1}^2 \Big[R_3^2 B_{21}(\varkappa)+R_4^2 B_{22}(\varkappa)\Big]\Psi_{12}(\varkappa).\label{eq:chi2nlin1OP3Ten}
\end{multline}
    
In accordance with these formulas, the total nonlinear absorption coefficient of the probe beam (\ref{eq:ChiNonLinTotal}) can be written as:
\begin{multline}\label{eq:chiNonLinTotal2}
    \chi_p^{\rm nlin} = \chi_{p,{\rm sc}}^{\rm nlin1}({\rm SAR}) + \chi_{p,{\rm ten}}^{\rm nlin1}({\rm SAR}) +\\+
    \chi_{p,{\rm sc}}^{\rm nlin1}({\rm OP}) + \chi_{p,{\rm ten}}^{\rm nlin1}({\rm OP}) + \chi_{p12}^{\rm inter}(z),
\end{multline}
where:
\begin{multline} \label{eq:chiNonLinScSAR}
    \chi_{p,{\rm sc}}^{\rm nlin1}({\rm SAR}) = \frac{1}{2}\Bigl[\chi_{p1,{\rm sc}}^{\rm nlin1}({\rm SAR}) +\\+ \chi_{p2,{\rm sc}}^{\rm nlin1}({\rm SAR})\Bigr],
\end{multline}
\begin{multline} \label{eq:chiNonLinTenSAR}
     \chi_{p,{\rm ten}}^{\rm nlin1}({\rm SAR}) = \frac{1}{2}\Bigl[\chi_{p1,{\rm ten}}^{\rm nlin1}({\rm SAR}) +\\+ \chi_{p2,{\rm ten}}^{\rm nlin1}({\rm SAR})\Bigr],
\end{multline}
\begin{equation} \label{eq:chiNonLinScOP}
     \chi_{p,{\rm sc}}^{\rm nlin1}({\rm OP}) = \frac{1}{2}\Bigl[\chi_{p1,{\rm sc}}^{\rm nlin1}({\rm OP}) + \chi_{p2,{\rm sc}}^{\rm nlin1}({\rm OP})\Bigr],
\end{equation} 
\begin{equation} \label{eq:chiNonLinTenOP}
     \!\chi_{p,{\rm ten}}^{\rm nlin1}({\rm OP})\! = \!\frac{1}{2}\Bigl[\chi_{p1,{\rm ten}}^{\rm nlin1}({\rm OP}) \!+ \!\chi_{p2,{\rm ten}}^{\rm nlin1}({\rm OP})\Bigr],
\end{equation}
and the absorption parameter $\chi_{p12}^{\rm inter}(z)$ is given by the expression (\ref{eq:chiinter}).

The coefficients (\ref{eq:chi1nlin1SAR3Scal})–(\ref{eq:chi2nlin1OP3Ten}) represent symmetric Lorentzian absorption profiles calculated in the $\delta_R \ll \gamma_{eg}$ approximation.
In the case where $\delta_R \sim \gamma_{eg}$, asymmetry may arise in the resonance profile described by the $\chi_{pj}^{\rm nlin1}(\rm SAR)$ terms.
However, for the sake of brevity, we do not consider this regime here; it warrants a separate study.

\subsection{Scalar Part of the Nonlinear Absorption Coefficient}

In this section, we analyze the behavior of the scalar contributions $\chi_{p,\,{\rm sc}}^{\rm nlin1}({\rm SAR})$ and $\chi_{p,\,{\rm sc}}^{\rm nlin1}({\rm OP})$ to the total absorption coefficient (\ref{eq:chiNonLinTotal2}).
For the sake of definiteness, in this and subsequent sections, we assume $R_3 = 0.05\gamma$ and $R_4 \approx 0.85R_3$, in accordance with formula (\ref{eq:RabiRatio}) with $F_3 = 4$.

\begin{figure}[!t]
    \centering
    \includegraphics[width=1\linewidth]{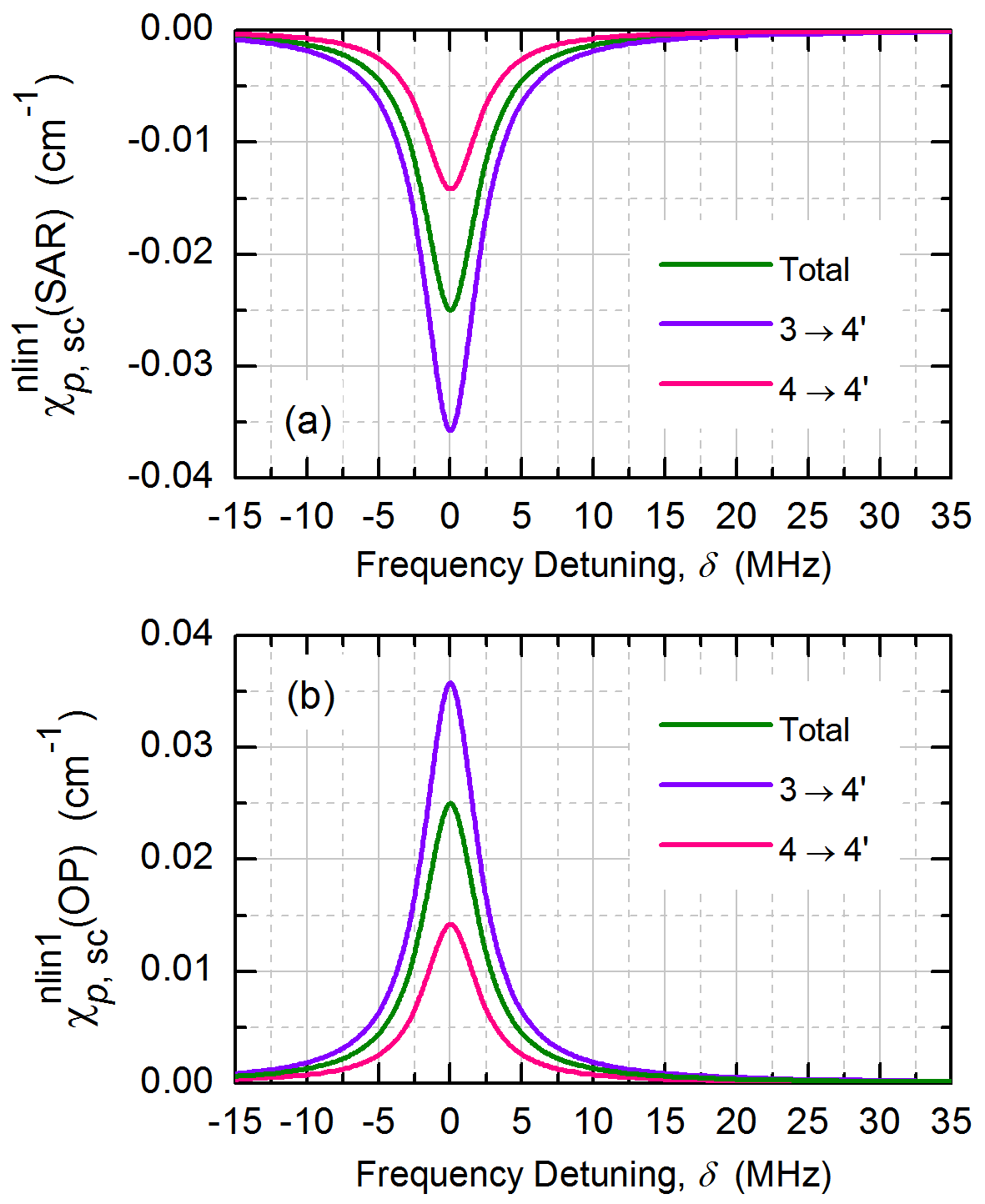}
    \caption{Scalar parts of the nonlinear absorption coefficients for two probe waves, ${\bf E}_1$ (purple curve) and ${\bf E}_2$ (pink curve), resonant with the dipole transitions $F_1=3 \to F_3=4$ and $F_2=4 \to F_3=4$, respectively (see Fig. \ref{fig:1}). (a) Contribution from saturated absorption resonance according to expressions (\ref{eq:chi1nlin1SAR3Scal}) and (\ref{eq:chi2nlin1SAR3Scal}); (b) contribution from optical pumping according to formulas (\ref{eq:chi1nlin1OP3Scal}) and (\ref{eq:chi2nlin1OP3Scal}). Also shown are the scalar parts of the total absorption coefficient for the two waves, i.e., the dual-frequency probe laser beam (middle green curve in both panels)—calculated according to formulas (\ref{eq:chiNonLinScSAR}) and (\ref{eq:chiNonLinScOP}).}
   \label{fig:5}
\end{figure}

Since all parameters in expressions (\ref{eq:chi1nlin1SAR3Scal}) and (\ref{eq:chi2nlin1SAR3Scal}) are positive and both expressions are preceded by a minus sign, it is evident that the contributions $\chi_{p1,\,{\rm sc}}^{\rm nlin1}({\rm SAR})$ and $\chi_{p2,\,{\rm sc}}^{\rm nlin1}({\rm SAR})$ to the scalar part of the absorption coefficient $\chi_{p,\,{\rm sc}}^{\rm nlin1}({\rm SAR})$ always describe a negative-sign resonance in the medium's absorption, specifically, a Lorentzian profile directed toward a reduction in probe-beam absorption (see Fig.~\ref{fig:5}a). 
This behavior is analogous to the saturated absorption resonance in a simple two-level atomic model with non-degenerate energy levels \cite{Haroche1972,Shirley1973}, where a counterpropagating pump wave reduces the population difference between the ground and excited levels, thereby rendering the medium more transparent to the probe beam in the region $\delta \lesssim \gamma_{eg}$.
It should be noted that, since the absorption coefficients $\chi_{pj}^{\rm nlin1}({\rm SAR})$ are associated with stimulated transitions in the atom, and given that the approximation under consideration lacks terms responsible for stimulated photon emission during transitions from $F_3$ to $F_1$ or $F_2$, the effect described here arises solely from the reduction in the total population of levels $F_1$ and $F_2$ due to the stimulated absorption of pump-wave photons.

The behavior of the scalar components of the absorption coefficients responsible for optical pumping, $\chi_{p1,\,{\rm sc}}^{\rm nlin1}({\rm OP})$ and $\chi_{p2,\,{\rm sc}}^{\rm nlin1}({\rm OP})$ from (\ref{eq:chi1nlin1OP3Scal}) and (\ref{eq:chi2nlin1OP3Scal}), is strictly opposite: they give rise to an upward-directed Lorentzian profile, i.e., an increase in absorption at the center of the resonance curve (Fig.~\ref{fig:5}b).
This behavior is also quite similar to what occurs in a two-level or three-level $\Lambda$-type atomic scheme with non-degenerate energy levels, where ground-state populations increase due to spontaneous relaxation from the excited state, leading to increased absorption of the probe beam \cite{Hernandez2009}.
For instance, in (\ref{eq:chi1nlin1OP3Scal}), the term proportional to $R_4^2$ describes the increased absorption of the probe wave ${\bf E}_1$ (resonant with the $F_1$$\,\to\,$$F_3$ transition) resulting from the transfer of atoms from level $F_2$ to level $F_1$ driven by the pump wave ${\bf E}_4$ and spontaneous relaxation $\gamma_{31}$ (see Fig.~\ref{fig:1}).

The resonances shown in Figs.~\ref{fig:5}a and \ref{fig:5}b have opposite signs but equal amplitudes.
This is because the depletion of the populations of levels $F_1$ and $F_2$ due to stimulated transitions ("SAR") is compensated by the reverse process -- optical pumping ("OP"), whereby atoms return to these levels via spontaneous radiative relaxation from level $F_3$.
This applies to either of the two excited-state levels ($F_3 = 3, 4$). Clearly, in the regime where $\gamma\tau \sim 1$ (which we do not consider here), the amplitude of the resonance in Fig.~\ref{fig:5}a would be slightly larger than that in Fig.~\ref{fig:5}b. 
This difference arises from additional terms within the summations in formulas (\ref{eq:chi1nlin1SAR}) and (\ref{eq:chi2nlin1SAR}), which feature the factor $(1 + \gamma\tau)$ in the denominator.
However, in the regime considered here ($\gamma\tau \gg 1$), these terms do not make a significant contribution to the sub-Doppler resonance.

It should be emphasized that all the arguments in this section pertain to the total populations of levels $F_1$ and $F_2$, since only the scalar components of the corresponding absorption coefficients are considered. 
As follows from the value of the coefficient $\Psi(\varkappa)$ at $\varkappa = 0$ given in Appendix B, the scalar components of the absorption coefficient are independent of the light beam polarizations; consequently, their behavior is identical for any polarization configuration of the light field within the cell. 
In contrast, the terms associated with the orientation ($\varkappa = 1$) and alignment ($\varkappa = 2$) of atomic spins exhibit polarization sensitivity; therefore, we shall examine them separately below for specific field polarization configurations.

Since the contributions from the scalar parts of the absorption coefficient, namely, $\chi_{p,\,\rm sc}^{\rm nlin1}({\rm SAR})$ and $\chi_{p,\,\rm sc}^{\rm nlin1}({\rm OP})$, cancel each other out, \emph{the formation of the sub-Doppler resonance is driven specifically by the tensor contributions to the absorption coefficient}.
These contributions are associated with the creation of various "dark" states in the atoms under the influence of the pump beam, both within the individual $F_1$ and $F_2$ levels and between these levels (the "microwave" CPT state).
These conclusions are consistent with earlier results obtained using a $\Lambda$-type atomic energy level scheme \cite{Hafiz2017,Mikhailov2021}.
However, unlike the multilevel atomic model considered here, the spectroscopic model based on the $\Lambda$-scheme does not account for the process of spontaneous transfer of anisotropy from an excited level to the ground level \cite{Rautian1994}, which is known to significantly influence light absorption in a medium \cite{Akulshin1998,Taichenachev1999,Goren2003,Zigdon2008}.
The impact of this effect on the formation of the sub-Doppler resonance observed via the frequency-modulated saturation-dispersion spectroscopy (DFSDS) technique has not been previously investigated.

\subsection{Tensor Part of the Nonlinear Absorption Coefficient}

\subsubsection{Configuration \texorpdfstring{$lin\,$$||$$\,lin$}{lin || lin}}

\begin{figure}[t!]
    \centering
    \includegraphics[width=1\linewidth]{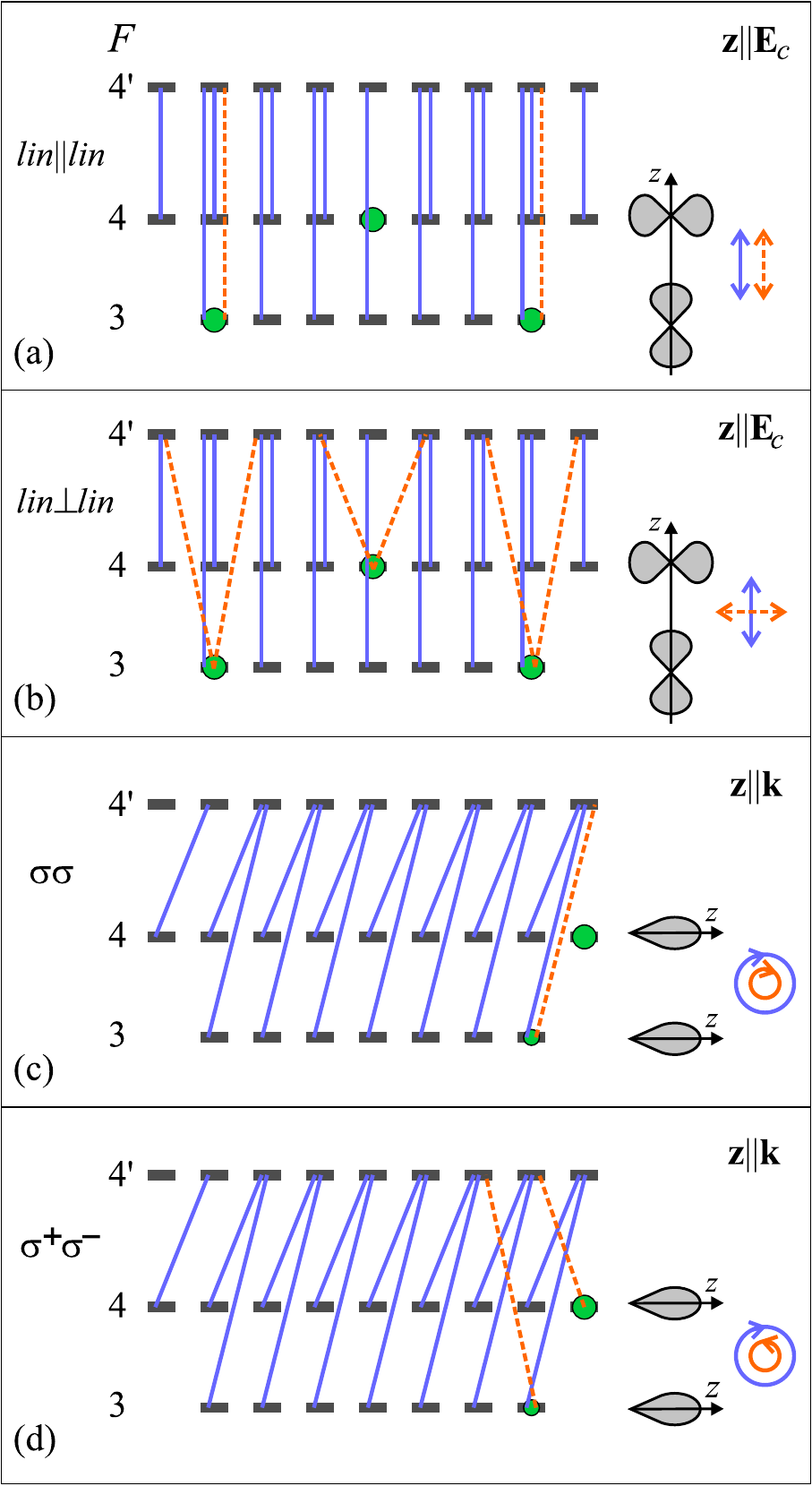}
    \caption{Optical dipole transitions in the D$_1$ line of an alkali metal atom for various polarization configurations of counterpropagating light beams: (a) parallel linear polarizations, (b) orthogonal linear polarizations, (c) identical circular polarizations, (d) opposite circular polarizations. Solid blue lines represent optical transitions induced by the pump beam. For the sake of diagram simplicity, only some of the transitions induced by the probe beam (dashed orange lines) are shown. Numbers without a prime correspond to the total angular momentum values of the atom in the ground state, while those with a prime correspond to the excited state.}
    \label{fig:6}
\end{figure}

As is evident from the form of the coefficients $\Psi_{12}(\varkappa)$ presented in Appendix B, in the case of linearly polarized beams ($lin$$\,||\,$$lin$ or $lin$$\,\perp\,$$lin$), no vector ($\varkappa$$\,=\,$$1$) terms appear in the absorption coefficients (\ref{eq:chi1nlin1SAR3Ten}), (\ref{eq:chi2nlin1SAR3Ten}), (\ref{eq:chi1nlin1OP3Ten}), and (\ref{eq:chi2nlin1OP3Ten}).
This is evidently due to the fact that the spin state of the photons in such a field is a superposition of spins directed along and opposite to the wave vector.
Consequently, the absorption of such photons by an atom cannot result in the orientation of its spin along or opposite to the wave vector, a process that would be described by a first-rank polarization moment (orientation). 
Instead, a quadrupole polarization state, alignment, is induced in the atom.

First, let us analyze the behavior of the probe beam absorption coefficient $\chi_p^{\rm nlin}(\delta)$ qualitatively.
To this end, for a linearly polarized pump beam, it is convenient to choose the quantization axis $z$ along the pump beam's polarization vector ($z \parallel {\bf E}_{3,4}$) rather than along the wave vectors of the counterpropagating beams, as shown in Fig.~\ref{fig:2}.
As seen in Fig.~\ref{fig:6}a, excitation of atoms to the $F_3 = 4$ level results in optical pumping into the extreme Zeeman sublevels with $m = \pm 3$ of the $F_1 = 3$ level and into the single Zeeman sublevel with $m = 0$ of the $F_2 = 4$ level. 
In the latter case, due to the absence of stimulated transitions $|F_2, m = 0\rangle \to |F_3, m = 0\rangle$, atoms in the $|F_2, m = 0\rangle$ state do not interact with the light at all. 
At the $F_1$ level, atomic spins are aligned along the linear polarization of the pump beam, whereas at the $F_2$ level, they are aligned transverse to it.
The geometry of the aligned spins and the beam polarization directions is schematically shown in Fig.~\ref{fig:6} next to each ground-state hyperfine energy level.
Note that in a basis where $z \parallel {\bf k}$, the aforementioned atomic states would correspond to Zeeman coherences within each of the $F_{1,2}$ levels. 
A particular case of such superposition states is the state of coherent population trapping \cite{Smirnov1989}.

As shown in Fig.~\ref{fig:6}a, the pump wave aligns the atomic spins at levels $F_1=3$ and $F_2=4$ such that the medium becomes transparent to this wave. Specifically, the state $|F_2, m=0\rangle$ is a "dark" state—that is, it does not participate in light absorption.
Population accumulates in this state to an extent that increases with the duration of the atom's residence in the light beam.
Although the other two quantum states, $|F_1, m=-3\rangle$ and $|F_1, m=3\rangle$, are not "dark" in the strict sense, they absorb light less strongly than other states at level $F_1$ (due to the properties of Clebsch-Gordan coefficients).
Since the polarization state of the probe beam matches that of the pump beam, the probe beam also experiences reduced absorption in the region $\delta=0$, where it interacts with atoms already prepared by the pump wave.

\begin{figure}[!t]
    \centering
    \includegraphics[width=1\linewidth]{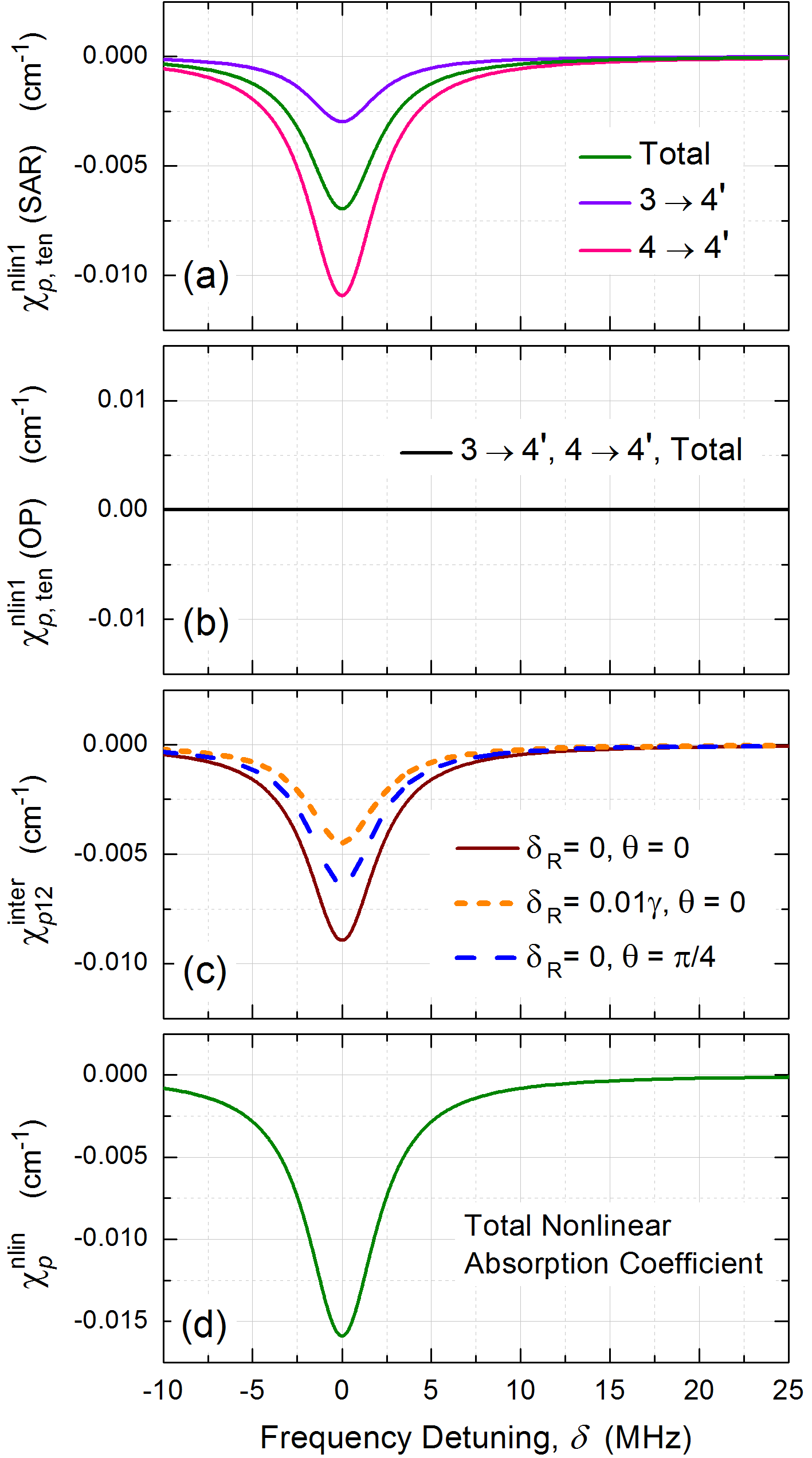}
    \caption{Tensor components of the nonlinear absorption coefficients for two probe waves, ${\bf E}_1$ (purple curve) and ${\bf E}_2$ (pink curve), resonant with the dipole transitions $F_1=3 \to F_3=4$ and $F_2=4 \to F_3=4$, respectively (see Fig. \ref{fig:1}). Polarization configuration: $lin||lin$. (a) Contribution of stimulated transitions from the atomic ground-state energy levels ($F_1$, $F_2$); (b) contribution from the spontaneous transfer of anisotropy from the excited level ($F_3$) to the ground levels; (c) contribution from the effect associated with the CPT state between levels $F_1$ and $F_2$; (d) total absorption coefficient of the dual-frequency probe beam.}
   \label{fig:7}
\end{figure}

In addition to the effects discussed above, which are associated with the redistribution of populations among Zeeman sublevels, low-frequency ("microwave") coherences between levels $F_1$ and $F_2$ are induced in the atom by the mutually coherent pump waves ${\bf E}_{3}$ and ${\bf E}_{4}$.
These superposition states describe the CPT effect, rendering the medium transparent to the pump beam (for the sake of simplicity, this effect is not shown in Fig.~\ref{fig:6}).
These states are also "dark" with respect to the probe beam, given the identical polarization states of the two counterpropagating beams (the influence of other parameters on this process will be discussed below).

Based on the foregoing, it is natural to expect that, in the polarization configuration of the light field under consideration, the sub-Doppler resonance will take the form of a Lorentzian profile corresponding to reduced absorption in the medium.
This conclusion is confirmed by the resonance curve in Fig.~\ref{fig:7}d, plotted using formula (\ref{eq:chiNonLinTotal2}).

Now let us analyze the behavior of the individual components of the total absorption coefficient for the probe beam, based on formulas (\ref{eq:chiNonLinScSAR})–(\ref{eq:chiNonLinTenOP}) and (\ref{eq:chiinter}), taking into account that, as established above, the scalar terms $\chi_{p,\,{\rm sc}}^{\rm nlin1}(\rm SAR)$ and $\chi_{p,\,{\rm sc}}^{\rm nlin1}(\rm OP)$ mutually cancel each other out and are therefore of no interest.
Figure~\ref{fig:7}a shows the contributions of the quadrupole ($\varkappa=2$) polarization moments to the sub-Doppler resonance. 
The pump wave aligns the atoms in both ground-state levels, $F_1=3$ and $F_2=4$, in such a way that the absorption of the probe beam decreases.

An interesting feature of the dual-frequency excitation scheme for the atom and its energy-level system in the D$_1$ line is that, for quadrupole ($\varkappa=2$) polarization moments, there is \emph{no contribution from spontaneous anisotropy transfer}, as shown in Fig.~\ref{fig:7}b.
This is because the individual contributions to this process from the different components of the pump beam cancel each other out. 
Specifically, in expression (\ref{eq:chi1nlin1OP3Ten}), the terms $R_3^2B_{11}(2)$ and $R_4^2B_{12}(2)$ are equal in absolute value but opposite in sign.
The same applies to the terms $R_3^2B_{21}(2)$ and $R_4^2B_{22}(2)$ in (\ref{eq:chi2nlin1OP3Ten}).
Thus, the process of spontaneous anisotropy transfer could make a significant contribution either if only a single dipole transition ($F_1 \to F_3$ or $F_2 \to F_3$) were excited, or if the pump wave intensities were unequal ($I_3 \ne I_4$).
This issue warrants more detailed theoretical study but is not examined at length here for the sake of brevity.
We note only that the characteristics of sub-Doppler resonances observed via the DFSDS method with unequal pump wave intensities were recently investigated experimentally in \cite{Sabakar2026}.

Figure~\ref{fig:7}c illustrates the contribution of the CPT effect to the formation of the sub-Doppler resonance, in accordance with formula (\ref{eq:chiinter}). 
As the figure shows, the superposition states created by the pump beam are also "dark" with respect to the probe beam, thereby rendering the medium transparent to it. 
These states are efficiently formed when the frequency difference between the ${\bf E}_{3}$ and ${\bf E}_{4}$ waves is sufficiently close to the ground-state hyperfine splitting frequency ($\Delta_g$), that is, when the Raman detuning ($\delta_R$) is close to zero \cite{Arimondo1996}.
Consequently, as this detuning increases, the contribution from the CPT effect diminishes, a trend reflected by the orange curve (short dashes) in Fig.~\ref{fig:7}c.
The influence of Raman detuning on the sub-Doppler resonance in the DFSDS method was experimentally confirmed in \cite{Hafiz2017}.

It is important to note that the CPT state formed by the pump beam between the hyperfine levels $F_1$ and $F_2$ may not, under certain conditions, be a "dark" state for the probe beam.
In our case, this depends on the phase $\theta$. In the experiment, this phase can be controlled, for example, by moving the cell along the $z$-axis. 
Figure~\ref{fig:7}c shows (blue dashed lines) that for $\theta \neq 0$, the contribution from the CPT effect decreases.
Clearly, this behavior is periodic as the phase $\theta$ is gradually increased, giving rise to \emph{spatial oscillations of the sub-Doppler resonance amplitude}, which have been repeatedly observed in experiments \cite{Hafiz2017,Brazhnikov2019,Brazhnikov2020,Velichansky2025}.

Figure~\ref{fig:7}d shows the sub-Doppler resonance resulting from all the aforementioned effects, as described by formula (\ref{eq:chiNonLinTotal2}).
Combining this resonance with the broad linear absorption profile of the probe beam (Fig.~\ref{fig:4}, middle green curve) yields the overall shape of the resonance curve shown in Fig.~\ref{fig:3}a.

\begin{figure}[!t]
    \centering
    \includegraphics[width=1\linewidth]{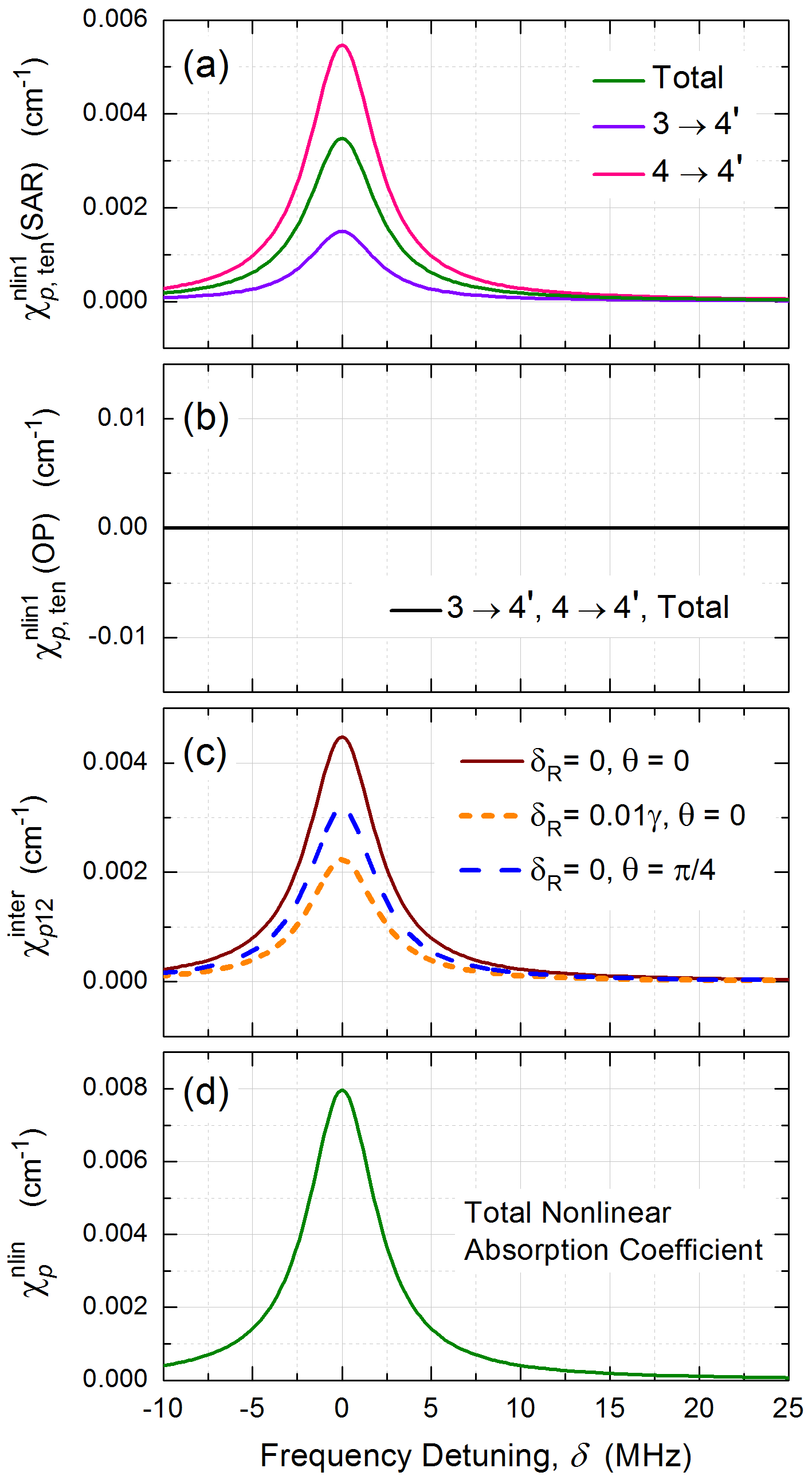}
    \caption{Tensor components of the nonlinear absorption coefficients for two probe waves, ${\bf E}_1$ (purple curve) and ${\bf E}_2$ (pink curve), resonant with the dipole transitions $F_1=3 \to F_3=4$ and $F_2=4 \to F_3=4$, respectively (see Fig.~\ref{fig:1}). Polarization configuration: $lin \perp lin$. (a) Contribution from stimulated transitions originating from the atomic ground-state levels ($F_1$, $F_2$); (b) contribution from the spontaneous transfer of anisotropy from the excited level ($F_3$) to the ground levels; (c) contribution from the effect associated with the CPT state between levels $F_1$ and $F_2$; (d) total absorption coefficient of the dual-frequency probe beam.}
   \label{fig:8}
\end{figure}

\subsubsection{Configuration \texorpdfstring{$lin\,$$\perp$$\,lin$}{lin ⊥ lin}}

It is precisely the $lin\perp lin$ field configuration that is used in the vast majority of experimental studies to observe sub-Doppler resonance via the frequency-modulated saturated absorption spectroscopy (DFSDS) technique, appearing as a high-contrast peak (or dip) in the medium's absorption (or transparency).
Qualitatively, the emergence of increased absorption of the probe beam in the medium can also be readily understood by analyzing Fig. \ref{fig:6}b.
Indeed, as in the case of Fig. \ref{fig:8}a, interaction with the pump beam causes atoms to accumulate in the $|F_1, m=-3\rangle$, $|F_1, m=3\rangle$, and $|F_2, m=0\rangle$ states.
However, unlike the previous configuration, the probe wave now interacts actively with these atoms, resulting in enhanced absorption within the medium.
In other words, the probe wave photons "attempt" to align the atomic spins in a direction perpendicular to the spin alignment imposed by the pump wave, and are consequently strongly absorbed by the medium.

This enhancement of probe-beam absorption is associated not only with population effects but also with the fact that the microwave CPT states formed by the pump wave between levels $F_1$ and $F_2$ are no longer "dark" for the probe wave; instead, they are "bright" states that absorb probe-beam photons.
Consequently, this effect further increases the absorption of the probe beam.

Let us now consider the behavior of the sub-Doppler resonance based on the derived analytical formulas.
Figure~\ref{fig:8}a shows the contributions of $\chi_{p1,\,{\rm ten}}^{\rm nlin1}(\rm SAR)$ and $\chi_{p2,\,{\rm ten}}^{\rm nlin1}(\rm SAR)$ to the total nonlinear absorption coefficient.
As expected, and in contrast to Fig.~\ref{fig:7}a, an absorption peak is now observed for the reasons mentioned earlier. 
As in the $lin\,$$||$$\,lin$ configuration, the spontaneous transfer of anisotropy, described by the term $\chi_{p,\,{\rm ten}}^{\rm nlin1}(\rm OP)$, does not contribute to the resonance (Fig.~\ref{fig:8}b).
Finally, the term $\chi_{p12}^{\rm inter}$, which is responsible for the CPT effect, leads to increased absorption in the medium in this case (Fig.~\ref{fig:8}c). 
As with the $lin\,$$||$$\,lin$ configuration, the strength of this effect can be controlled either by adjusting the Raman detuning (orange dashed curve) or by moving the cell along the $z$-axis (blue dashed curve).
Ultimately, the combined effects result in the total resonance shown in Fig.~\ref{fig:8}d manifesting as a light-induced absorption resonance.

\begin{figure}[!t]
    \centering
    \includegraphics[width=1\linewidth]{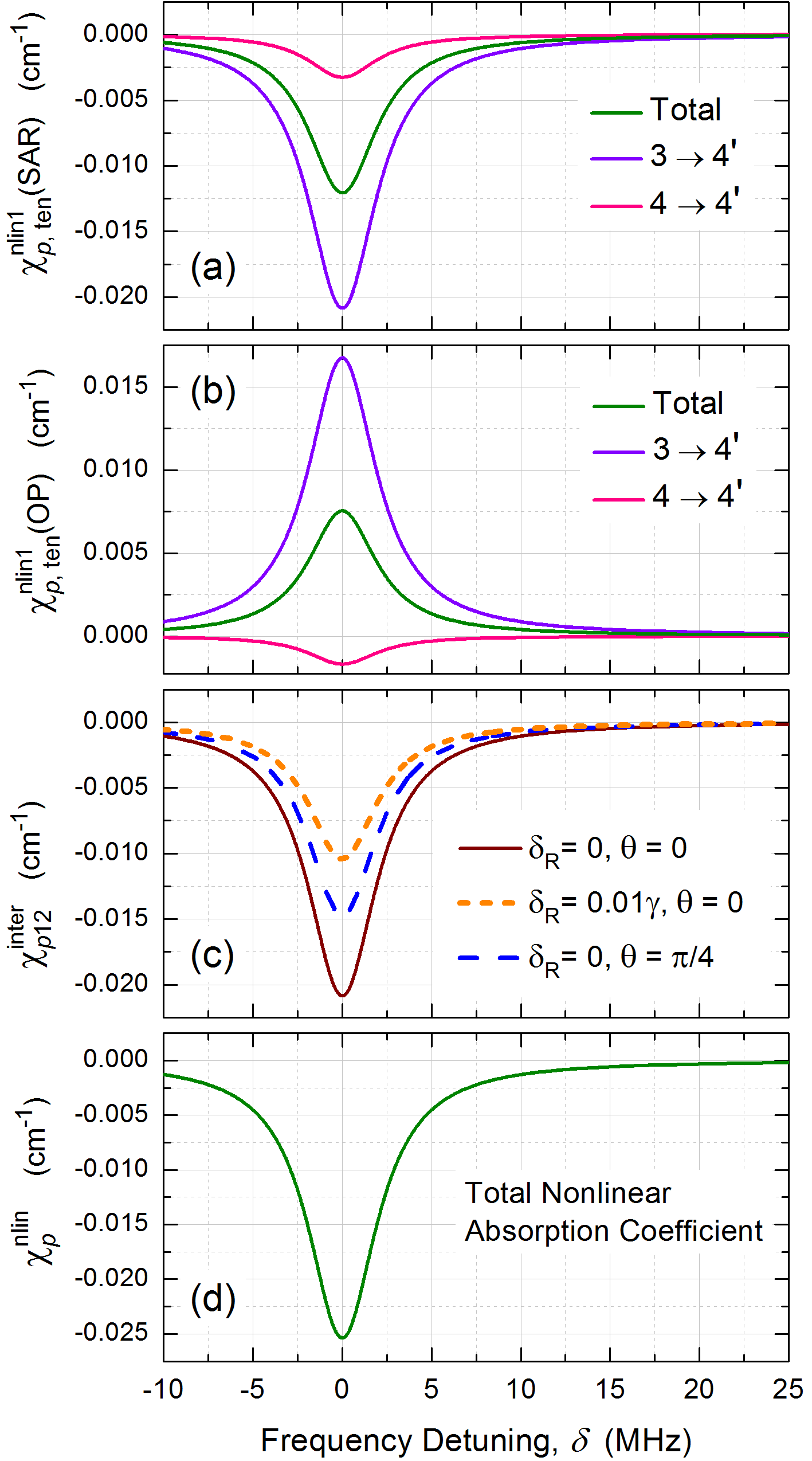}
    \caption{Tensor components of the nonlinear absorption coefficients for two probe waves, ${\bf E}_1$ (purple curve) and ${\bf E}_2$ (pink curve), resonant with the dipole transitions $F_1=3 \to F_3=4$ and $F_2=4 \to F_3=4$, respectively (see Fig. \ref{fig:1}). Polarization configuration: $\sigma\sigma$. (a) Contribution from stimulated transitions originating from the atomic ground-state levels ($F_1$, $F_2$); (b) contribution from the spontaneous transfer of anisotropy from the excited level ($F_3$) to the ground levels; (c) contribution from the effect associated with the CPT state between levels $F_1$ and $F_2$; (d) total absorption coefficient for the dual-frequency probe beam.}
   \label{fig:9}
\end{figure}

\subsubsection{Configuration \texorpdfstring{$\sigma$$\sigma$}{σσ}}

As can be seen from Figs.~\ref{fig:6}a and \ref{fig:6}c, the configuration with identical circular polarizations ($\sigma$$\sigma$) is largely similar to the $lin$$||$$lin$ configuration.
Indeed, just as in the $lin$$||$$lin$ case, the $\sigma$$\sigma$ configuration features only one state that does not interact with the laser field at all—namely, $|F_2,$$\,m$$=$$F_2$$\rangle$ (or $|F_2,$$\,m$$=$$-F_2$$\rangle$, if one considers left-circular polarization of the pump beam).

At the $F_1$ level, as a result of multiple excitation and spontaneous relaxation events, the population drifts toward the extreme sublevel $|F_1, m=F_1\rangle$ (or $|F_1, m=-F_1\rangle$ if the pump beam has left-circular polarization).
Once in this sublevel, the atoms absorb light from both beams with the highest probability compared to the other Zeeman sublevels of the $F_1$ level.
This differs from the $lin||lin$ configuration, in which the two extreme states $|F_1, m=\pm F_1\rangle$ interacted weakly with the light.
It is important to note that, since the optical transition $|F_1, m=F_1\rangle \to |F_3, m=F_3\rangle$ is not cyclic (closed), the population of the $|F_1, m=F_1\rangle$ sublevel, which interacts strongly with the resonant light, is partially transferred to the $|F_2, m=F_2\rangle$ sublevel, which does not interact with the light.

As can be seen from Fig.~\ref{fig:9}a, the terms $\chi_{p1,\,{\rm ten}}^{\rm nlin1}(\rm SAR)$ and $\chi_{p2,\,{\rm ten}}^{\rm nlin1}(\rm SAR)$ create a dip in the probe beam absorption, since the probe beam photons possess the same spin state as the pump beam photons and do not attempt to reorient the atomic spins in a different direction.

Unlike the previous two configurations ($lin$$||$$lin$ and $lin$$\,\perp\,$$lin$), in this case, not only the quadrupole ($\varkappa$$\,=\,$$2$) polarization moment but also the dipole ($\varkappa$$\,=\,$$1$) moment participates in the spontaneous transfer of anisotropy.
This follows from the fact that for the $\sigma$$\sigma$ configuration, $\Psi$$(\varkappa$$\,=\,$$1)$$\,\ne\,$$0$ (see Appendix B).
At the same time, the contribution from the spontaneous transfer of the quadrupole moment from the excited level to the hyperfine levels $F_1$ and $F_2$ remains absent for the same reasons as in the $lin$$||$$lin$ and $lin$$\,\perp\,$$lin$ configurations. 
Figure~\ref{fig:9}b shows the contribution to the sub-Doppler resonance arising from the spontaneous transfer of the dipole polarization moment (orientation).

As in the case of the $lin$$||$$lin$ configuration, in the $\sigma$$\sigma$ configuration, the pump beam induces a microwave CPT state in the atoms, which also serves as a CPT state for the probe beam. 
Therefore, the resonance in Fig.~\ref{fig:9}c is inverted (directed downwards).
The total sub-Doppler resonance shown in Fig.~\ref{fig:9}d is also inverted, that is, light-induced transparency of the medium is observed.

\subsubsection{Configuration \texorpdfstring{$\sigma^+$$\sigma^-$}{σ⁺σ⁻} }

The distinction between the $\sigma^+\sigma^-$ and $\sigma\sigma$ configurations bears a strong resemblance to the difference between the $lin\perp lin$ and $lin|| lin$ configurations.
As in the previous case, the pump beam in the $\sigma^+\sigma^-$ configuration orients the atomic spins along the $z$-axis.
Regarding the distribution of sublevel populations, as shown in Fig.~\ref{fig:6}d, atoms accumulate primarily in the "dark" state $|F_2, m=F_2\rangle$. 
However, unlike in the $\sigma\sigma$ configuration, this state is now "bright" for the probe beam. 
Consequently, when the probe and pump beams interact with the same atoms in the gas, i.e., when $\delta \lesssim \gamma_{eg}$, the probe beam undergoes enhanced absorption within the medium. 
This effect is also influenced by the CPT state formed by the pump beam between levels $F_1$ and $F_2$; much like in the $lin\perp lin$ configuration, this state is now "bright" for the probe beam.

\begin{figure}[!t]
    \centering
    \includegraphics[width=1\linewidth]{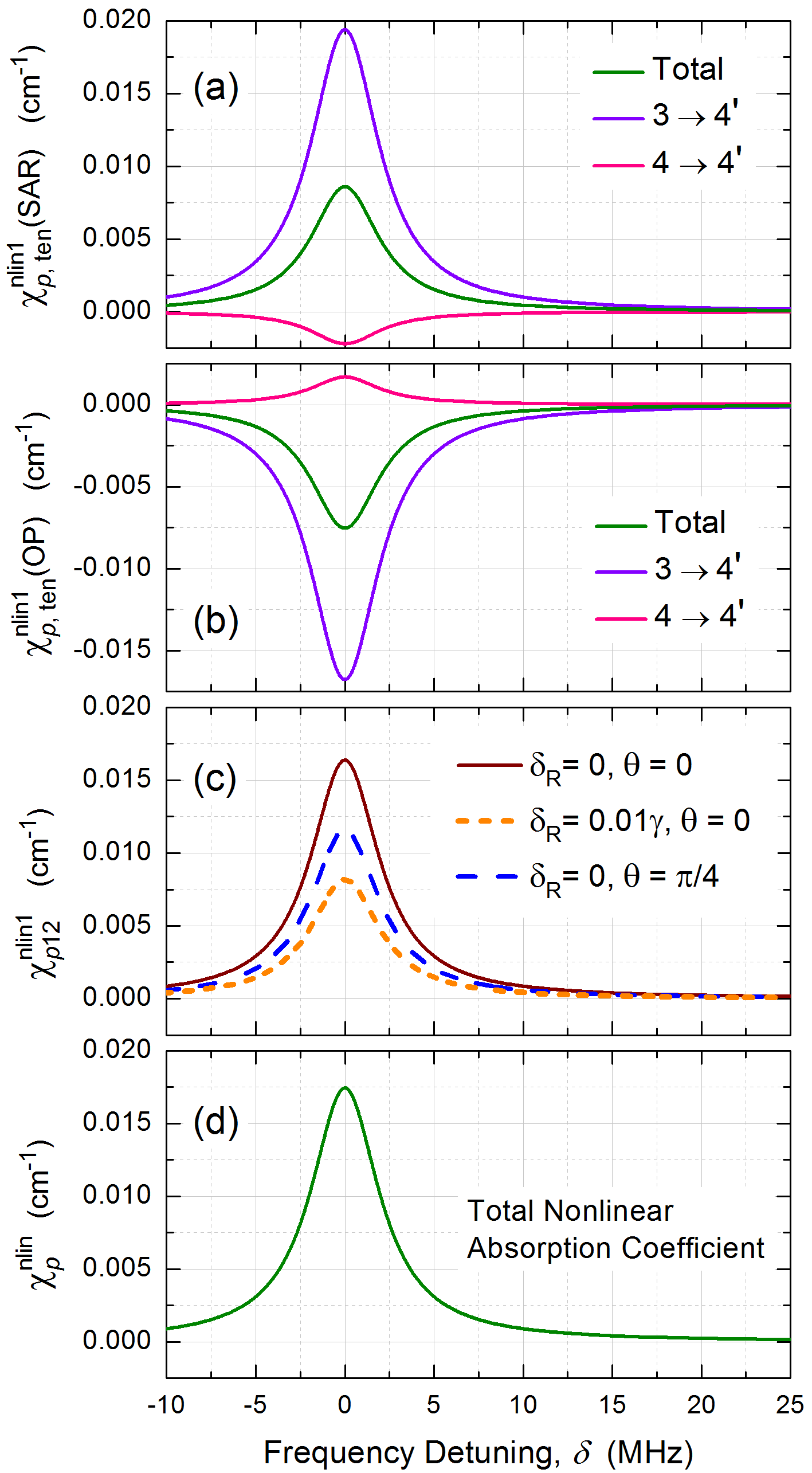}
    \caption{Tensor components of the nonlinear absorption coefficients for two probe waves, ${\bf E}_1$ (purple curve) and ${\bf E}_2$ (pink curve), resonant with the dipole transitions $F_1=3 \to F_3=4$ and $F_2=4 \to F_3=4$, respectively (see Fig. 1). Polarization configuration: $\sigma^+\sigma^-$. (a) Contribution from stimulated transitions originating from the atomic ground-state levels ($F_1$, $F_2$); (b) contribution from the spontaneous transfer of anisotropy from the excited level ($F_3$) to the ground levels; (c) contribution from the effect associated with the CPT state between levels $F_1$ and $F_2$; (d) total absorption coefficient of the dual-frequency probe beam.}
   \label{fig:10}
\end{figure}

Let us analyze the various contributions to the probe-beam absorption coefficient.
As shown in Fig.~\ref{fig:10}a, the tensor component of the absorption coefficient associated with stimulated transitions leads to slight bleaching of the medium at the $F_2=4 \to F_3=4$ transition and a much larger increase in absorption at the $F_2=3 \to F_3=4$ transition. 
Together, these two effects produce a resonance in the form of an absorption peak (green curve). 
The opposite behavior is observed in the contribution to the absorption coefficient describing the spontaneous transfer of anisotropy (Fig.~\ref{fig:10}b). 
As in the $\sigma\sigma$ configuration, the process of spontaneous transfer of atomic spin orientation (dipole polarization moment) contributes to the sub-Doppler resonance.

Finally, the CPN effect makes a significant contribution to the formation of the composite peak in the probe beam absorption (Fig.~\ref{fig:10}d), just as in the case of the $lin$$\perp$$lin$ configuration.
As with other configurations, the strength of this effect can be effectively controlled by varying the Raman frequency detuning ($\delta_R$) or the position of the atomic cell along the light beams (the phase $\theta$). 

An important feature arising from a comparison of the resonances shown in Figs.~\ref{fig:8}d and \ref{fig:10}d is that, in the $\sigma^+$$\sigma^-$ configuration, the resonance amplitude can be approximately twice as large as in the $lin$$\perp$$lin$ configuration commonly used in experiments.
This issue is of practical interest and requires experimental verification.

\section{Conclusion}

In this work, an analytical theory of two-frequency sub-Doppler spectroscopy (DFSDS) of the D$_1$ line of alkali metal atoms ($^{87}$Rb and $^{133}$Cs) is developed, accounting for energy level degeneracy with respect to the projection of the atomic angular momentum, arbitrary elliptical light polarization, and various relaxation processes involving the atom's internal degrees of freedom.
Using the density matrix formalism, irreducible tensor operators, and perturbation theory, analytical expressions were derived for the absorption coefficient of a probe light beam passing through a gas of resonant atoms in the presence of a counterpropagating pump beam.
These expressions make it possible to isolate the various nonlinear optical effects that shape the overall resonant absorption profile and to compare their contributions to the observed sub-Doppler resonances. 
Particular emphasis was placed on polarization configurations of the counterpropagating beams commonly used in practice: linear parallel ($lin$$||$$lin$) and linear orthogonal ($lin$$\perp$$lin$) polarizations, as well as identical circular ($\sigma$$\sigma$) and opposite circular ($\sigma^+$$\sigma^-$) polarizations.

In particular, it was shown that in the $lin$$\perp$$lin$ configuration, the observation of a peak in the probe-beam absorption is governed by two main effects: the formation, by the pump beam, of a quadrupole polarization moment (alignment) on each of the hyperfine energy levels of the atomic ground state ($F_1$ and $F_2$), and the induction of microwave-frequency coherence between the $F_1$ and $F_2$ levels, which actively absorbs probe-beam photons. 

The vast majority of experiments employ the $lin$$\perp$$lin$ light configuration.
However, the present work demonstrates that a peak in probe-beam absorption can be observed with arbitrary elliptical polarizations of the light beams as well as with opposite circular polarizations ($\sigma^+$$\sigma^-$).
Notably, the latter result is of particular practical interest, as the amplitude of the sub-Doppler resonance in this configuration is significantly larger than in the $lin$$\perp$$lin$ configuration.
This theoretical prediction requires further experimental verification.

Particular attention was paid to the influence of the spontaneous anisotropy transfer effect on the formation of the sub-Doppler resonance, an aspect not previously investigated in studies of the frequency-modulated saturated absorption spectroscopy (DFSDS) method.
Specifically, it was demonstrated that with equal pump wave intensities ($I_3 = I_4$), this effect has no impact on nonlinear resonance formation in the $lin||lin$ and $lin\perp lin$ configurations. 
In contrast, it makes a significant contribution in the $\sigma\sigma$ and $\sigma^+\sigma^-$ field configurations.
For instance, in the $\sigma^+\sigma^-$ configuration, this process almost entirely offsets the influence of atomic spin alignment and orientation induced by pump-beam-driven transitions on the $F_1$ and $F_2$ levels.
Consequently, unlike the $lin\perp lin$ configuration, the primary cause of the sub-Doppler absorption peak in the probe beam in the $\sigma^+\sigma^-$ configuration is the formation of CPT (Coherent Population Trapping) states between the $F_1$ and $F_2$ levels by the pump beam.

Overall, the theory developed here contributes to the field of sub-Doppler laser spectroscopy of atomic gases, a cornerstone of modern quantum metrology and an indispensable tool in numerous precision experiments.
It is worth noting that, unlike many other problems in sub-Doppler spectroscopy where individual nonlinear optical effects are investigated, the scenario considered here involves the simultaneous manifestation and competition of multiple nonlinear effects, owing to the complex configuration of the laser field and the atomic optical transitions involved.
Previous studies on the influence of various nonlinear effects on the formation of the probe-beam absorption peak have examined these effects only in isolation, relying on a simplified three-level ($\Lambda$) atomic scheme \cite{Hafiz2017,Mikhailov2021}.

Beyond its purely academic interest, this work also holds practical significance.
Specifically, the results make it possible to determine the optimal configuration of counterpropagating beams for maximizing the amplitude of the sub-Doppler resonance, a factor crucial for improving the short-term stability of optical frequency standards currently being developed using the frequency-modulated saturated dispersion spectroscopy (DFSDS) method \cite{Brazhnikov2019,Brazhnikov2020,Gusching2021,Zhao2021,Gusching2023}.
\vspace{\baselineskip}

\textbf{Funding.} This work was supported by the Russian Science Foundation (Project Nos.: 22-12-00279, 26-12-00132).

\clearpage

\appendix

\section*{Appendix}

\section{Bloch’s Optical Equations}

\makeatletter
\renewcommand{\theequation}{\thesection.\arabic{equation}}
\makeatother
\setcounter{equation}{0} % Обнуляем счетчик для этого приложения

In the present work, we consider the quantum kinetic equation (\ref{eq:DM1}) in the steady-state regime, where the diagonal matrix blocks $\hat{\rho}_{ii}$ ($i=1, 2, 3$) in formula (\ref{eq:rhoblocks}) are time-independent.
Then, for the spatial Fourier harmonics of the density matrix introduced in (\ref{eq:Harmonics1})–(\ref{eq:Harmonics4}), and taking into account formulas (\ref{eq:H0}), (\ref{eq:V}), and (\ref{eq:Relax}), we obtain the following system of algebraic equations:
\begin{multline}
    \Gamma\hat{\rho}_{11}^{(0)} = \Gamma\hat{\Upsilon}_{11} + \hat{\Gamma}_{31}\bigl\{\hat{\rho}_{33}^{(0)}\bigr\} + \\+
    i\Biggl(R_1\widehat{\tilde{V}}_{13}^{(1)}\hat{\rho}_{31}^{(+1)} + R_3\widehat{\tilde{V}}_{13}^{(3)}\hat{\rho}_{31}^{(-1)} - \\ - R_1\hat{\rho}_{13}^{(-1)}\widehat{\tilde{V}}_{31}^{(1)} - R_3\hat{\rho}_{13}^{(+1)}\widehat{\tilde{V}}_{31}^{(3)}\Biggr),\label{eq:sysro110}
\end{multline}
\begin{multline}
    \Gamma\hat{\rho}_{22}^{(0)} = \Gamma\hat{\Upsilon}_{22} + \hat{\Gamma}_{32}\bigl\{\hat{\rho}_{33}^{(0)}\bigr\} +\\+i\Biggl(R_2\widehat{\tilde{V}}_{23}^{(2)}\hat{\rho}_{32}^{(+2)} + R_4\widehat{\tilde{V}}_{23}^{(4)}\hat{\rho}_{32}^{(-2)} -\\- R_2\hat{\rho}_{23}^{(-2)}\widehat{\tilde{V}}_{32}^{(2)} - R_4\hat{\rho}_{23}^{(+2)}\widehat{\tilde{V}}_{32}^{(4)}\Biggr)\,,\label{eq:sysro220}
    \end{multline}
\begin{multline}
    \bigl(\Gamma+\gamma\bigr)\hat{\rho}_{33}^{(0)} =\\= i\Biggl(R_1\widehat{\tilde{V}}_{31}^{(1)}\hat{\rho}_{13}^{(-1)} + R_2\widehat{\tilde{V}}_{32}^{(2)}\hat{\rho}_{23}^{(-2)} +\\+ R_3\widehat{\tilde{V}}_{31}^{(3)}\hat{\rho}_{13}^{(+1)} + R_4\widehat{\tilde{V}}_{32}^{(4)}\hat{\rho}_{23}^{(+2)} - \\ - R_1\hat{\rho}_{31}^{(+1)}\widehat{\tilde{V}}_{13}^{(1)}  - R_2\hat{\rho}_{32}^{(+2)}\widehat{\tilde{V}}_{23}^{(2)}-\\ - R_3\hat{\rho}_{31}^{(-1)}\widehat{\tilde{V}}_{13}^{(3)} - R_4\hat{\rho}_{32}^{(-2)}\widehat{\tilde{V}}_{23}^{(4)}\Biggr),\label{eq:sysro330}
\end{multline}
\begin{multline}
    \Bigl(\Gamma+2ik_{12}\upsilon\Bigr)\hat{\rho}_{11}^{(+)} = \hat{\Gamma}_{31}\bigl\{\hat{\rho}_{33}^{(+)}\bigr\} +\\+
    i\Biggl(R_3\widehat{\tilde{V}}_{13}^{(3)}\hat{\rho}_{31}^{(-21)} - R_1\hat{\rho}_{13}^{(-21)}\widehat{\tilde{V}}_{31}^{(1)}\Biggr),\label{eq:sysro11p}
    \end{multline}
\begin{multline}
    \Bigl(\Gamma+2ik_{12}\upsilon\Bigr)\hat{\rho}_{22}^{(+)} = \hat{\Gamma}_{32}\bigl\{\hat{\rho}_{33}^{(+)}\bigr\} +\\+
    i\Biggl(R_2\widehat{\tilde{V}}_{23}^{(2)}\hat{\rho}_{32}^{(+12)} - R_4\hat{\rho}_{23}^{(+12)}\widehat{\tilde{V}}_{32}^{(4)}\Biggr),\label{eq:sysro22p}
    \end{multline}
\begin{multline}
    \Bigl(\Gamma+\gamma+2ik_{12}\upsilon\Bigr)\hat{\rho}_{33}^{(+)} =\\= i\Biggl(R_1\widehat{\tilde{V}}_{31}^{(1)}\hat{\rho}_{13}^{(-21)} + R_4\widehat{\tilde{V}}_{32}^{(4)}\hat{\rho}_{23}^{(+12)} -\\- R_2\hat{\rho}_{32}^{(+12)}\widehat{\tilde{V}}_{23}^{(2)} - R_3\hat{\rho}_{31}^{(-21)}\widehat{\tilde{V}}_{13}^{(3)}\Biggr),\label{eq:sysro33p}
    \end{multline}
\begin{multline}
    \Gamma\hat{\rho}_{12}^{(+)} = iN^{(+)}\Biggl(R_1\widehat{\tilde{V}}_{13}^{(1)}\hat{\rho}_{32}^{(+12)}  - R_2\hat{\rho}_{13}^{(-21)}\widehat{\tilde{V}}_{32}^{(2)}+\\+
    R_3\widehat{\tilde{V}}_{13}^{(3)}\hat{\rho}_{32}^{(-2)}e^{i\phi_{12}} - R_4\hat{\rho}_{13}^{(+1)}\widehat{\tilde{V}}_{32}^{(4)}e^{i\phi_{12}}\Biggr),\label{eq:sysro12p}
    \end{multline}
\begin{multline}
    \Gamma\hat{\rho}_{12}^{(-)} = iN^{(-)}\Biggl(R_1\widehat{\tilde{V}}_{13}^{(1)}\hat{\rho}_{32}^{(+2)} - R_2\hat{\rho}_{13}^{(-1)}\widehat{\tilde{V}}_{32}^{(2)} +\\\!+\!
    R_3\widehat{\tilde{V}}_{13}^{(3)}\hat{\rho}_{32}^{(-12)}e^{i\phi_{12}}\! - \!R_4\hat{\rho}_{13}^{(+21)}\widehat{\tilde{V}}_{32}^{(4)}e^{i\phi_{12}}\Biggr).\label{eq:sysro12m}
\end{multline}
The remaining equations for the matrix blocks $\hat{\rho}_{11}^{(-)}$, $\hat{\rho}_{22}^{(-)}$, $\hat{\rho}_{33}^{(-)}$, $\hat{\rho}_{21}^{(-)}$, and $\hat{\rho}_{12}^{(+)}$ can be easily obtained by taking the Hermitian conjugates of equations (\ref{eq:sysro11p})–(\ref{eq:sysro12m}), respectively.
The system employs the notations $\phi_{12} = \phi_1 - \phi_2$ and $k_{12} = k_1 - k_2$. 
The functional operators $\hat{\Gamma}_{31}$ and $\hat{\Gamma}_{32}$ are defined by formula (\ref{eq:SponPrih}), while the Lorentzians $N^{\pm}$ are defined by formula (\ref{eq:NLorentz}). 
The matrices $\hat{\Upsilon}_{ii}$ are proportional to the identity matrix and, using Dirac bra and ket vectors, can be written in the form ($i = 1, 2$):
\begin{equation}\label{eq:Upsilon}
    \hat{\Upsilon}_{ii} = \sum_{m=-F_i}^{F_i}\frac{\big|F_i,m\big\rangle \big\langle F_i,m\big|}{(2J_g+1)(2I_n+1)}\,.
\end{equation}

On the right-hand side of the equations in the system (\ref{eq:sysro110})–(\ref{eq:sysro12m}), various spatial harmonics of optical coherences—i.e., the off-diagonal matrix blocks $\hat{\rho}_{i \ne j}^{(n)}$ are employed.
Just as for equations (\ref{eq:sysro110})–(\ref{eq:sysro12m}), we can derive the following solutions for these matrices from (\ref{eq:DM1}): 
\begin{multline}
        \hat{\rho}_{13}^{(-1)} = iR_1L_1^{(-)}\Bigl(\widehat{\tilde{V}}_{13}^{(1)}\hat{\rho}_{33}^{(0)}-\hat{\rho}_{11}^{(0)}\widehat{\tilde{V}}_{13}^{(1)}\Bigr) -\\- iR_2L_1^{(-)}\hat{\rho}_{12}^{(-)}\widehat{\tilde{V}}_{23}^{(2)},\label{eq:sysro13m1}
\end{multline}
\begin{multline}
        \hat{\rho}_{13}^{(+1)} = iR_3L_1^{(+)}\Bigl(\widehat{\tilde{V}}_{13}^{(3)}\hat{\rho}_{33}^{(0)}-\hat{\rho}_{11}^{(0)}\widehat{\tilde{V}}_{13}^{(3)}\Bigr) - \\-iR_4L_1^{(+)}\hat{\rho}_{12}^{(+)}\widehat{\tilde{V}}_{23}^{(4)}e^{-i\phi_{12}},\label{eq:sysro13p1}
\end{multline}
\begin{multline}
        \hat{\rho}_{13}^{(-21)} = iR_1M_1^{(-)}\Bigl(\widehat{\tilde{V}}_{13}^{(1)}\hat{\rho}_{33}^{(+)}-\hat{\rho}_{11}^{(+)}\widehat{\tilde{V}}_{13}^{(1)}\Bigr) -\\- iR_2M_1^{(-)}\hat{\rho}_{12}^{(+)}\widehat{\tilde{V}}_{23}^{(2)},\label{eq:sysro13m21}
\end{multline}
\begin{multline}
        \hat{\rho}_{13}^{(+21)} = iR_3M_1^{(+)}\Bigl(\widehat{\tilde{V}}_{13}^{(3)}\hat{\rho}_{33}^{(-)}-\hat{\rho}_{11}^{(-)}\widehat{\tilde{V}}_{13}^{(3)}\Bigr) -\\-iR_4M_1^{(+)}\hat{\rho}_{12}^{(-)}\widehat{\tilde{V}}_{23}^{(4)}e^{-i\phi_{12}},\label{eq:sysro13p21}
\end{multline}
\begin{multline}
        \hat{\rho}_{23}^{(-2)} = iR_2L_2^{(-)}\Bigl(\widehat{\tilde{V}}_{23}^{(2)}\hat{\rho}_{33}^{(0)}-\hat{\rho}_{22}^{(0)}\widehat{\tilde{V}}_{23}^{(2)}\Bigr) -\\- iR_1L_2^{(-)}\hat{\rho}_{21}^{(+)}\widehat{\tilde{V}}_{13}^{(1)},\label{eq:sysro23m2}
\end{multline}
\begin{multline}
        \hat{\rho}_{23}^{(+2)} = iR_4L_2^{(+)}\Bigl(\widehat{\tilde{V}}_{23}^{(4)}\hat{\rho}_{33}^{(0)}-\hat{\rho}_{22}^{(0)}\widehat{\tilde{V}}_{23}^{(4)}\Bigr) -\\- iR_3L_2^{(+)}\hat{\rho}_{21}^{(-)}\widehat{\tilde{V}}_{13}^{(3)}e^{i\phi_{12}},\label{eq:sysro23p2}
\end{multline}
\begin{multline}
        \hat{\rho}_{23}^{(-12)} = iR_2M_2^{(-)}\Bigl(\widehat{\tilde{V}}_{23}^{(2)}\hat{\rho}_{33}^{(-)}-\hat{\rho}_{22}^{(-)}\widehat{\tilde{V}}_{23}^{(2)}\Bigr) -\\- iR_1M_2^{(-)}\hat{\rho}_{21}^{(-)}\widehat{\tilde{V}}_{13}^{(1)},\label{eq:sysro23m12}
\end{multline}
\begin{multline}
        \hat{\rho}_{23}^{(+12)} = iR_4M_2^{(+)}\Bigl(\widehat{\tilde{V}}_{23}^{(4)}\hat{\rho}_{33}^{(+)}-\hat{\rho}_{22}^{(+)}\widehat{\tilde{V}}_{23}^{(4)}\Bigr) -\\- iR_3M_2^{(+)}\hat{\rho}_{21}^{(+)}\widehat{\tilde{V}}_{13}^{(3)}e^{i\phi_{12}}.\label{eq:sysro23p12}
\end{multline}
The Lorentzians $L_j^{(\pm)}$ and $M_j^{(\pm)}$ used in these expressions are defined by formulas (\ref{eq:lorenziansL}) and (\ref{eq:MLorentz}), respectively.

It should be noted that expressions (\ref{eq:sysro110})–(\ref{eq:sysro12m}) account for the influence of not only the pump beam (Rabi frequencies $R_3$ and $R_4$) but also the probe beam (Rabi frequencies $R_1$ and $R_2$) on the populations of Zeeman sublevels, Zeeman coherences, and microwave coherences.
However, when considering the linear regime with respect to the probe beam, as is done in the present work, all terms in (\ref{eq:sysro110})–(\ref{eq:sysro12m}) containing $R_1$ and $R_2$ should be set to zero.

\clearpage

\section{Values of the Coefficients}

\makeatletter
\renewcommand{\theequation}{\thesection.\arabic{equation}}
\renewcommand{\thetable}{\thesection.\arabic{table}}
\makeatother
\setcounter{equation}{0} % Обнуляем счетчик для этого приложения
\setcounter{table}{0}

\begin{table}[ht]
\centering
\renewcommand{\arraystretch}{1.4}
\setlength{\tabcolsep}{6pt}
\caption{Relative transition strengths in the D$_1$ line of the $^{87}$Rb atom ($I_n = 3/2$).}
\label{tab:1}
\centering
\begin{tabularx}{\linewidth}{||Y|Y||}
 \hline
 Transition & Strength \\ [0.5ex] 
 \hline\hline
 $F_1$$\,=\,1$$\,\to\,$$F_3$$\,=\,$$1$ & $1/16$ \\ [0.5ex]
 \hline
  $F_1$$\,=\,1$$\,\to\,$$F_3$$\,=\,$$2$ & $5/16$ \\ [0.5ex]
 \hline
  $F_2$$\,=\,2$$\,\to\,$$F_3$$\,=\,$$1$ & $5/16$ \\ [0.5ex]
 \hline
 $F_2$$\,=\,2$$\,\to\,$$F_3$$\,=\,$$2$ & $5/16$ \\ [1ex] 
 \hline
\end{tabularx}
\end{table}

\begin{table}[ht]
\centering
\renewcommand{\arraystretch}{1.4}
\setlength{\tabcolsep}{6pt}
\caption{Relative transition strengths in the D$_1$ line of the $^{133}$Cs atom ($I_n = 7/2$).}
\label{tab:2}
\centering
\begin{tabularx}{\linewidth}{||Y|Y||} 
 \hline
 Transition & Strength \\ [0.5ex] 
 \hline\hline
 $F_1$$\,=\,3$$\,\to\,$$F_3$$\,=\,$$3$ & $7/64$ \\ [0.5ex]
 \hline
  $F_1$$\,=\,3$$\,\to\,$$F_3$$\,=\,$$4$ & $21/64$ \\ [0.5ex]
 \hline
  $F_2$$\,=\,4$$\,\to\,$$F_3$$\,=\,$$3$ & $21/64$ \\ [0.5ex]
 \hline
 $F_2$$\,=\,4$$\,\to\,$$F_3$$\,=\,$$4$ & $15/64$ \\ [1ex] 
 \hline
\end{tabularx}
\end{table}

Tables \ref{tab:1} and \ref{tab:2} present the relative strengths of the $F_{1,2} \to F_3$ transitions in the D$_1$ line of $^{87}$Rb and $^{133}$Cs atoms, calculated using formula (\ref{eg:strength}).

Using the formulas from \cite{Varshalovich}, it can be shown that the coefficients $\Psi_{12}(\varkappa)$ take the following form in an invariant representation:
\begin{multline}
    \Psi_{12}(\varkappa) = -\frac{\delta_{\varkappa 0}}{\sqrt{3}} + \frac{\delta_{\varkappa 1}}{2}\bigl[|\xiB_{(1)}\xiB_{(2)}|^2 - |\xiB_{(1)} \xiB_{(2)}^*|^2\bigr] +\\+
    \frac{\delta_{\varkappa 2}}{2\sqrt{15}}\bigl[2-3|\xiB_{(1)}\xiB_{(2)}^*|^2-3|\xiB_{(1)}\xiB_{(2)}|^2\bigr].
\end{multline}

Based on the light polarization parametrization introduced in Fig.~\ref{fig:2} and taking into account expressions (\ref{eq:xi2}), we arrive at explicit expressions:
\begin{multline}
    |\xiB_{(1)}\xiB_{(2)}|^2 = \sin^2(\alpha)\sin^2(\epsilon_1-\epsilon_2) +\\+ \cos^2(\alpha)\cos^2(\epsilon_1+\epsilon_2),
\end{multline}
\begin{multline}
    |\xiB_{(1)}\xiB_{(2)}^*|^2 = \sin^2(\alpha)\sin^2(\epsilon_1+\epsilon_2) +\\+ \cos^2(\alpha)\cos^2(\epsilon_1-\epsilon_2).
\end{multline}

Table \ref{tab:3} lists the values of the coefficients $\Psi_{12}(\varkappa)$ responsible for the polarization features of the sub-Doppler resonance. The values are presented for light-field polarization configurations commonly used in spectroscopic experiments.

Tables \ref{tab:5},\ref{tab:6},\ref{tab:4},\ref{tab:7} list the values of the coefficients $A_j$, $B_{ij}$, and $C$ used to formulate the expressions for the absorption coefficients (\ref{eq:chiinter}) and (\ref{eq:chi1nlin1SAR3Scal})--(\ref{eq:chi2nlin1OP3Ten}).
\begin{table}[thb]
\centering
\renewcommand{\arraystretch}{1.4}
\setlength{\tabcolsep}{6pt}
\caption{Coefficients $B_{ij}(\varkappa)$ for various values of the excited-state angular momentum $F_3$ in the D$_1$ line of the $^{87}$Rb atom.}
\label{tab:5}
\centering
\begin{tabularx}{\linewidth}{||Y|Y|Y||} 
 \hline
 Coefficient & $^{87}$Rb & $^{87}$Rb \\
 {} & $F_3$$\,=\,$$1$ & $F_3$$\,=\,$$2$ \\ [0.5ex] 
 \hline\hline
 $B_{11}(0)$ & $-\frac{1}{27}$ & $-\frac{1}{45}$ \\ [0.5ex]
 \hline
  $B_{11}(1)$ & $\frac{1}{72\sqrt{3}}$ & $-\frac{1}{40\sqrt{3}}$ \\ [0.5ex]
 \hline
 $B_{11}(2)$ & $\frac{\sqrt{5}}{216}$ & $-\frac{7}{1800\sqrt{5}}$ \\ [0.5ex]
 \hline
  $B_{12}(0)$ & $-\frac{1}{27}$ & $-\frac{1}{45}$ \\ [0.5ex]
 \hline
 $B_{12}(1)$ & $-\frac{1}{72\sqrt{3}}$ & $-\frac{1}{120\sqrt{3}}$ \\ [0.5ex]
 \hline
 $B_{12}(2)$ & $-\frac{1}{216\sqrt{5}}$ & $\frac{7}{1800\sqrt{5}}$ \\ [0.5ex]
 \hline
 $B_{21}(0)$ & $-\frac{1}{45}$ & $-\frac{1}{75}$ \\ [0.5ex]
 \hline
 $B_{21}(1)$ & $\frac{1}{40\sqrt{3}}$ & $\frac{1}{120\sqrt{3}}$ \\ [0.5ex]
 \hline
 $B_{21}(2)$ & $\frac{7}{360\sqrt{5}}$ & $\frac{7}{600\sqrt{5}}$ \\ [0.5ex]
 \hline
 $B_{22}(0)$ & $-\frac{1}{45}$ & $-\frac{1}{75}$ \\ [0.5ex]
 \hline
 $B_{22}(1)$ & $-\frac{1}{40\sqrt{3}}$ & $\frac{1}{360\sqrt{3}}$ \\ [0.5ex]
 \hline
 $B_{22}(2)$ & $-\frac{7}{1800\sqrt{5}}$ & $-\frac{7}{600\sqrt{5}}$ \\ [0.5ex] 
 \hline
\end{tabularx}
\centering
\renewcommand{\arraystretch}{1.4}
\setlength{\tabcolsep}{6pt}
\caption{Coefficients $B_{ij}(\varkappa)$ for various values of the excited-state angular momentum $F_3$ in the D$_1$ line of the $^{133}$Cs atom.}
\label{tab:6}
\centering
\begin{tabularx}{\linewidth}{||Y|Y|Y||} 
 \hline
 Coefficient & $^{133}$Cs & $^{133}$Cs \\
 {} & $F_3$$\,=\,$$3$ & $F_3$$\,=\,$$4$
 \\ [0.5ex] 
 \hline\hline
 $B_{11}(0)$ & $-\frac{1}{147}$ & $-\frac{1}{189}$ \\ [0.5ex]
 \hline
  $B_{11}(1)$ & $\frac{11}{14122\sqrt{3}}$ & $-\frac{5}{672\sqrt{3}}$ \\ [0.5ex]
 \hline
 $B_{11}(2)$ & $-\frac{3\sqrt{5}}{1568}$ & $-\frac{55\sqrt{5}}{127008}$ \\ [0.5ex]
 \hline
  $B_{12}(0)$ & $-\frac{1}{147}$ & $-\frac{1}{189}$ \\ [0.5ex]
 \hline
 $B_{12}(1)$ & $-\frac{11}{4704\sqrt{3}}$ & $-\frac{1}{672\sqrt{3}}$ \\ [0.5ex]
 \hline
 $B_{12}(2)$ & $\frac{\sqrt{5}}{1568}$ & $\frac{11\sqrt{5}}{18144}$ \\ [0.5ex]
 \hline
 $B_{21}(0)$ & $-\frac{1}{189}$ & $-\frac{1}{243}$ \\ [0.5ex]
 \hline
 $B_{21}(1)$ & $\frac{5}{2016\sqrt{3}}$ & $\frac{19}{12960\sqrt{3}}$ \\ [0.5ex]
 \hline
 $B_{21}(2)$ & $\frac{55\sqrt{5}}{42336}$ & $\frac{187}{38880\sqrt{5}}$ \\ [0.5ex]
 \hline
 $B_{22}(0)$ & $-\frac{1}{189}$ & $-\frac{1}{243}$ \\ [0.5ex]
 \hline
 $B_{22}(1)$ & $-\frac{5}{672\sqrt{3}}$ & $\frac{19}{64800\sqrt{3}}$ \\ [0.5ex]
 \hline
 $B_{22}(2)$ & $-\frac{55\sqrt{5}}{127008}$ & $-\frac{1309}{194400\sqrt{5}}$
 \\ [0.5ex] 
 \hline
\end{tabularx}
\end{table}

\clearpage

\begin{table*}[t!]
\centering
\renewcommand{\arraystretch}{1.5}
\setlength{\tabcolsep}{6pt}
\caption{Coefficients $A_{ij}^2(\varkappa)$ appearing in formulas (\ref{eq:chi1nlin1SAR2}) and (\ref{eq:chi2nlin1SAR2}) for various values of the excited-state angular momentum $F_3$ in the D$_1$ line of $^{87}$Rb and $^{133}$Cs atoms.}
\label{tab:4}
\centering
\begin{tabularx}{\linewidth}{||c|Y|Y|Y|Y||} 
 \hline
 Coefficient & $^{87}$Rb & $^{87}$Rb & $^{133}$Cs & $^{133}$Cs \\
 {} & $F_3$$\,=\,$$1$ & $F_3$$\,=\,$$2$ & $F_3$$\,=\,$$3$ & $F_3$$\,=\,$$4$
 \\ [0.5ex] 
 \hline\hline
 $A_{13}^2(0)$ & $\frac{1}{9}$ & $\frac{1}{9}$ & $\frac{1}{21}$ & $\frac{1}{21}$ \\ [0.5ex]
 \hline
  $A_{13}^2(1)$ & $\frac{1}{12\sqrt{3}}$ & $\frac{1}{12\sqrt{3}}$ & $\frac{1}{168\sqrt{3}}$ & $\frac{\sqrt{3}}{56}$ \\ [0.5ex]
 \hline
 $A_{13}^2(2)$ & $\frac{\sqrt{5}}{36}$ & $\frac{1}{180\sqrt{5}}$ & $\frac{\sqrt{5}}{56}$ & $\frac{\sqrt{5}}{504}$\\ [0.5ex]
 \hline
  $A_{23}^2(0)$ & $\frac{1}{15}$ & $\frac{1}{15}$ & $\frac{1}{27}$ & $\frac{1}{27}$ \\ [0.5ex]
 \hline
 $A_{23}^2(1)$ & $\frac{\sqrt{3}}{20}$ & $\frac{1}{60\sqrt{3}}$ & $\frac{5}{72\sqrt{3}}$ & $\frac{1}{360\sqrt{3}}$ \\ [0.5ex]
 \hline
 $A_{23}^2(2)$ & $\frac{7}{60\sqrt{5}}$ & $\frac{7}{60\sqrt{5}}$ & $\frac{11\sqrt{5}}{1512}$ & $\frac{77}{1080\sqrt{5}}$
 \\ [0.5ex] 
 \hline
\end{tabularx}
\centering
\renewcommand{\arraystretch}{1.5}
\setlength{\tabcolsep}{6pt}
\caption{Coefficients $C(\varkappa)$ for various values of the angular momentum of the excited state $F_3$ in the D$_1$ line of $^{87}$Rb and $^{133}$Cs atoms.}
\label{tab:7}
\centering
\begin{tabularx}{\linewidth}{||c|Y|Y|Y|Y||} 
 \hline
 Coefficient & $^{87}$Rb & $^{87}$Rb & $^{133}$Cs & $^{133}$Cs \\
 {} & $F_3$$\,=\,$$1$ & $F_3$$\,=\,$$2$ & $F_3$$\,=\,$$3$ & $F_3$$\,=\,$$4$
 \\ [0.5ex] 
 \hline\hline
 $C(0)$ & 0 & 0 & 0 & 0 \\ [0.5ex]
 \hline
  $C(1)$ & $\frac{1}{12\sqrt{3}}$ & $\frac{\sqrt{3}}{20}$ & $\frac{\sqrt{3}}{56}$ & $\frac{5}{72\sqrt{3}}$ \\ [0.5ex]
 \hline
 $C(2)$ & $\frac{1}{4\sqrt{5}}$ & $\frac{1}{20\sqrt{5}}$ & $\frac{\sqrt{5}}{56}$ & $\frac{1}{24\sqrt{5}}$ \\ [0.5ex] 
 \hline
\end{tabularx}
\centering
\renewcommand{\arraystretch}{1.5}
\setlength{\tabcolsep}{6pt}
\caption{Coefficients $\Psi_{12}(\varkappa)$ for simple cases of counterpropagating light beam polarizations (independent of the coordinate system).}
\label{tab:3}
\centering
\begin{tabularx}{\linewidth}{||c|Y|Y|Y||} 
 \hline
 Configuration & $\Psi_{12}(0)$ & $\Psi_{12}(1)$ & $\Psi_{12}(2)$ \\ [0.5ex] 
 \hline\hline
 {\it lin}$\,||\,${\it lin} & $-\frac{1}{\sqrt{3}}$ & 0 & $-\frac{2}{\sqrt{15}}$ \\ [0.5ex]
 \hline
  {\it lin}$\,\perp\,${\it lin} & $-\frac{1}{\sqrt{3}}$ & 0 & $\frac{1}{\sqrt{15}}$ \\ [0.5ex]
 \hline
  $\sigma$$\sigma$ & $-\frac{1}{\sqrt{3}}$ & $-\frac{1}{2}$ & $-\frac{1}{2\sqrt{15}}$ \\ [0.5ex]
 \hline
 $\sigma^+$$\sigma^-$ & $-\frac{1}{\sqrt{3}}$ & $\frac{1}{2}$ & $-\frac{1}{2\sqrt{15}}$ \\ [1ex] 
 \hline
\end{tabularx}
\end{table*}

\end{document}